\documentclass{jaa}
\usepackage{natbib}
\usepackage{caption}

\usepackage{graphicx}
\usepackage{mathtools}
\usepackage{tabularx}
\usepackage[colorlinks=true, linkcolor=blue, citecolor=blue, urlcolor=blue]{hyperref}
\usepackage{makecell}

\newcounter{rowcount}

\newcommand\swift{{\it Swift}}

\newcommand\xte{{\it RXTE}}

\newcommand\maxi{{\it MAXI}}

\begin{document}\sloppy

%%paper title
%%For line breaks \\ can be used within title
\title{Black hole-Neutron star distinction based on long-term {\it MAXI} and {\it Swift} study of 42 low mass X-ray binaries}

\author{Prakhar Maheshwari\textsuperscript{1}, Mayukh Pahari\textsuperscript{2}, Anish Sarkar\textsuperscript{2}, Saurabh Sharma\textsuperscript{2}}
\affilOne{\textsuperscript{1}Department of Physics, Indian Institute of Technology, Madras, 600036, India.\\}
\affilTwo{\textsuperscript{2}Department of Physics, Indian Institute of Technology, Hyderabad, 502285, India.}

\twocolumn[{

\maketitle

%%include \corres to print the corresponding author Email id
\corres{mayukh@phy.iith.ac.in}

%%include \msinfo for
%%manuscript information such as
%%received, revised and accepted dates
%%
%\msinfo{22 November 2024}{22 November 2024}

%%abstract
\begin{abstract}
In this study, we analysed about $\sim$13 years of publicly available data from \maxi{} and \swift{}/BAT to examine the long-term source evolution of 42 transient low-mass X-ray binaries.  The sample consists of 11 confirmed black hole X-ray binaries (BHXBs), 10 black hole candidates (BHC), and 21 neutron star X-ray binaries (NSXBs). Outbursts and flaring activities studied over 13 years show that 19/21 NSXBs spend significantly longer time in the hard state (observations for which hardness ratio is $\geq$ 0.2) while 15/21 BHXB+XRC spend substantially longer time in the soft state (observations for which hardness ratio is $<$ 0.2). The frequency distribution of the hardness ratio clearly shows two distinct distributions for BHXBs and NSXBs, with their peaks separated: NSXBs prefer harder values, while BHXBs prefer softer values of hardness. Our model-independent analysis for 42 transient sources shows that statistically NSXBs do not prefer to show a canonical high soft state as observed in BHXBs. Additionally, the probability distribution of the duration of the 2-20 keV X-ray outburst is observed to peak at a significantly longer duration ($>$100 days) for BHXBs than for NSXBs (15-60 days). 

Our analysis shows that among candidate sources, Swift J1728.9-3613, MAXI J1535-571, MAXI J1659-152, EXO 1846-031 show a `q' diagram in the HID and prefer to align with the HID frequency distribution of BHXBs that show `q' diagram, MAXI J1305-704 and MAXI J1836-194 align with frequency distribution of black hole sources without `q' diagram while MAXI J1848-015 shows the HID distribution similar to NSXBs, indicating a neutron star accretor. Therefore, a long-term statistical study of \maxi{} and \swift{}/BAT X-ray outbursts from a large sample of transient sources may be used to distinguish BHXB from NSXB.   
\end{abstract}


}]

\setcounter{page}{1}
\lp{36}

\section{Introduction}
X-ray binaries can be broadly classified into two types: High-mass X-ray binaries (HMXBs) and low-mass X-ray binaries (LMXBs). LMXBs emit most of their radiation in the X-ray, and companion stars have low masses and are significantly faint in the V band \citep{remillard2006}. They host either a black hole or a neutron star, and their emission primarily comes from the accretion disk. HMXBs, on the other hand, are usually brighter in hard X-rays and usually host a neutron star at its centre.
Among a few methods proposed for distinguishing black holes from neutron stars binaries, a potential one is to compare their power-law indices when the mean spectra are fitted with powerlaw model during the soft state, where the quasi-thermal, soft component dominates their luminosity ~\citep{titarchuk1998extended}. According to this approach, black hole systems are expected to show a distinct high-energy tail in their X-ray spectra, typically represented by a steep power-law component which can extend up to a few hundred keV ~\citep{gierlinski2003xtej1550}. Such a steep powerlaw may originate from the bulk motion Comptonization process, where soft X-ray photons gain energy by scattering off rapidly infalling material ~\citep{ebisawa1996spectral}. In contrast, neutron star systems are not expected to show such a high-energy tail. This is because the intense radiation pressure from the neutron star’s solid surface counteracts the infalling matter, significantly reducing the speed of the bulk motion and thereby suppressing the associated Compton upscattering of photons~\citep{farinelli2007hardtails}. Thus, the presence or absence of a steep power-law component in X-ray spectra is suggested to be associated with the absence or presence of a hard surface, respectively ~\citep{banerjee2020imprints}.

Neutron Star Soft X-ray Transients (NSSXTs) can also be distinguished from Black Hole Soft X-ray transients (BHSXTs) by analysing their respective luminosities \citep{menou1999black}. \citet{narayan97,narayan02,garcia01} suggested that in quiescence, BH accretes via radiatively inefficient advection-dominated accretion flow (ADAF), which is not the case for NSs; therefore, NS binaries, with their radiating boundary layers, are much more luminous at the same accretion rate. Observations confirm NSs are brighter, but the luminosity gap is far smaller than ADAF predictions \citep{menou1999black}. However, the low-hard spectral state cannot be uniquely associated with BHXBs. Interestingly, when observed in low luminosity states, NSXBs exhibit spectra that are remarkably similar to the low-hard state of BHXBs. The same  has been well demonstrated through observations of X-ray bursters during periods of low intensity ~\citep{titarchuk2000black}. 

Early spectral studies \citet{shrader98,shrader99,shrader03,borozdin99} identified a BH signature in the high soft state: $\sim$1 keV thermal emission along with a steep power-law tail (photon index 2.5$-$2.7), consistent with predictions from radiative transfer models of disk photon Comptonization in converging flows \citep{chakrabarti95,titarchuk97,titarchuk98,laurent99,laurent01}. \citet{done03}, analysing extensive \xte{}/PCA data, further showed that spectral evolution differs markedly between BHs and low magnetic field NSs, and reported a BH-specific bright-state spectrum.  The observed spectral state, dominated by cool accretion disk emission with a steep high-energy power-law tail, is proposed as a unique BH diagnostic. Using relation between the Compton y parameter and electron temperature (kT$_e$) deduced from \textit{RXTE} analysis, \citet{ban20} showed that NS and BH occupy different regions in the y-kT$_e$ plane.

Attempts were made to distinguish NS from BH systems based on power spectral properties. \citet{sunyaev2000fourier,ti05} show that the PDS of NSXBs had significant broad noise components at frequencies 0.5--1 kHz, sometimes coupled with well-defined kHz quasi-periodic oscillations (QPOs). However, BHXBs showed a rapid decrease in the noise power during a low-hard state for frequencies above 10 Hz \citep{kl08}. Such a difference indicates that any source that demonstrates significant variability (continuum or QPO) at frequencies close to 1 kHz could be considered an NS. However, the occurrence of kHz QPOs is unpredictable.

A neutron star system can also be potentially identified using a temperature-luminosity relation. A low bolometric luminosity and a high blackbody temperature show more evidence of a neutron star rather than a black hole ~\citep{torrejon2004evidence}. The rate of pair production, which depends on the luminosity and the dimensions of the region in question, can also be used to determine the nature of the compact object ~\citep{titarchuk2006observed}.
While several spectral and timing features have been proposed to distinguish neutron stars from black holes in X-ray binaries-including power-law tails ~\citep{sunyaev2000fourier}, QPOs ~\citep{psaltis1999qpo}, and temperature-luminosity relations$-$these diagnostics often rely on complex spectral modelling or the detection of transient features, which may not always be present or conclusive~\citep{titarchuk2005distinguish}. Additionally, degeneracies in spectral signatures between black hole and neutron star systems, especially in their low-hard states, make it difficult to uniquely classify them based solely on short-term or model-dependent characteristics \citep{pszota2024highenergy, burke2016dichotomy}.

What remains lacking is a more accessible, model-independent, long-term approach that can offer complementary insight into the nature of the compact object. In particular, the statistical behaviour of sources over long timescales, such as the time they spend in different spectral states, has not been systematically exploited for this purpose across a large sample. In this work, we aim to fill this gap by using about 13 years of publicly available data to systematically analyse the long-term variability patterns and spectral state transitions of 42 low-mass X-ray binaries (LMXBs), comprising both confirmed and candidate neutron stars and black holes. 
The largest sample and regular monitoring data in both soft (2-4 keV) and hard bands (10-20 keV) can only be obtained from the \maxi{} survey. Therefore, we restrict ourselves to \maxi{} observations only. Since \maxi{} is still operational, any outburst in the future will be uniformly monitored by \maxi{}, and if the source nature is not known, an attempt can be made to identify it by using the proposed method.
We focus on the use of X-ray light curves and Hardness-Intensity Diagrams (HIDs) to quantify state occupation times and outburst properties in a model-independent manner.

Data reduction, selection criteria, and analysis method are discussed in section \ref{MAXI data reduction} while analysis and result of individual sources are provided in the section \ref{Analysis of X-ray Black-Hole Binary Sources}, \ref{Analysis of Neutron Star Sources}, \textbf{Appendix A} and \textbf{Appendix B}.

\begin{figure*}[!ht]
    \centering
    \includegraphics[scale=0.60]{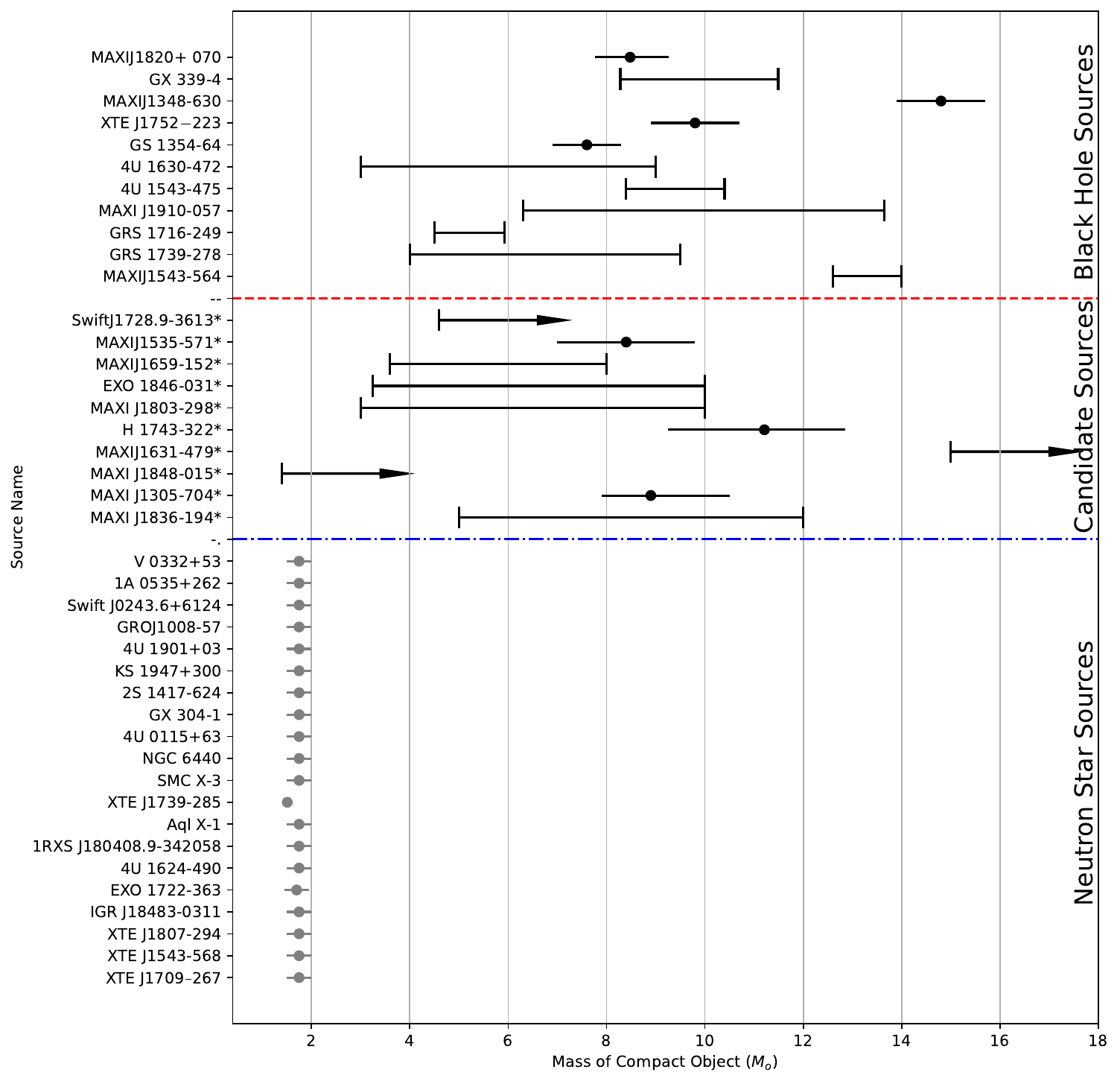}
    \caption{ {\it Source name vs mass of compact objects}. Top panel: Black holes. Middle panel: Candidate sources. Bottom panel: Neutron stars. References for masses of individual objects are provided in the respective sections. }
    \label{fig: Masses of Compact Object}
\end{figure*}

\section{Data reduction and analysis}
\label{MAXI data reduction}
The study employed the \href{http://maxi.riken.jp/top/lc.html}{MAXI} (Monitor of All-sky X-ray Image) and \href{https://swift.gsfc.nasa.gov/results/transients/}{Swift/BAT Hard X-ray Transient monitor} data for the purpose of generating light curves and conducting subsequent analysis. The sample was chosen such that it contains sources consisting of at least one outburst during the entire observational period of \maxi{}. 

\subsection{MAXI data reduction}

\maxi{} provides X-ray lightcurves in multiple energy bands: 2--20 KeV, 2--4 keV, 4--10 KeV and 10--20 KeV. Additionally, \maxi{} offers count rate data in two temporal resolutions: one-day bin and one-orbit bin and the data is updated daily. For this work, the one-day bin data have been utilised to generate the light curves. The uncertainty on the count rate of each band is minimal for one-day-averaged data. All \maxi{} data are reduced using the latest pipeline (V7L).

\subsection{Swift BAT data reduction}

\swift{}/BAT provide X-ray count rates within the 15--50 KeV band from MJD 53414 to present. Again, the data is available in two bin sizes: one-day bins and one-orbit bins. The one-day bins have been used for the present work. Both statistical and systematic errors are taken into account. For detailed data analysis and uncertainty measurements, see \citet{krimm2013swift}.

\subsection{Data selection and analysis}
A list of transient BHXBs and NSXBs chosen for the present work are shown in Table \ref{black_hole_details} and Table \ref{neutron_star_details}, while their estimated masses are provided in Figure \ref{fig: Masses of Compact Object}. The selection criteria for choosing the transient sample are that at least one outburst is present during the entire \maxi{} observation period. We found 42 sources fulfilling the criteria.
The light curves were generated from the \maxi{} data in the energy bands of 2--4 KeV (soft band) and 10--20 KeV (hard band), while the same were generated from the \swift{}/BAT data in the 15--50 KeV band for the chosen sample. Count rate data points which are negative or have uncertainty equal to or larger than the count rate have been removed from the analysis in each energy band for accurate analysis. For all analyses, we have used a one-day averaged X-ray count rate. The hardness ratio is defined as the count rate ratio of the hard band to the soft band obtained from \maxi{} data. A hard state is defined by a hardness value higher than 0.2, while a soft state is defined as a hardness value less than 0.2. The hardness boundary is chosen such that it is consistent with the hard and soft state identification of confirmed black hole X-ray binaries in their `q' diagram \citep{fender2004,remillard2006,homan2005}.

For better visibility in outburst lightcurves and easy comparison with \maxi{} data, the \swift{}/BAT count rate is multiplied by different factors for each source depending on \maxi{} count rate. It is important to note that our work does not use the absolute count rates of each source for the interpretation of the results. Rather, we are interested in count rate variations, highlighting features like outburst peaks, the time of occurrences, and total outburst duration. Therefore, our conclusions are independent of the instruments used but depend on the X-ray bands used to compute HID. For better visibility in the lightcurves, the 10-20 keV band \maxi{} count rate has been multiplied by a factor to highlight the hard X-ray features. We note that all analyses and results presented here are performed using the original count rate from the respective instruments.

All left (for two-panel figure) and top left (for four-panel figure) panels from Fig. \ref{fig:MAXI J1820+070} to Fig.\ref{fig:KS 1947+300}, show the \maxi{} soft and hard band lightcurve as well \swift{}/BAT lightcurve if available. If multiple outbursts are observed, then zoomed lightcurves to a couple of outbursts and one single outburst are provided in the top right and bottom left (for four-panel figure) panels of Fig. \ref{fig:GX 339-4} to Fig. \ref{fig:KS 1947+300}. Similar notations have been used in \textbf{Appendix A} and \textbf{Appendix B}.
The hollow circles in the light curves represent the soft band, while the solid circles denote the hard band count rates obtained from MAXI. \swift{}/BAT data, wherever available, are shown by stars. 
To find the error in the Hardness ratio, we have used $ \delta(\frac{x}{y}) = \sqrt{(\frac{\delta x}{x})^2 + (\frac{\delta y}{y})^2}\times\frac{x}{y}$ where x, $\delta x$, y and $\delta y$ represents the soft band count rate and its error while y and $\delta y$ represents hard band count rate and its error. The HID values for which errors are $<2\sigma$ are not taken into account for further analysis. The right panel (for a two-panel figure) and bottom right panel (for a four-panel figure) from Fig. \ref{fig:MAXI J1820+070} to Fig. \ref{fig:KS 1947+300} show the HID with \maxi{} data. The soft state in the HID is represented by inverted triangles, while the hard state is represented by grey triangles. Similar notations have been used in \textbf{Appendix A} and \textbf{Appendix B}.
 
To ensure a high signal-to-noise ratio, we have removed lightcurve data points with the \maxi{} X-ray intensity below 0.05 photon $cm^2 s^{-1}$ in the 2--20 KeV band. Such a value is consistent with the \maxi{} background count rate. The HID is calculated with soft and hard X-ray intensity values for which at least 2$\sigma$ error is constrained. Since the data points have a one-day bin size, a count of the number of data points in the hard state of the HID gives us a count of the number of days the source spends in the hard state and vice versa for the soft state. A count of the number of days spent in the hard state vs those spent in the soft state can be found in Table \ref{black_hole_details} and \ref{neutron_star_details} for BH and NS sources, respectively. Below, we provide analysis and results from individual sources.

\section{Analysis of Black-Hole X-ray Binary Sources}
\label{Analysis of X-ray Black-Hole Binary Sources}
This section presents an analysis of five black hole X-ray binaries. Discussions of the remaining 16 sources are provided in Appendix A
\subsection{MAXI J1820+070}
\begin{figure*}
    \centering
    \includegraphics[scale=0.60]{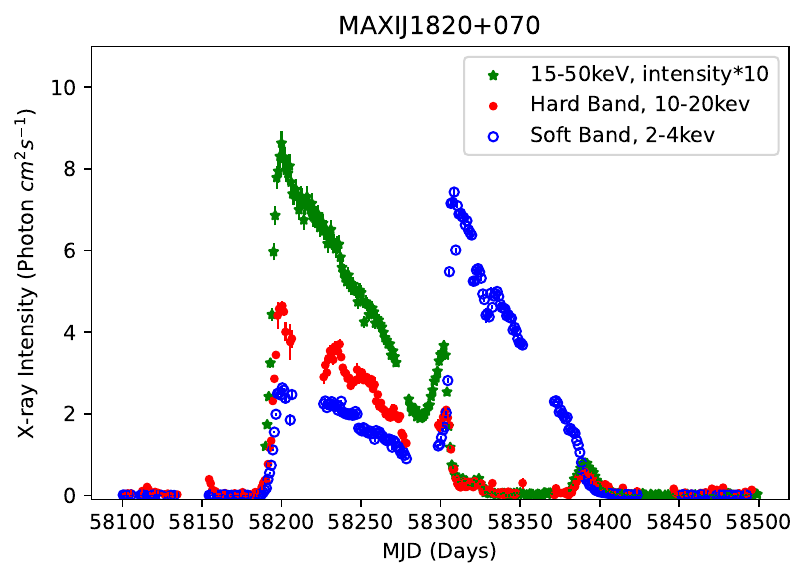}
    \includegraphics[scale=0.60]{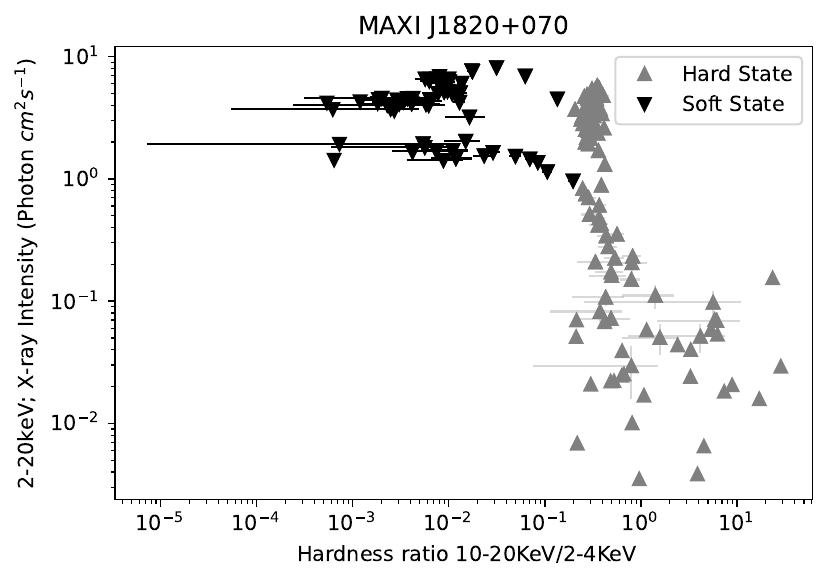}
    \caption{{\it Lightcurve and HID of MAXI J1820+070} Left panel: The only outburst of MAXI J1820+070 in 2018 observed with \maxi{} are shown in the 2-4 keV soft band (blue open circles), 10-20 keV hard bands (red solid circles) respectively. The same outburst as observed with \swift{}/BAT in 15-50 keV is shown using green stars. A multiplication factor of 10 is used for plotting \swift{} lightcurve for easy comparison. Right panel: Hardness Intensity Diagram (HID) of the same outburst is shown where hardness ratio is defined as the ratio of X-ray intensity in 10-20 keV and 2-4 keV while the intensity is defined as 2-20 keV count rate. Grey triangles denote hard state (when Hardness ratio $>$ 0.2) while black inverted triangles denote soft state (when Hardness ratio $<$ 0.2).}
    \label{fig:MAXI J1820+070}
\end{figure*}

The black hole mass is estimated within a range of 5.73-8.34 $M_\odot$ \citep{torres2020binary}.
The light curve provided in Fig. \ref{fig:MAXI J1820+070} indicates an outburst in the hard band and another in the soft band separated by $\sim$50 days. We can say that the outburst transitioned from the hard band at the inner disc to the soft band in the outer disc during two-month intervals. The HID in the right panel of Fig. \ref{fig:MAXI J1820+070} shows the typical q-diagram observed from BHXB HIDs.

\subsection{GX 339-4}  
\begin{figure*}
    \centering
    \includegraphics[scale=0.59]{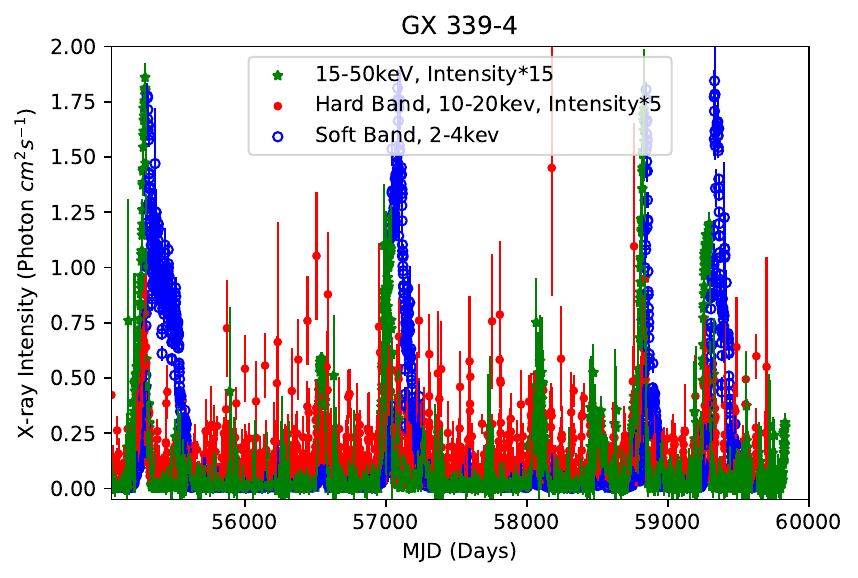}
    \includegraphics[scale=0.60]{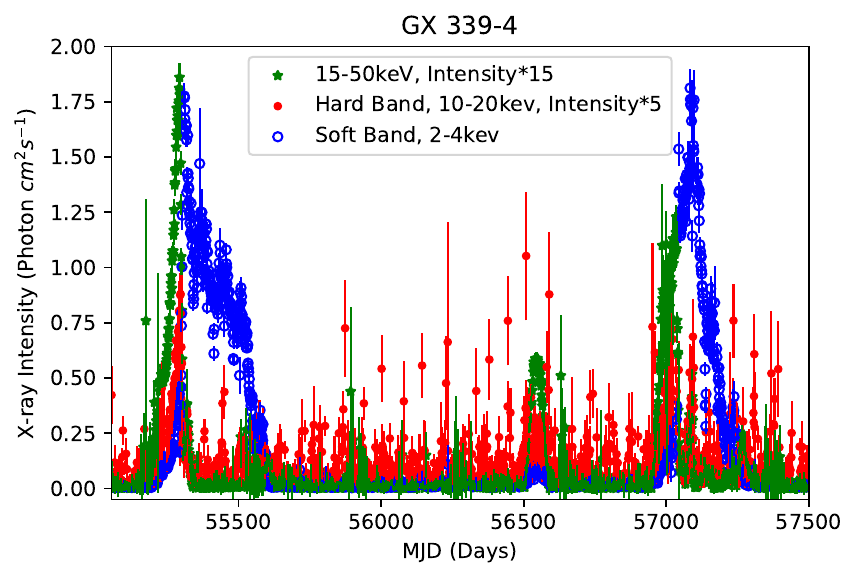}
    \includegraphics[scale=0.60]{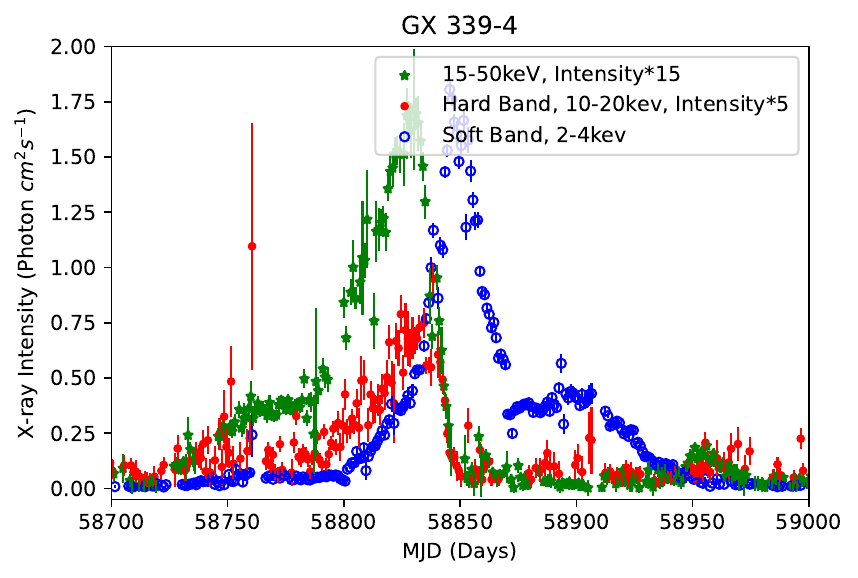}
    \includegraphics[scale=0.60]{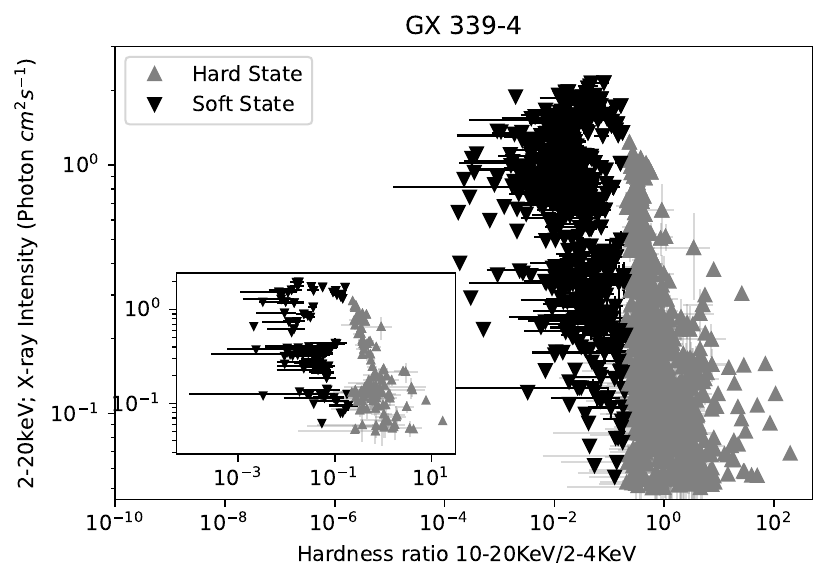}
   \caption{{\it Lightcurve and HID of GX 339-4}: Top left panel: Lightcurves throughout the entire observation duration with \maxi{} are shown in the 2-4 keV soft band (blue open circles), 10-20 keV hard bands (red solid circles) respectively while the same as observed with \swift{}/BAT in 15-50 keV is shown using green stars. A multiplication factor of 15 is used for plotting \swift{} lightcurve for easy comparison. Top right panel: Zoomed-in view of the first two outbursts and a failed outburst as observed with \maxi{} and \swift{} are shown. Bottom left panel: for the clarity of outburst features, further zoomed-in view of the third outburst is shown. Bottom right panel: Hardness Intensity Diagram (HID) of all outbursts are shown where hardness ratio is defined as the ratio of X-ray intensity in 10-20 keV and 2-4 keV while the intensity is defined as 2-20 keV count rate. Grey triangles denote hard state (when Hardness ratio $>$ 0.2) while black inverted triangles denote soft state (when Hardness ratio $<$ 0.2). Inset demonstrates the HID during the outburst shown in the bottom left panel.}
    \label{fig:GX 339-4}
\end{figure*}

\begin{figure*}
    \centering
    \includegraphics[scale=0.60]{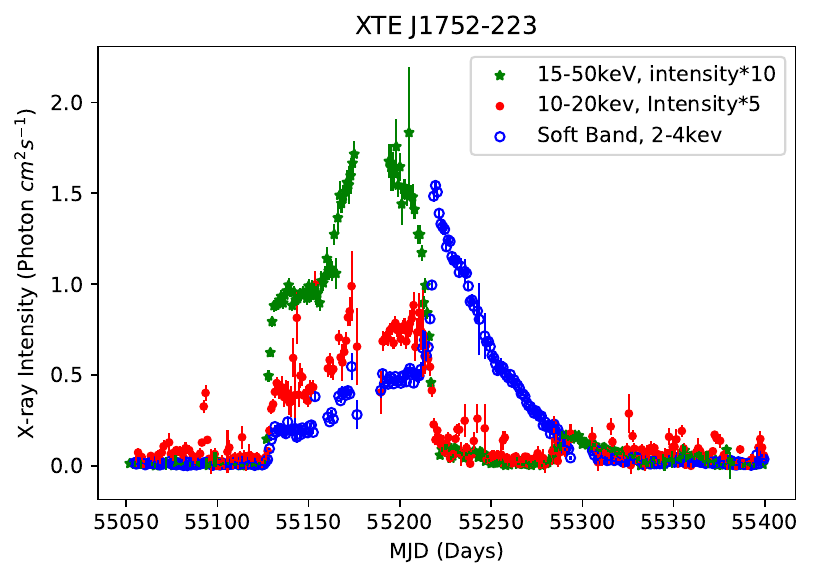}
    \includegraphics[scale=0.60]{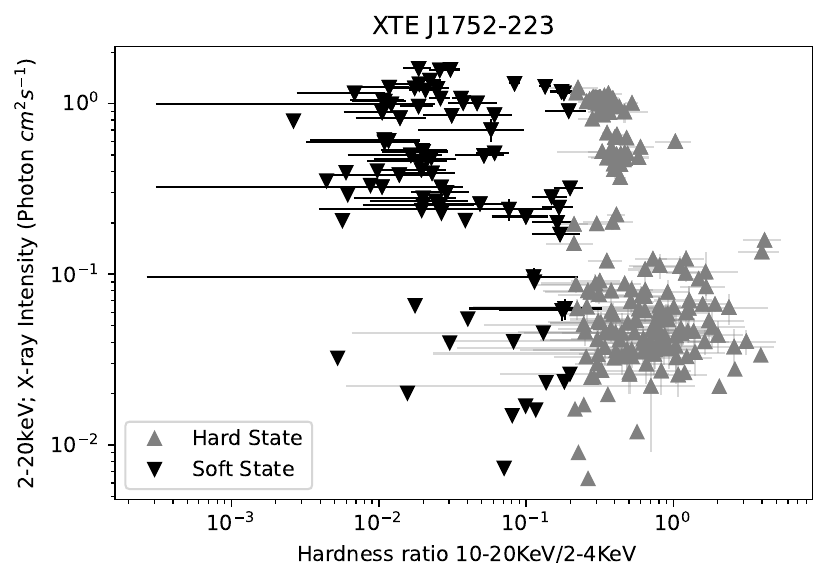}
    \caption{{\it Lightcurve and HID of XTE J1752–223}: 
    Left panel: Zoomed view of the outburst observed with \maxi{} in the 2–4 keV (blue open circles) and 10–20 keV (red solid circles) bands, along with \swift{}/BAT 15–50 keV data (green stars). The \swift{} lightcurve is scaled by a factor of 10 for clarity. The hard-band flux rises prior to the soft band. 
    Right panel: Corresponding Hardness–Intensity Diagram (HID), where hardness is defined as the ratio of 10–20 keV to 2–4 keV intensity and intensity as the 2–20 keV count rate. Grey triangles denote the hard state and black inverted triangles denote the soft state.}

    \label{fig: XTE J1752−223}
\end{figure*}

\begin{figure*}
    \centering
    \includegraphics[scale=0.60]{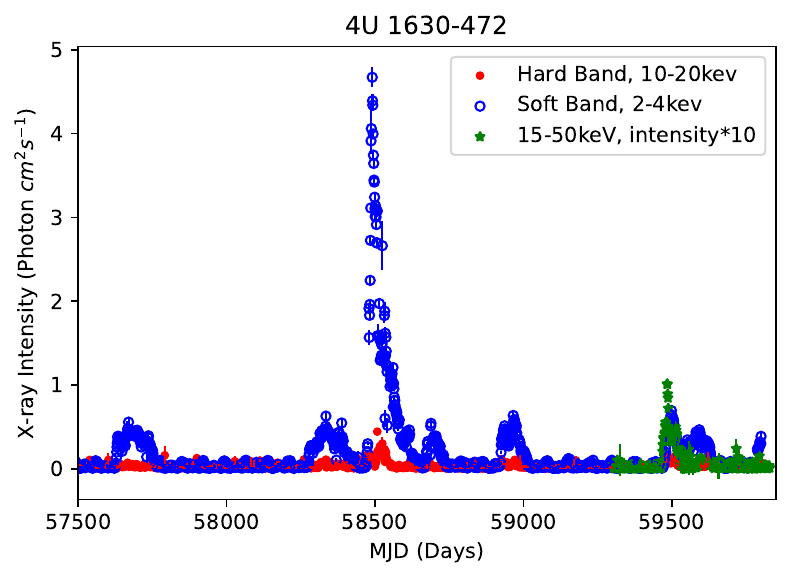}
    \includegraphics[scale=0.60]{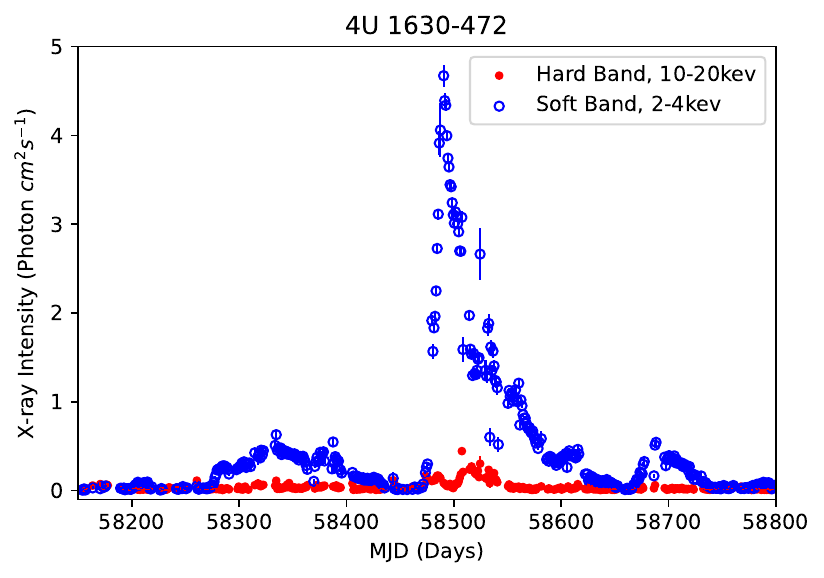}
    \includegraphics[scale=0.60]{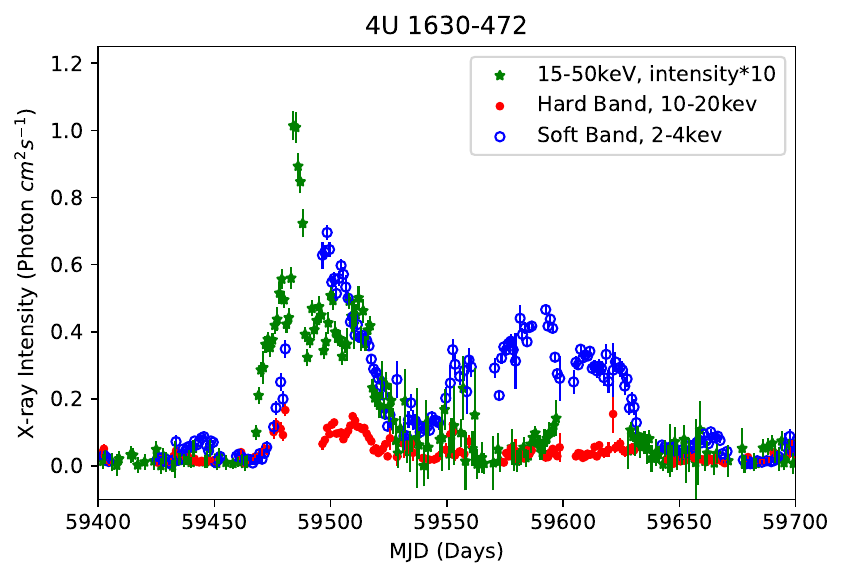}
    \includegraphics[scale=0.60]{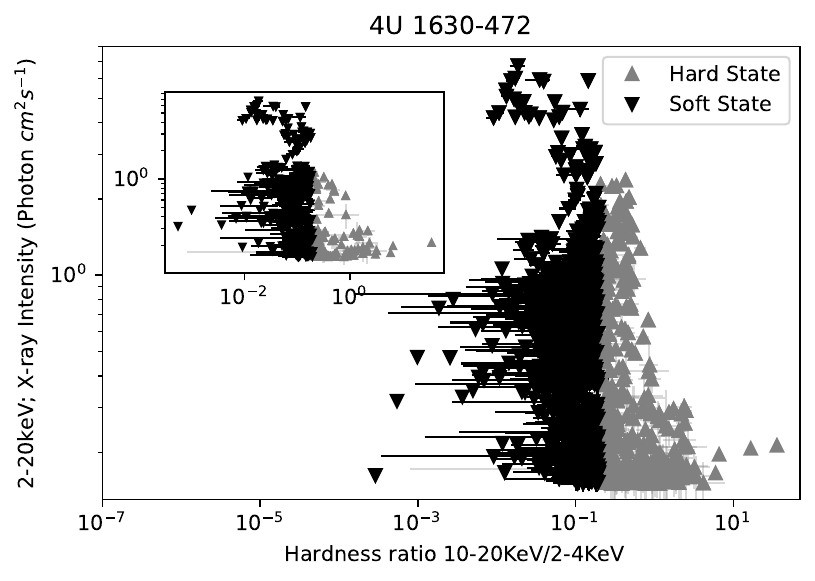}
    \caption{{\it Lightcurves and HID of 4U 1630–472}: 
    Top left panel: Lightcurves during the full observation period obtained with \maxi{} in the 2–4 keV (blue open circles) and 10–20 keV (red solid circles) bands. The \swift{}/BAT 15–50 keV data (green stars), available are also shown. The \swift{} lightcurve is scaled by a factor of 10 for easy comparison.
    Top right panel: Zoomed view of the hard-band dominated (failed) outburst. 
    Bottom left panel: Zoomed view of a full outburst around 59500 MJD.  
    Bottom right panel: Corresponding Hardness–Intensity Diagram (HID), where hardness is defined as the ratio of 10–20 keV to 2–4 keV intensity and intensity as the 2–20 keV count rate. Grey triangles represent the hard state (when Hardness ratio $>$ 0.2) and black inverted triangles the soft state (when Hardness ratio $<$ 0.2).}

    \label{fig:4U 1630-472}
\end{figure*}

The black hole mass in the X-ray binary GX 339$-$4 is estimated within the range of 5.8--12 $M_\odot$ through various methods \citep{hynes2003dynamical, parker2016nustar, heida2017mass, sreehari2019constraining}. A total of four outbursts are visible during the observation period (top left panel of Fig. \ref{fig:GX 339-4}). Upon zooming in (top right panel of Fig. \ref{fig:GX 339-4}), we are able to see a failed outburst, which is only present in the hard band. In the entire light curve, there are multiple failed outbursts. Upon zooming in further onto a particular outburst (bottom left panel of Fig. \ref{fig:GX 339-4}), we see a similar trend where the outbursts in the hard band lag behind that in the soft band, indicating that it has transitioned to the outer part of the accretion disk (the soft band) within a matter of $\sim$50 days.

\subsection{XTE J1752-223}
XTE J1752-223 is identified as a BHXB \citep{reis2011multistate} with a mass estimate of  8.1-11.9 $M_\odot$ \citep{chatterjee2020inference}. Again, we observe an outburst in the hard band, which precedes the soft band, transitioning within a time range of $\sim$50 days (Left panel of Fig. \ref{fig: XTE J1752−223}).

\subsection{4U 1630-472}
This X-ray source is classified as a BHXB ~\citep{baby2020astrosat}. The mass of the BH is estimated to be 3-10$M_\odot$ \citep{baby2020astrosat, capitanio2015missing}. Two outbursts are visible during the observation period. However, Swift data are available between MJD 59230 to present (top left panel of Fig. \ref{fig:4U 1630-472}).

\subsection{H 1743-322}
\begin{figure*}
    \centering
    \includegraphics[scale=0.60]{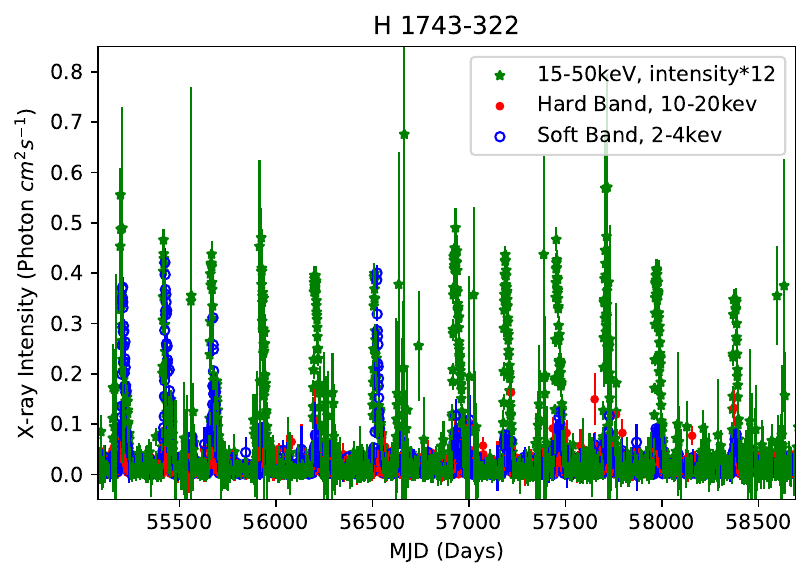}
    \includegraphics[scale=0.60]{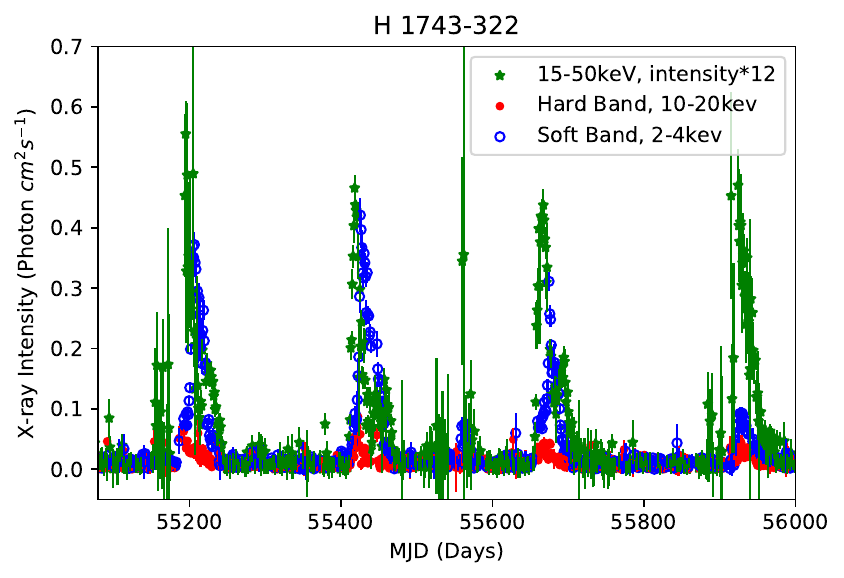}
    \includegraphics[scale=0.60]{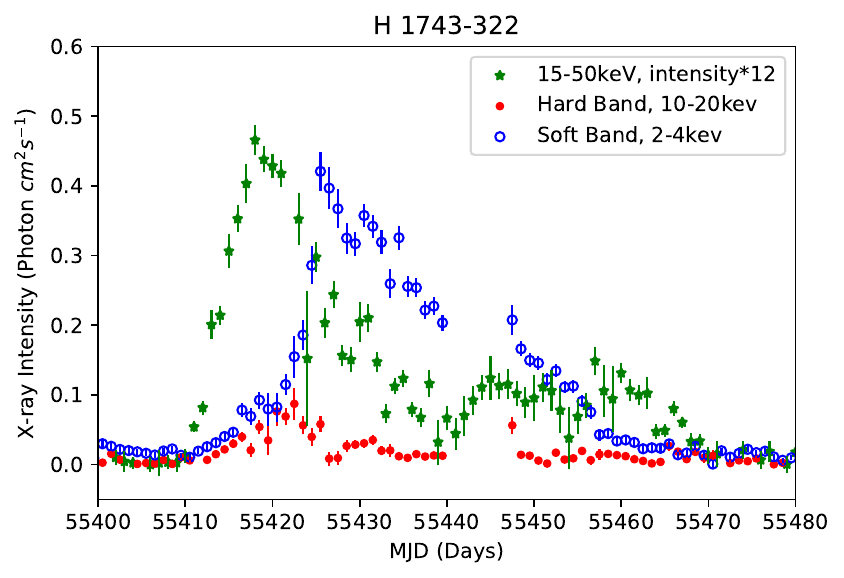}
    \includegraphics[scale=0.60]{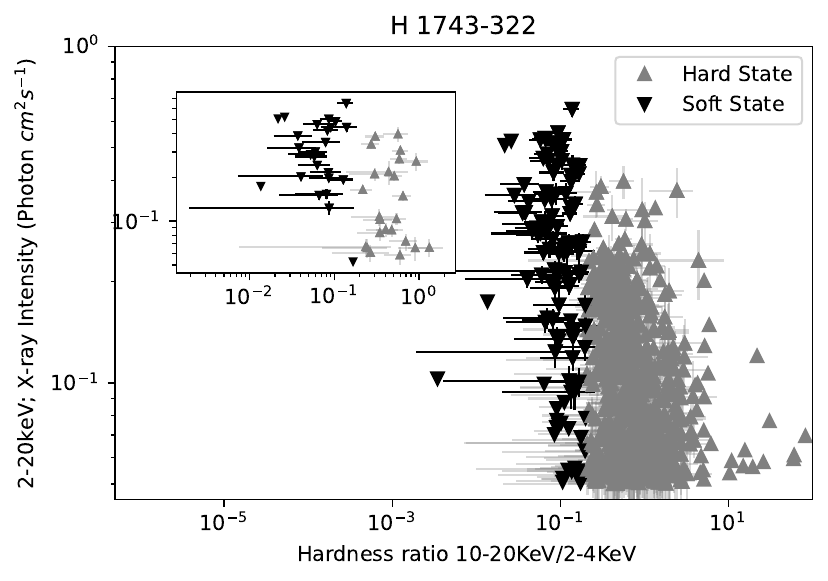}
    \caption{{\it Lightcurves and HID of H 1743–322}: 
    Top left panel: Lightcurves obtained with \maxi{} in the 2–4 keV (blue open circles) and 10–20 keV (red solid circles) bands, along with \swift{}/BAT 15–50 keV data (green stars), showing a series of quasi-periodic outbursts. Several of the later outbursts are prominent only in the hard band, with minimal enhancement in the soft band. The \swift{} lightcurve is scaled by a factor of 12 for easy comparison.
   Top right panel: Zoomed view of the first four outbursts; the right-most outburst is dominated by hard-band emission. 
   Bottom left panel: Zoomed view of a single outburst, where the rise in the hard band precedes the soft band. 
   Bottom right panel: Hardness Intensity Diagram (HID) of the full observation period. Hardness is defined as the ratio of 10–20 keV to 2–4 keV intensity, and intensity as the 2–20 keV count rate. Grey triangles denote the hard state (when Hardness ratio $>$ 0.2) and black inverted triangles denote the soft state (when Hardness ratio $<$ 0.2). The inset shows the HID during one representative outburst in the bottom left panel.}

    \label{fig:H 1743-322}
\end{figure*}

H 1743-322 is identified as a candidate for BHXB \citep{parmar2003integral}. Multiple outbursts have been observed during the observation period (top left panel of Fig. \ref{fig:H 1743-322}). The top right panel of Fig. \ref{fig:H 1743-322} presents a zoomed-in view where we observe that some of the outbursts are dominant in the hard band but have not completely transitioned into the soft band, indicating that the energy has been dissipated within that period. The zoomed-in view of a particular outburst (bottom left panel of Fig. \ref{fig:H 1743-322}) indicates that the outbursts in the hard band always precede the detection in the soft band, once again signifying that it originates in the inner part of the disk and propagates outwards.

\section{Analysis of Neutron Star Sources}
In this section, we have discussed five neutron star sources. The discussion of the remaining 16 sources can be found in Appendix B.
\label{Analysis of Neutron Star Sources}
\subsection{Swift J0243.6+6124}
\begin{figure*}
    \centering
    \includegraphics[scale=0.60]{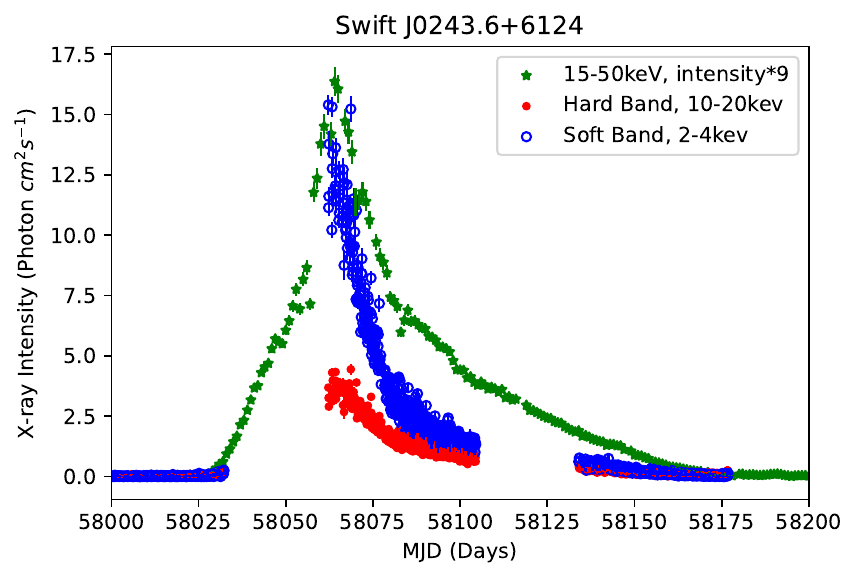}
    \includegraphics[scale=0.60]{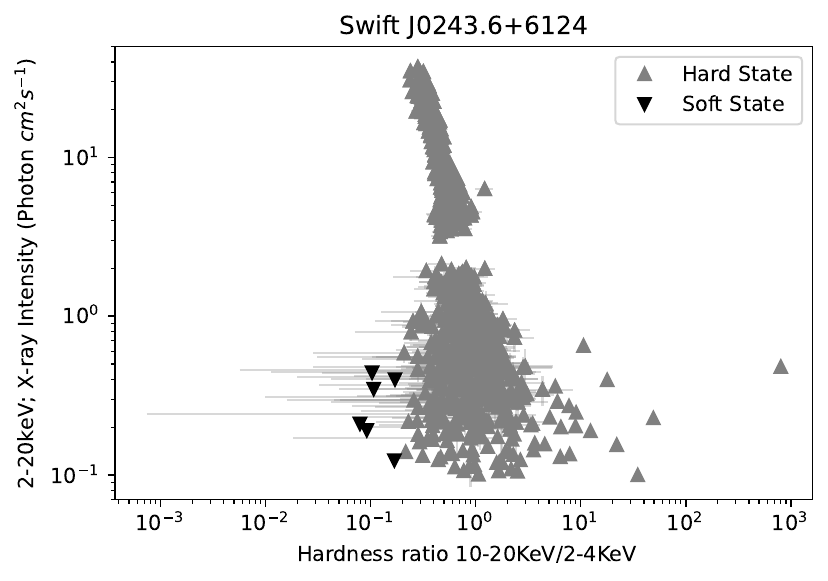}
   \caption{{\it Lightcurves and HID of Swift J0243.6+6124}: 
  Left panel: Lightcurves obtained with \maxi{} in the 2–4 keV (blue open circles) and 10–20 keV (red solid circles) bands, along with \swift{}/BAT 15–50 keV data (green stars), during the outburst. The \swift{} lightcurve is scaled by a factor of 9 for clarity. The rise and peak of the outburst are observed nearly simultaneously across all energy bands. 
  Right panel: Hardness Intensity Diagram (HID) of the outburst, where hardness is defined as the ratio of 10–20 keV to 2–4 keV intensity and intensity as the 2–20 keV count rate. The distribution of points indicates predominantly high hardness values during the outburst.}

    \label{fig:Swift J0243.6+6124}
\end{figure*}

\cite{kong2020two} categorizes Swift J0243.6+6124 as a NSXB. A zoomed-in view of the light curve during the outburst shows that the outburst occurs simultaneously in all bands (left panel of Fig. \ref{fig:Swift J0243.6+6124}). The HID (right panel of Fig. \ref{fig:Swift J0243.6+6124}) shows a peak in the intensity in the hard band. Thus, the spectrum of the source is primarily dominated by hard X-rays.

\subsection{Aql X-1} 
\begin{figure*}
    \centering
    \includegraphics[scale=0.60]{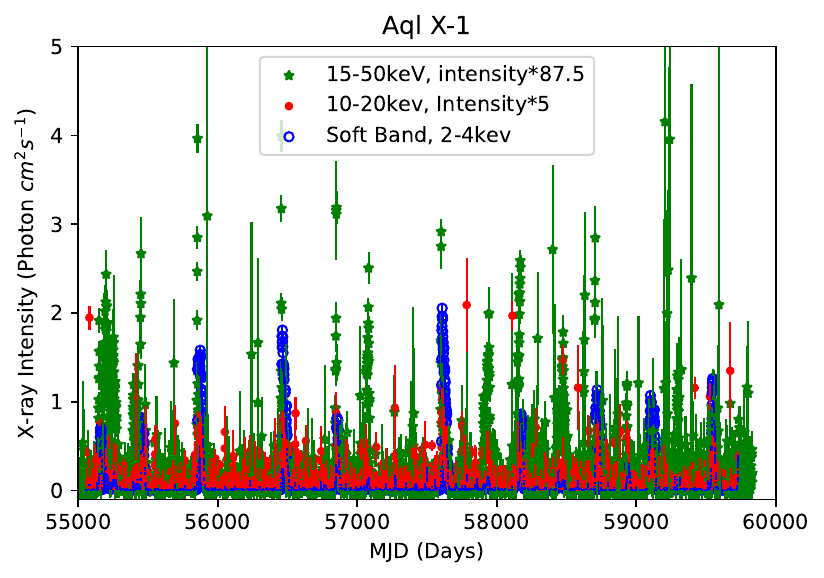}
    \includegraphics[scale=0.60]{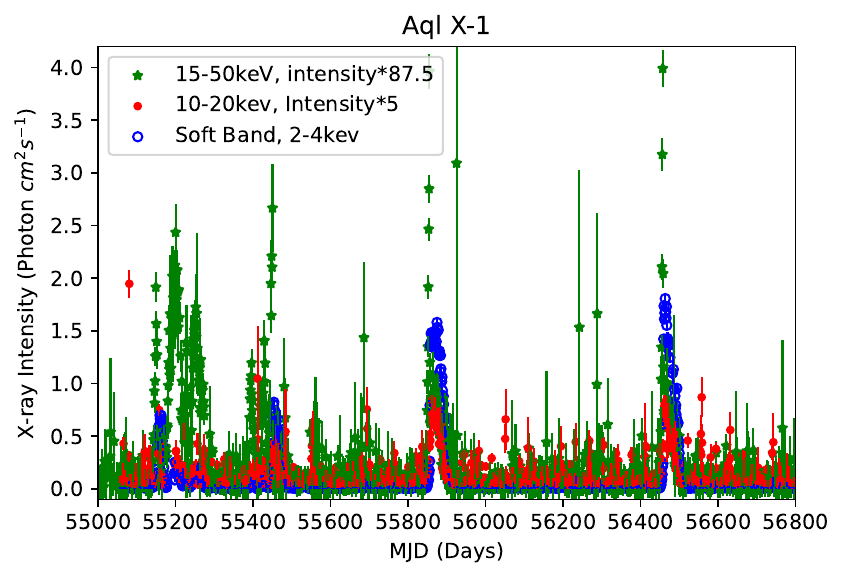}
    \includegraphics[scale=0.60]{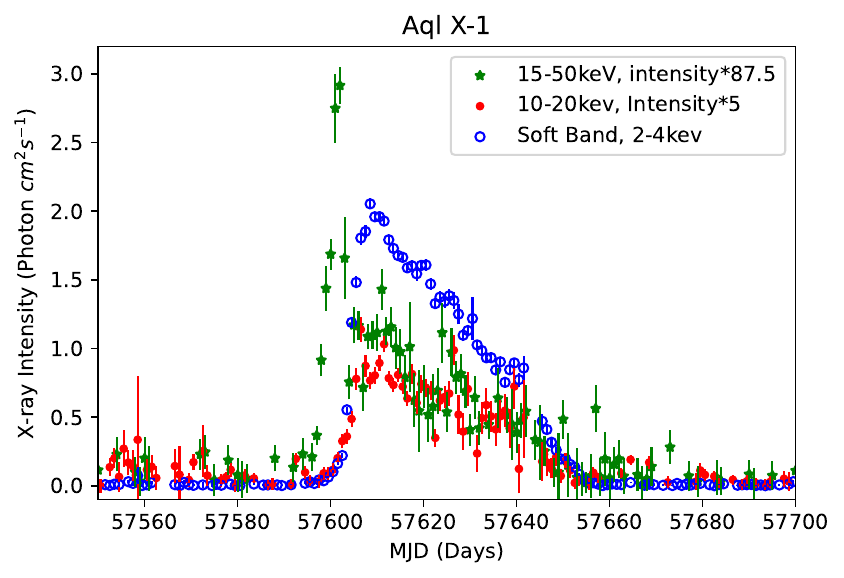}
    \includegraphics[scale=0.60]{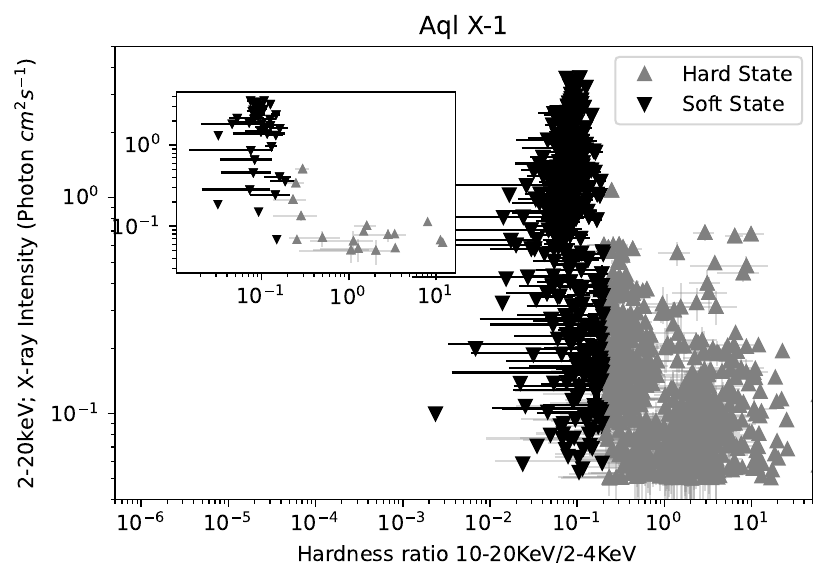}
    \caption{{\it Lightcurves and HID of Aql X–1}: 
    Top left panel: Lightcurves obtained with \maxi{} in the 2–4 keV (blue open circles) and 10–20 keV (red solid circles) bands, along with \swift{}/BAT 15–50 keV data (green stars), showing multiple outbursts and several failed peaks. Some outbursts are prominent only in the hard band, with no significant enhancement in the 2–4 keV soft band. The \swift{} lightcurve is scaled by factor of 87.5  for clarity. 
    Top right panel: Zoomed view of the first four outbursts. 
    Bottom left panel: Zoomed view of a representative outburst characterized by a rapid rise and gradual decay. 
    Bottom right panel: Hardness Intensity Diagram (HID) of the full observation period, where hardness is defined as the ratio of 10–20 keV to 2–4 keV intensity and intensity as the 2–20 keV count rate. Grey triangles represent the hard state (when Hardness ratio $>$ 0.2) and black inverted triangles the soft state (when Hardness ratio $<$ 0.2). The inset shows the HID during a particular outburst in the bottom left panel.}

    \label{fig:Aql X-1}
\end{figure*}

Aquila X-1 (Aql X-1) is a low-mass NSXB ~\citep{trigo2018evolving}. Multiple outbursts are observed in the light curve during the observation period (top left panel of Fig. \ref{fig:Aql X-1}). A few outbursts have propagated successfully from the hard band to the soft band. However, in some cases, the outbursts remain purely in the hard band (inner part of the accretion disk). The zoomed-in outburst is asymmetric with sharp ascent, signifying that the disk extends to near the compact object, where the spectrum is primarily dominated by the hard band. Furthermore, the intensity peak is observed in the soft band rather than in the hard band (bottom right panel of Fig. \ref{fig:Aql X-1}).

\begin{figure*}
    \centering
    \includegraphics[scale=0.60]{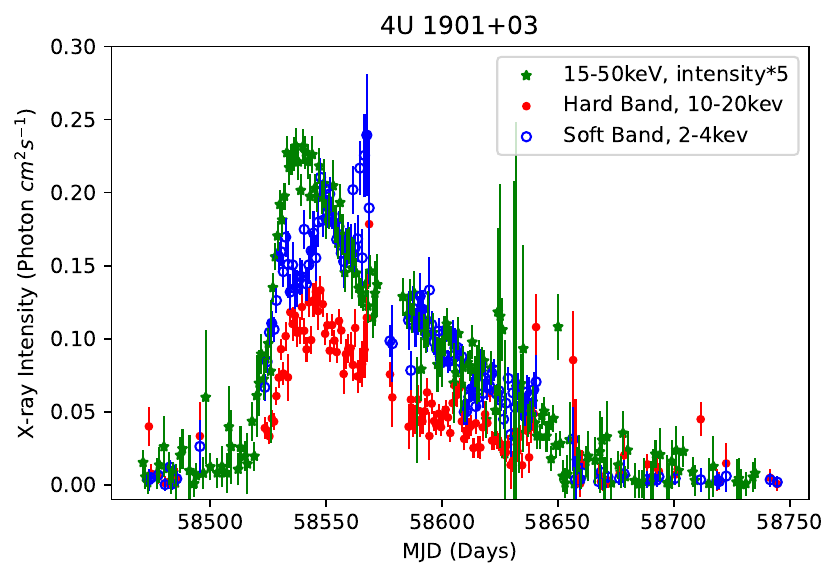}
    \includegraphics[scale=0.60]{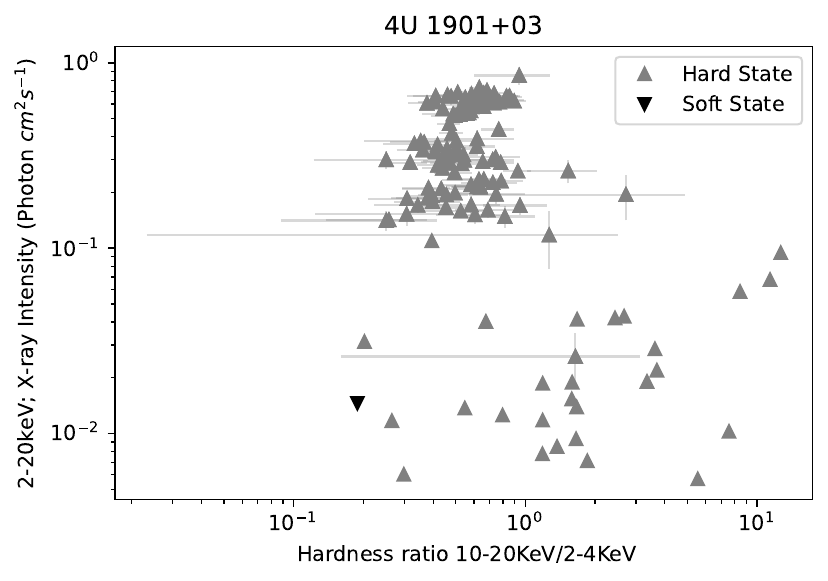}
    \caption{{\it Lightcurves and HID of 4U 1901+03}: 
    Left panel: Lightcurves obtained with \maxi{} in the 2–4 keV (blue open circles) and 10–20 keV (red solid circles) bands, along with \swift{}/BAT 15–50 keV data (green stars), showing the outburst peak. The intensity profile appears asymmetric, with a comparatively rapid rise and slower decay. The \swift{} lightcurve is scaled by a factor of 5 for clarity. 
    Right panel: Corresponding Hardness–Intensity Diagram (HID), where hardness is defined as the ratio of 10–20 keV to 2–4 keV intensity and intensity as the 2–20 keV count rate. Grey triangles denote the hard state (when Hardness ratio $>$ 0.2) and black inverted triangles denote the soft state (when Hardness ratio $<$ 0.2).}

    \label{fig:4U 1901+03}
\end{figure*}

\subsection{4U 1901+03}
4U 1901+03 is classified as an NSXB ~\citep{tuo2020insight}. An outburst with a distinct intensity peak can be observed in the light curve in all bands (left panel of Fig. \ref{fig:4U 1901+03}). However, in this case, the intensity peak is not symmetric with a sharp ascend. The HID again indicates that the spectra are primarily dominated by the hard X-ray photons.

\subsection{V 0332+53}

\begin{figure*}
    \includegraphics[scale=0.60]{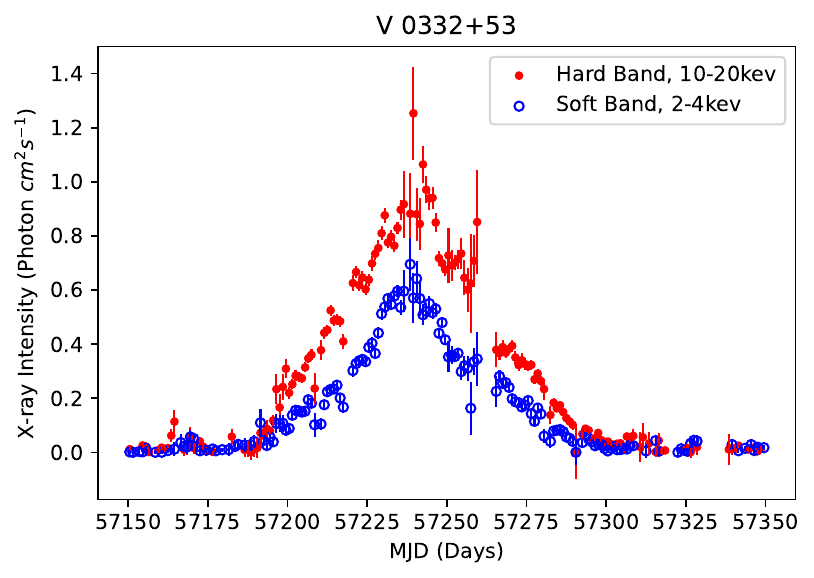}
    \includegraphics[scale=0.60]{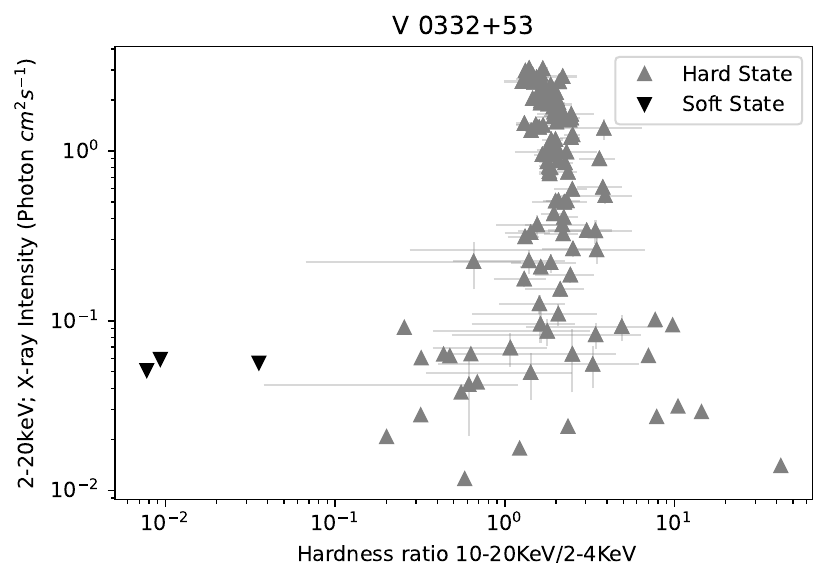}
    \caption{{\it Lightcurves and HID of V 0332+53}: 
  Left panel: Zoomed view of the outburst observed with \maxi{} in the 2–4 keV (blue open circles) and 10–20 keV (red solid circles) bands. The intensity increases gradually, and the outburst is observed in both the energy bands over the same time interval.
  Right panel: Corresponding Hardness–Intensity Diagram (HID), where hardness is defined as the ratio of 10–20 keV to 2–4 keV intensity and intensity as the 2–20 keV count rate. The distribution of points shows a greater concentration in the hard-state (when Hardness ratio $>$ 0.2) region compared to the soft state (when Hardness ratio $<$ 0.2).}

    \label{fig:V 0332+53}
\end{figure*}

V 0332+53 is classified as an NSXB \citep{hemphill2016x}. The zoomed-in view of the outburst (left panel of Fig. \ref{fig:V 0332+53}) shows a gradual increase in the intensity in both the soft and hard bands during the outburst, unlike the sudden rise in intensity in the case of black holes. The rise in intensity also occurs at the same time in both bands, indicating that the inner part of the disk might possess some black body component in the form of a thin disk along with the hot corona, unlike what we saw in the case of black holes, where there may not be any blackbody component in the inner parts of the accretion disk. The hard band dominates over the soft band, as can be seen from the HID (right panel of Fig.\ref{fig:V 0332+53}), and the intensity peak primarily comes from the hard band.

\subsection{KS 1947+300} 
\begin{figure*}[!ht]
    \centering
    \includegraphics[scale=0.60]{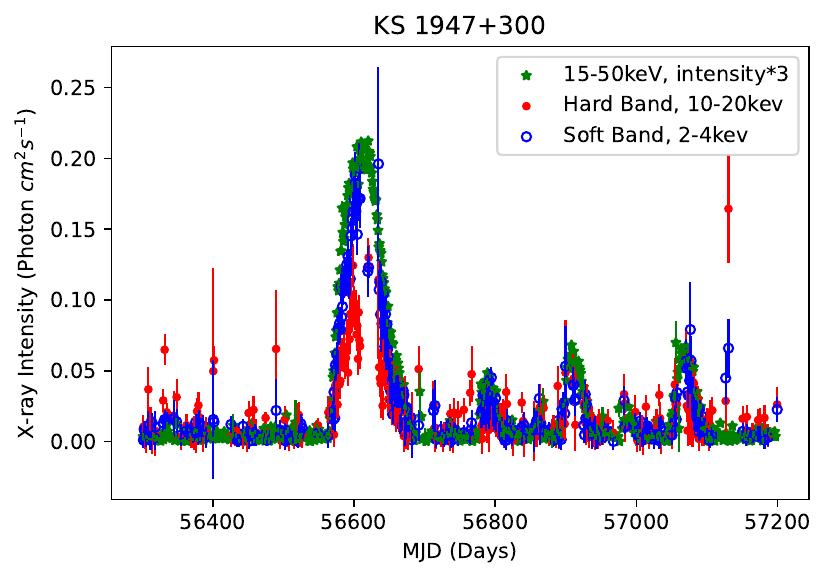}
    \includegraphics[scale=0.60]{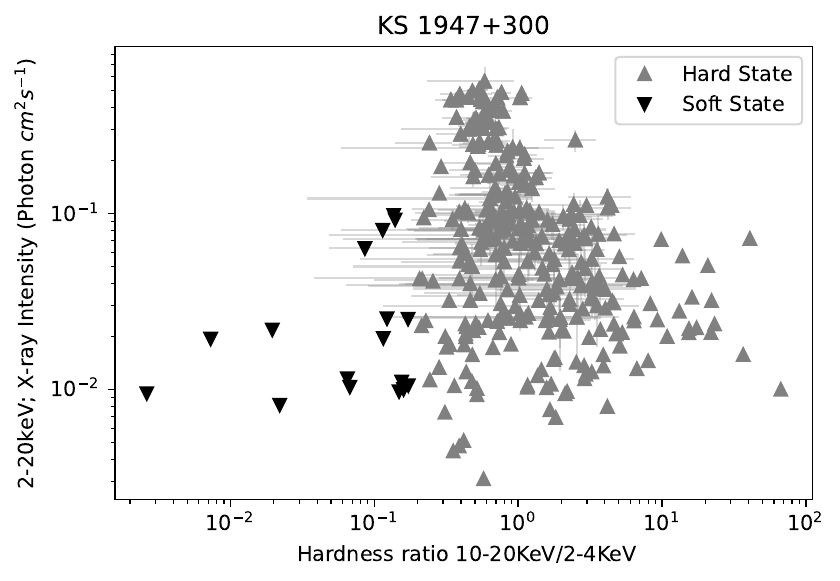}
    \caption{{\it Lightcurves and HID of KS 1947+300}: 
    Left panel: Lightcurves obtained with \maxi{} in the 2–4 keV (blue open circles) and 10–20 keV (red solid circles) bands, along with \swift{}/BAT 15–50 keV data (green stars). The intensity peaks are observed over the same time interval in both the hard and soft bands. The \swift{} lightcurve is scaled by a factor of 3 for clarity. 
    Right panel: Corresponding Hardness–Intensity Diagram (HID), where hardness is defined as the ratio of 10–20 keV to 2–4 keV intensity and intensity as the 2–20 keV count rate. Grey triangles represent the hard state (when Hardness ratio $>$ 0.2) and black inverted triangles the soft state (when Hardness ratio $<$ 0.2). The distribution of points indicates predominantly higher hardness values during the outburst.}

    \label{fig:KS 1947+300}
\end{figure*}

KS 1947+300 is classified as an NSXB ~\cite{furst2014nustar}. The observation shows an outburst with a few failed outbursts (left panel of Fig. \ref{fig:KS 1947+300}). The intensity peaks in the hard and soft bands occur at the same time. The HID indicates that the spectrum is dominated by hard X-rays (right panel of Fig. \ref{fig:KS 1947+300}).

\begin{figure*}
    \centering
    \includegraphics[scale=0.34]{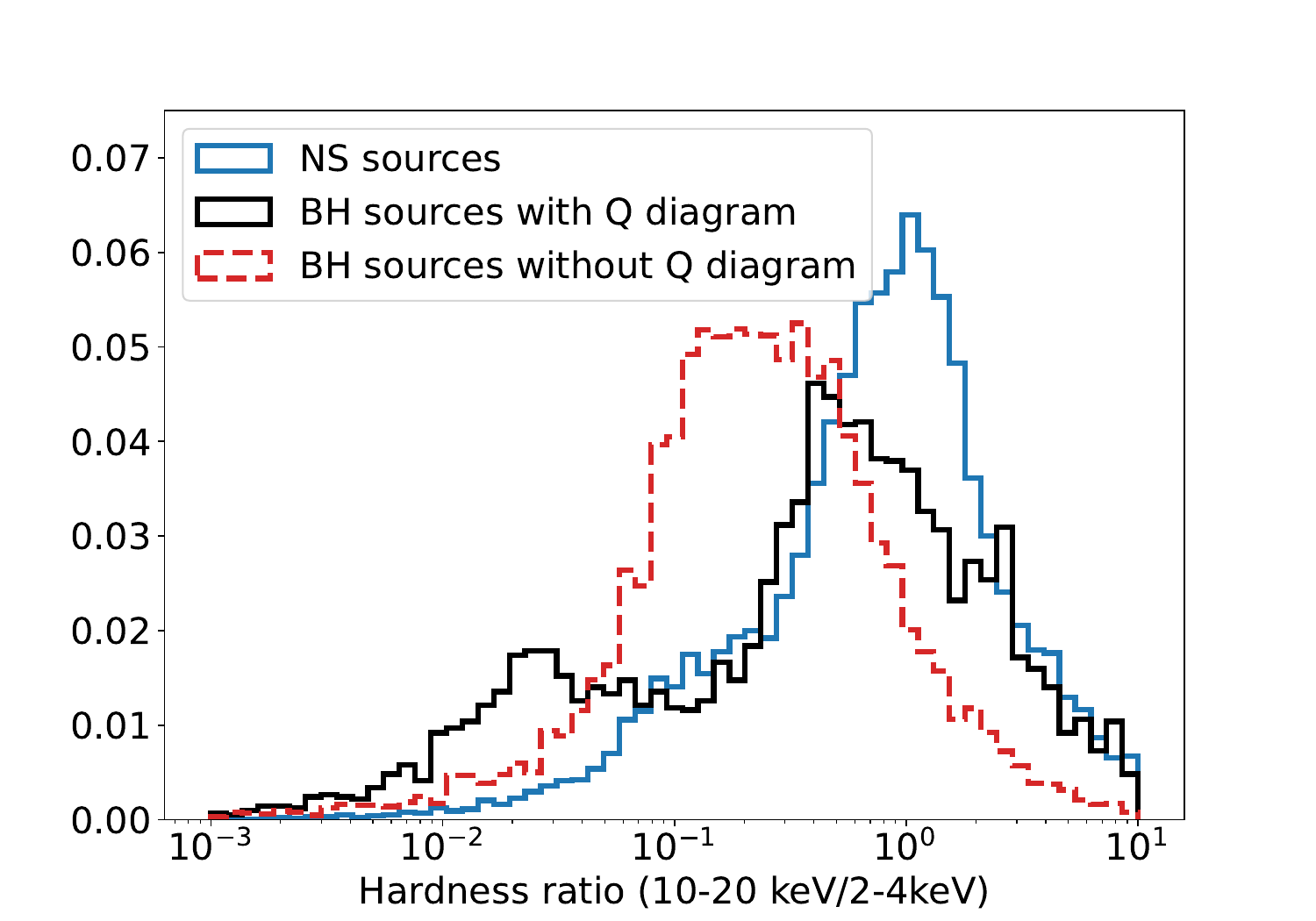}
    \includegraphics[scale=0.34]{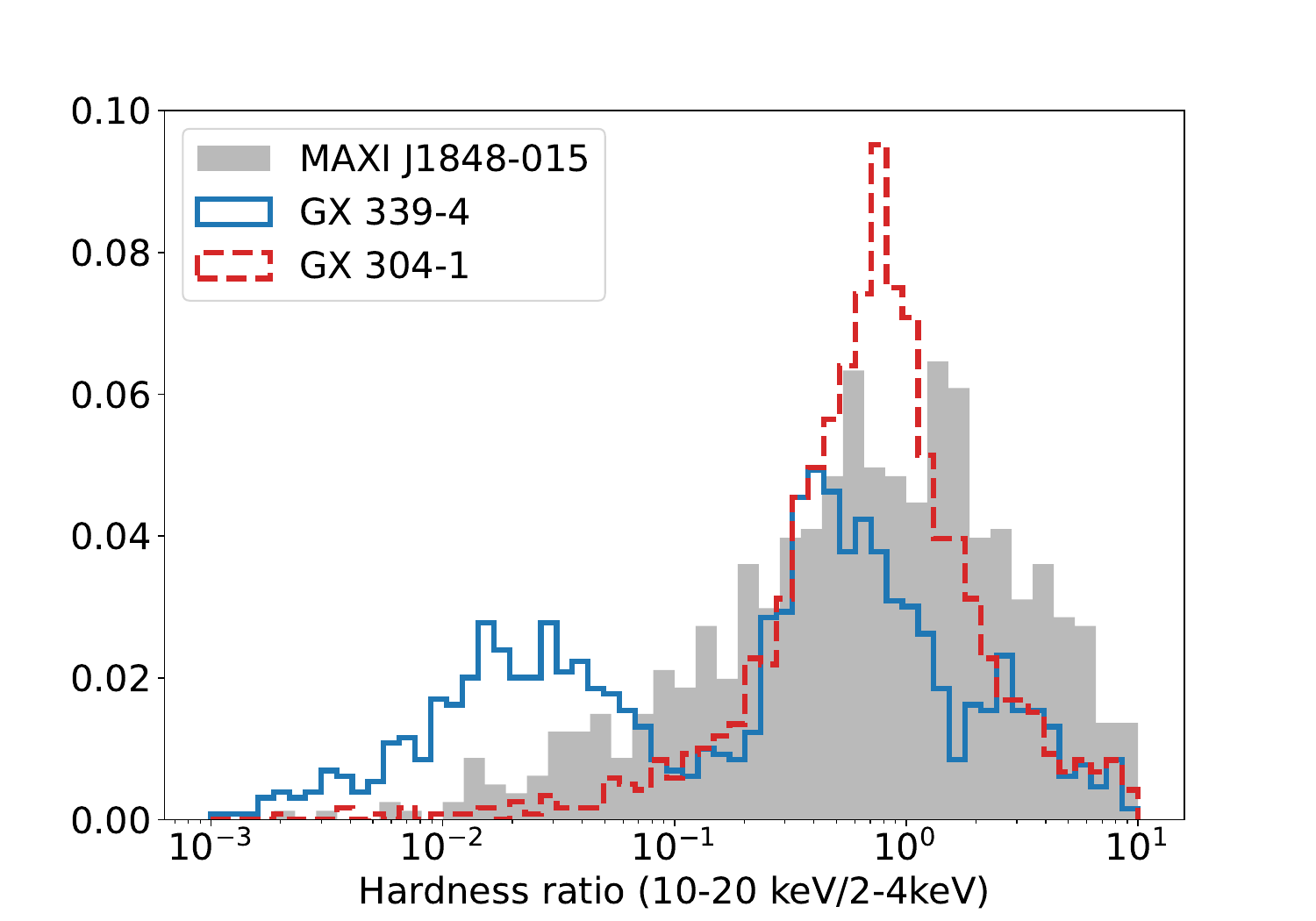}
    \caption{Frequency distribution of hardness ratio for (normalised using total number of hardness points). Left panel: neutron star X-ray binaries, black hole X-ray binaries with q diagrams and black hole X-ray binaries sources without q diagrams are represented in blue, black, and red, respectively. Right panel: the frequency distributions of HID are shown for known black hole and neutron star sources: GX 339-4 (green) and GX 304-1 (red), respectively, while the same for the candidate source MAXI J1848-015 is shown in grey. }
    \label{fig: Histogram}
\end{figure*}

\begin{figure*}
    \centering
    \includegraphics[scale=0.34]{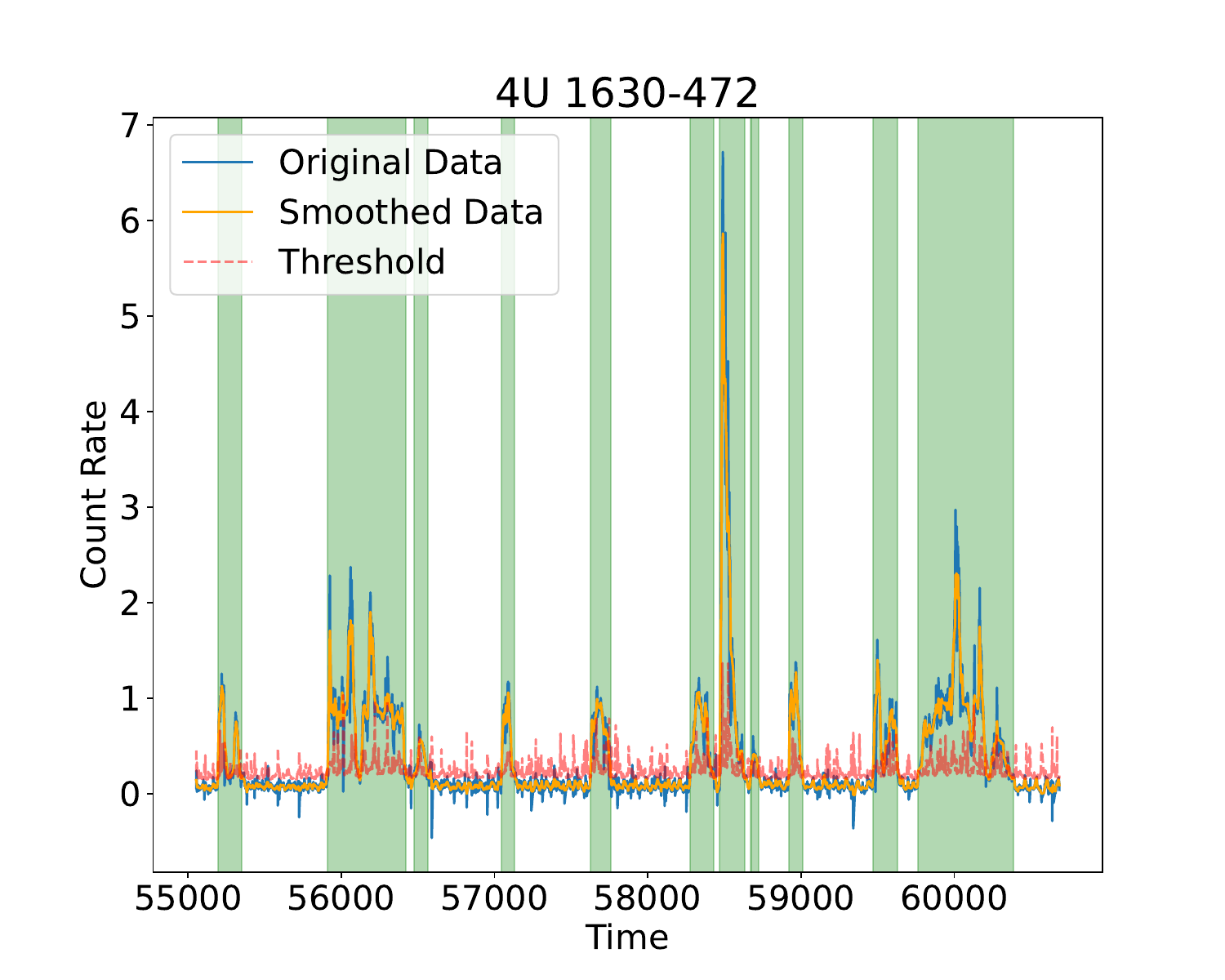}
    \includegraphics[scale=0.34]{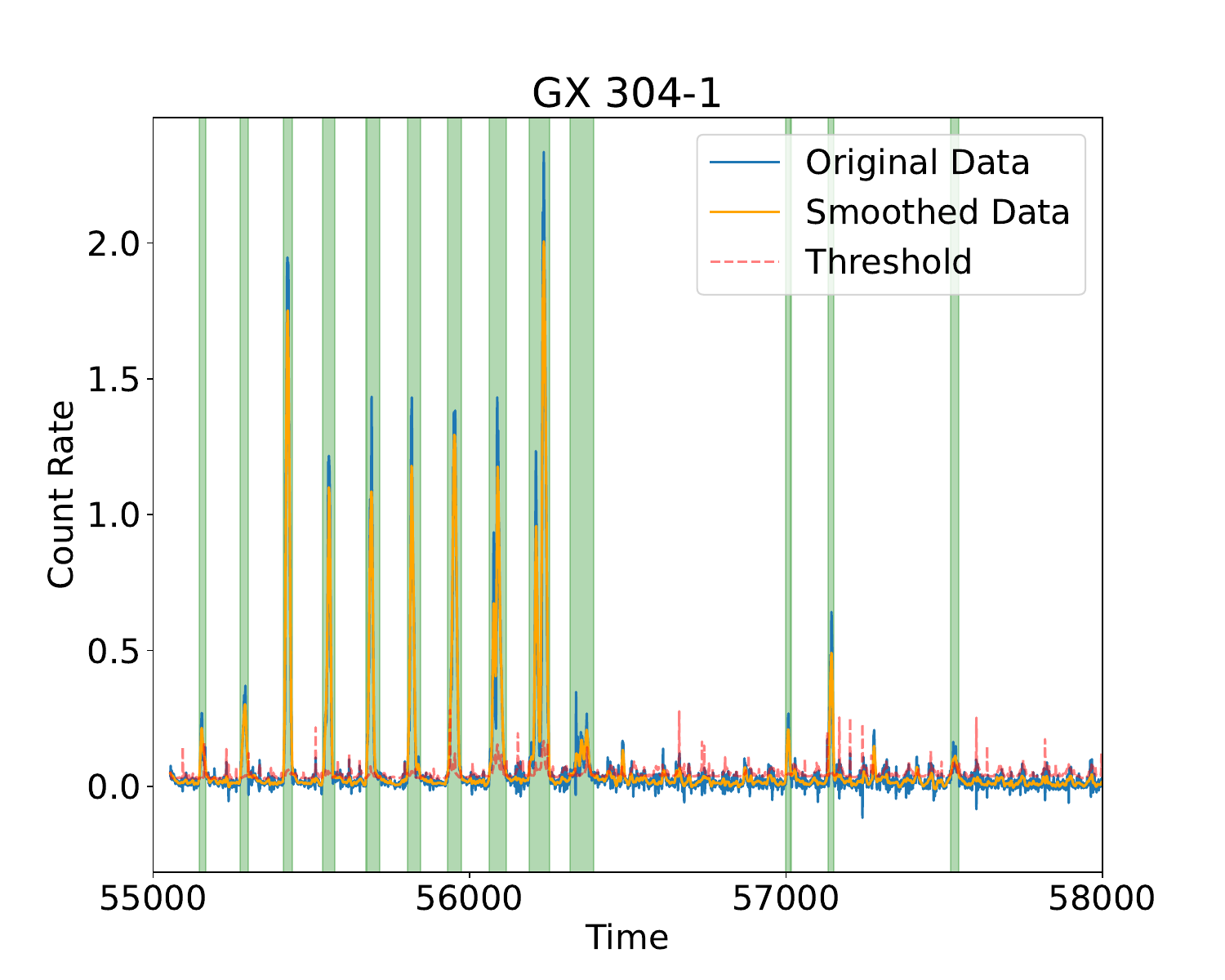}
    \caption{Demonstration of outburst/flare selection for left panel: 4U 1630-472 and right panel: GX 304-1. The original data is plotted in blue, while the smoothed-out data is plotted in yellow. The threshold is marked in red, and the outburst/flare regions are marked in green.}
    \label{fig:selection}
\end{figure*}

\begin{figure}
    \centering
    \includegraphics[scale=0.35]{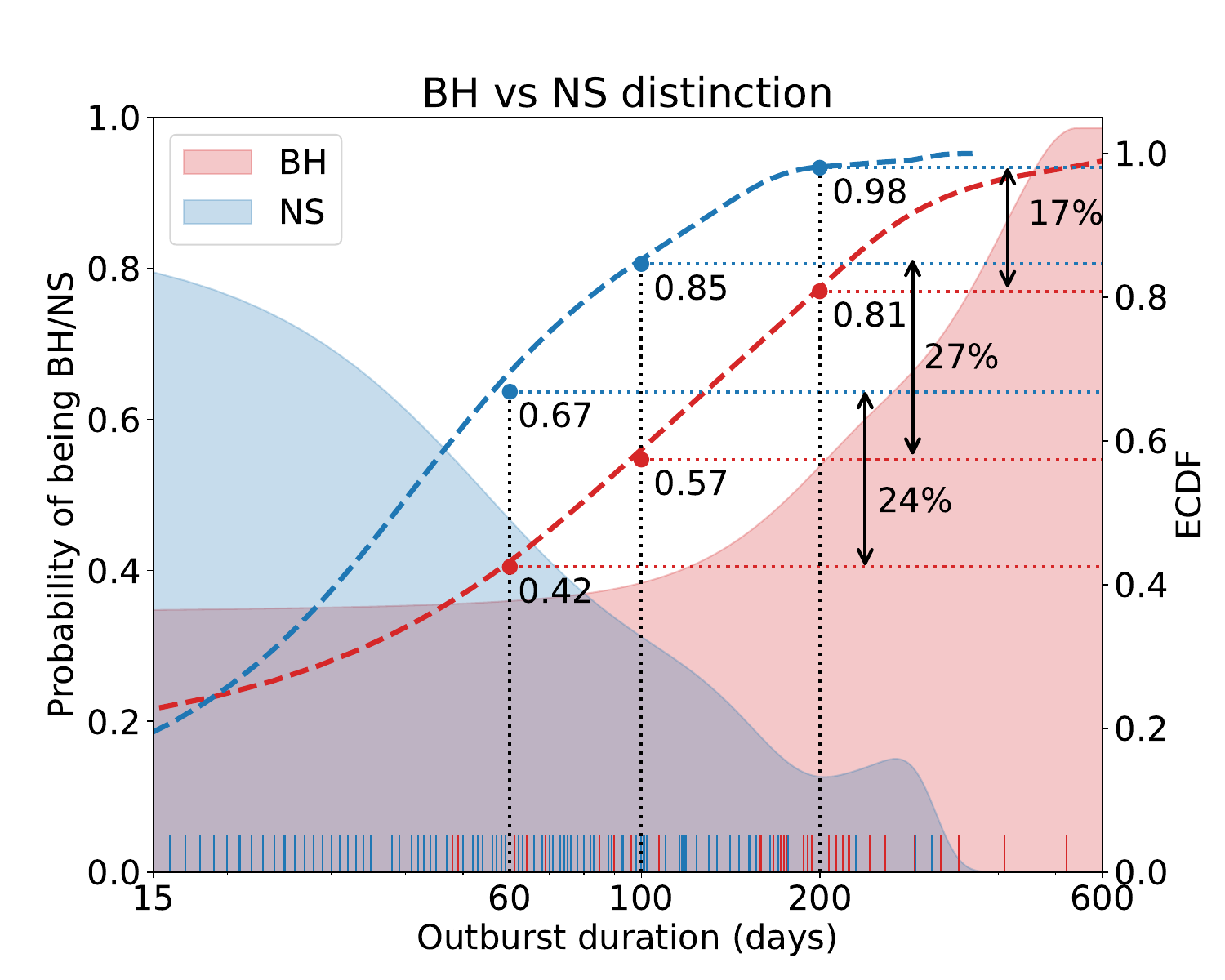}
    \caption{Probability and ECDF plot with respect to outburst/flare duration. Filled colours indicate the probability that a source having a certain outburst duration is a BH or NS. The vertical dotted line at 60 days marks the end of region 1. The region between the vertical dotted lines at 60 and 100 days is region 2. The regions on the right of the vertical dotted lines at 100 days and 200 days are regions 3 and 4, respectively. The percentage difference in the ECDF at the beginning of each region is denoted by the arrows joining the dotted lines. The rug plot at the base of the x-axis depicts the BH outbursts/flares in red and NS outbursts/flares in blue.}
    \label{fig:ecdf}
\end{figure}

\begin{table*}[!ht]
\centering
\begin{tabular}{cccccc}
\hline
\thead{{\bf Source name}} & \thead{{\bf Days spent} \\ {\bf in soft band} \\ {\bf (2--4 keV)}} & \thead{{\bf Days spent} \\ {\bf in hard band} \\ {\bf (10--20 keV)}}  & \thead{{\bf Outburst} \\ {\bf time (\%)}} & \thead{{\bf Days spent} \\ {\bf in soft state} \\ {\bf in HID }} & \thead{{\bf Days spent} \\ {\bf in hard state} \\ {\bf in HID }} \\
\hline
\hline
  MAXIJ1820+070 & 51.74\%&48.26\% & 8.12\%&49 &100\\
  GX 339-4 & 51.85\%& 48.15\% &28.70\% &328 &590\\
  MAXIJ1348-630& 52.77\%& 47.23\% & 4.91\%&76 &90\\
  XTE J1752-223&  56.19\%& 43.81\%& 6.86\%&58&166\\ 
 GS 1354-64& 58.61\%& 41.39\%& 3.81\%&34 &161 \\ 
 4U 1630-472 & 54.03\%& 45.97\%& 12.56\%& 989 &373\\ 
 4U 1543-475& 65.39\%& 34.66\%& 8.3\% &51 & 11\\ 
 MAXI J1910-057& 60\%& 40\%& 6.37\% &59 &68\\ 
 GRS 1716-249& 48.92\%& 51.08\%& 13.79\%&12&235\\
 GRS 1739-278& 62.49\%& 37.51\%& 8.92\%& 452& 1603\\ 
 MAXIJ1543-564&57.72\%& 48.28\%&8.38\% & 37 &113\\ 
 SwiftJ1728.9-3613* &  60.71\%& 39.29\%& 6.19\% &52 &45\\ 
 MAXIJ1535-571*  & 61.11\%& 38.89\%& 7.06\%&83&32 \\ 
 MAXIJ1659-152* & 49.75\%& 50.25\%& 4.13\%& 36&33 \\
 EXO 1846-031* & 63.73\%& 36.27\%&7.18\% &74 &51 \\
 MAXI J1803-298* & 65.5\%& 34.5\%&32.18\% &32 &13 \\
 H 1743-322* & 60.45\%& 39.55\%&  6.20\%&83&564\\
 MAXIJ1631-479* & 44.01\% &55.09\% &9.96\% & 731 &359\\
 MAXI J1848-015* & 63.95\%& 36.05\%& 8.86\%&38 &84\\
 MAXI J1305-704* & 70.80\%& 29.2\%& 7.91\%& 29& 54\\ 
 MAXI J1836-194* & 54.62\%& 45.38\%& 11.87\%&13&63\\
\hline
\end{tabular}
\caption{Details of results from BHXB outburst analysis. Sources marked with an asterisk * correspond to candidate black hole binary systems. }
\label{black_hole_details}
\end{table*}

\begin{table*}
\centering
\begin{tabular}{cccccc}
\hline
\thead{{\bf Source name}} & \thead{{\bf Days spent} \\ {\bf in soft band} \\ {\bf (2--4 keV)}} & \thead{{\bf Days spent} \\ {\bf in hard band} \\ {\bf (10--20 keV)}}  & \thead{{\bf Outburst} \\ {\bf time (\%)}} & \thead{{\bf Days spent} \\ {\bf in soft state} \\ {\bf in HID }} & \thead{{\bf Days spent} \\ {\bf in hard state} \\ {\bf in HID }} \\
\hline
\hline
V 0332+53   & 50.79\%& 49.21\%& 4.21\%& 0 &106\\
1A 0535+262  & 46.66\%& 53.34\%& 3.18\%&3 & 2850 \\ 
Swift J0243.6+6124  & 51.97\%& 48.03\%& 3.79\%& 0 &968 \\
GROJ1008-57 & 53.75\%& 46.24\%& 17.80\%& 1&909 \\
4U 1901+03  & 47.31\%& 52.69\%& 5.39\%&0 &106 \\
KS 1947+300   & 51.65\%& 48.35\%& 17.85\%&0 &170 \\
2S 1417-624   & 50.9\%& 49.1\%&9.31\% &1 &504 \\
GX 304-1   & 60.49\%& 39.51\%& 6.06\%&1 &57\\
4U 0115+63  & 48.07\%& 51.92\%& 1.48\%&0 &224 \\
NGC 6440  & 49.05\%& 50.95\%& 3.24\%&83 &274 \\
SMC X-3   & 53.67\%& 46.33\%& 3.88\%& 0&68\\
XTE J1739-285  & 64.13\%& 35.87\%& 27.23\%& 359 & 762\\
Aql X-1  & 52.53\%& 47.46\%&11.24\% & 357&489 \\
1RXS J180408.9-342058   & 53.85\%& 46.15\%& 3.78\%& 26& 57\\
4U 1624-490  & 53.25\%& 46.75\%& 3.85\%&71 &70\\
EXO 1722-363  & 53.74\%&  46.25\%& 4.15\%&42 &40\\
IGR J18483-0311   & 61.02\%& 38.98\%& 7.89\%& 80& 64\\
XTE J1807-294  & 63.98\%& 36.02\%& 3.74\%& 41& 25\\
XTE J1543-568  & 54.53\%& 45.47\%&7.64\% &54 &195\\
Terzan 5  & 64.34\%& 35.66\%&12.10\% &382 &504\\
XTE J1709–267   & 51.23\%& 48.77\%& 6.50\%&125 &251\\
\hline
\end{tabular}
\caption{Details of results from NSXBs outburst analysis.}
\label{neutron_star_details}
\end{table*}

%\clearpage
%\newpage

\section{Distinction based on statistical analysis}
\label{sec:results}
To understand the collective HID behaviour of different X-ray binary systems, we calculated the cumulative HID frequency distribution of NSXB sources, BH sources with and without `q' diagrams and show them in green, black and red stepped lines in the left panel of  \ref{fig: Histogram}. Each data point in the HID frequency distribution represents the fraction of hard-colour values at a given hardness for a specific class of objects, normalized by the total number of hard-colour values for that class. The normalized frequency distributions of neutron stars and black holes show a distinct separation: Neutron star binaries exhibit a single, skewed Gaussian-like distribution that peaks at a significantly harder value (green curve in the left panel of Fig. \ref{fig: Histogram}). 
Black holes with a canonical `q'-diagram display a more complex distribution, resembling the convolution of two skewed Gaussians. One peak lies at very low hardness (the softest part of the `q'-diagram), and the second peak occurs at a value softer than, but close to, the neutron star peak (red curve in Fig. \ref{fig: Histogram}). 
Black holes without a ‘q’-diagram show a simpler, single-peaked Gaussian distribution, with the peak located at a significantly softer value than that of neutron stars, further enhancing the contrast between the two classes (black curve in Fig. \ref{fig: Histogram}).

For further assessment, we plot the cumulative distribution of HID of all \maxi{} outbursts of one known black hole, GX 339-4 and a known neutron star source, GX 304-1, in the right panel of Figure \ref{fig: Histogram}.  As expected, the HID of GX 339-4 shows a double-peaked distribution with peaks at significantly low HID values ($\sim$0.025 and $\sim$0.52), while the HID distribution of GX 304-1 peaks at a higher HID value ($\sim$0.96). To predict the nature of the candidate source MAXI J1848-015, upon plotting the normalised hardness frequency distribution of the source, it is clear that the distribution (shown in grey shaded region in the right panel of \ref{fig: Histogram}) coincides with that of a neutron star rather than that of a black hole. Such an observation supports earlier studies that the X-ray binary MAXI J1848-015 hosts a neutron star as the accreting compact object \citep{pike2022maxi}.

In order to segregate outbursts and flares from the non-variable part of the light curve of each source, we have considered the original light curves and smoothed them using a Gaussian smoothing function with a $\pm$2 standard deviation. Such a filtering effect is shown in the \maxi{} lightcurve of 4U 1630-472 (plotted in yellow in fig. \ref{fig:selection}). The standard deviation was chosen so that the low, persistent count rate in the light curve can be skipped to avoid the overfitting of the light curve. A dynamic threshold for identifying outbursts, plotted in red in fig. \ref{fig:selection} is defined by combining two components: (1) the median of the smoothed light curve, which captures the baseline emission level of the source while suppressing short-term fluctuations. (2) An uncertainty term, given by the measurement error multiplied by a user-defined scaling factor. This term ensures that the threshold adapts to variations in data quality and persistent level. By adding these two terms, the threshold is not fixed but adjusts dynamically with both the intrinsic baseline flux and the statistical uncertainty of the observations. This allows genuine outbursts to be distinguished from small fluctuations in the light curve. The regions of the smoothed light curve above this threshold with a minimum duration of 15 days are selected as outbursts or flares (green regions in fig. \ref{fig:selection}). 

The durations obtained for all outbursts and flares in each source are further used to determine the percentage of sources present within a particular duration range. This is accomplished by breaking the flare/outburst duration into three primary regions. Region 1 is selected between 15--60 days, which typically corresponds to short and long X-ray flare durations for both black holes and neutron stars; region 2 is selected between 60--100 days, which corresponds to typical outbursts in the case of neutron stars but short outbursts for black holes, and region 3 for $>$100 days which corresponds to outbursts for both types of sources. However, for the sake of completeness, we have taken an additional region (region 4), which marks all outbursts with a duration $>$200 days. Since the number of outbursts is low, and owing to the inhomogeneity in the sample sizes for both sources, we have used a probability function to qualitatively visualise the nature of the distribution and a smoothed empirical cumulative distribution function (ECDF) to calculate the percentages (fig. \ref{fig:ecdf}). These quantities are independent of the sample size and are robust approaches. The probability of a source with a certain outburst duration being an NS or a BH was estimated using the Bayes theorem. The probability of a source with a certain outburst duration being a NS or a BH was estimated using the Bayes theorem as follows:
\begin{equation}
    P(S\ |\ O) = \dfrac{P(O\ |\ S)\cdot P(S)}{P(O)}
\end{equation}
where $P(S\ |\ O)$ is the probability of a source being of a certain type given the outburst duration, $P(O\ |\ S)$ is the probability of outburst duration given the type of source, $P(S)$ is the probability of the source being of a particular type and $P(O)$ is the probability of source-independent outburst duration. $P(S)$ is essentially a constant over the outburst duration range, estimated by dividing the number of outbursts/flares in a particular category by the total number of outburst/flare detections. $P(O\ |\ S)$ is estimated by performing a 1-D integration of the Gaussian kernel density estimate (KDE) for that particular source. The normalisation ($P(O)$) was carried out by integrating the KDE over the combined range of BH and NS detections. Thus, we have
\begin{equation}
    P(NS\ |\ O) = \dfrac{P(O\ |\ NS)\cdot P(NS)}{P(O)}
\end{equation}
\begin{equation}
    P(BH\ |\ O) = \dfrac{P(O\ |\ BH)\cdot P(BH)}{P(O)}
\end{equation}

where $P(NS\ |\ O)$ and $P(BH\ |\ O)$ are the respective probabilities of a source being a blackhole or a neutron star given the outburst duration.

The percentage in a region was extracted by integrating the ECDF within that region. In the case of a neutron star, a broad KDE mount is observed in the 15--100 days region, with a rapidly falling probability. The probability falls below 20\% above 100 days of outburst duration. A smaller, more spread-out bump-like structure is observed around 280 days due to the observation of longest NSXB outburst from XTE J1739-285 (Shown in Figure 38 in the Appendix). In the case of black holes, the distribution is more spread out, with some areas being covered in Region 4 (which was absent for NS). There are 24\%  more neutron star flares in the flaring region 1 compared to black hole flares. In region 2, neutron star flares are 3\% more than black hole outbursts. Thus, within the 15-100 days duration region, there are 27\% more neutron stars than black holes. However, 43\% of the black hole outbursts lie in region 3, with 15\% neutron star outbursts. Additionally, 19\% of the black hole outbursts (almost 1/5th the total number of outbursts) lie in region 4, while only 2\% of the neutron star outbursts lie in this region, forming a clear distinction between black hole and neutron star sources. Moreover, the black hole outburst duration has a mean value of 107.25 days and a median value of 61 days, while the neutron star outburst duration has a mean value of 52.62 days and a median value of 32 days.

\section{Discussion and Conclusion}

With the motivation of finding a model-independent scheme to distinguish BHXBs and NSXBs, we have analysed the entire archival data of \maxi{} and \swift{}/BAT lightcurves from 11 confirmed BHXB sources, 10 black hole candidates for which the nature and mass of the central accretor is not determined through reliable methods and 21 known NSXBs. We did so by considering \maxi{} lightcurves in soft (2--4 keV), medium (4--10 keV) and hard (10-20 keV) X-ray bands separately. We have processed and filtered individual one-day averaged lightcurves by removing negative counts, counts with errors higher than 100\% and following the steps suggested by \maxi{}. For each source, we calculate HID (the ratio of hard X-ray count rate to that of soft X-ray) values, flare and outburst durations, and perform statistical analysis from the entire sample. 
HID analysis consists of 4241 data points from NSXBs, BHXBs with and without `q' diagrams, shows that the cumulative frequency distribution of all NSXBs peaks around 1.72 $\pm$0.58 (1$\sigma$ width of the distribution) while the same for confirmed BHXBs with and without `q' diagrams peaks around 0.86 $\pm$0.45 and 0.31 $\pm$0.19, respectively. For further confirmation of our results, we have randomly picked a confirmed BHXB GX 339-4 and known NSXB GX 304-1 from our sample and studied their HID distribution as shown in the right panel of Fig. \ref{fig: Histogram}. Both distributions are clearly separated: GX 304-1 prefers harder values, while GX 339-4 prefer softer values. 

The distinct, harder peak in the HID frequency distribution of NSXBs may be consistent with the persistent existence of a Comptonized boundary layer close to the surface of the NS \citep{popham1995accretion,popham2001accretion}. A boundary layer consists of low-density, hot ($\ge$ 10$^8$ K) gas, with marginal radial (few hundreds of meters) and vertical extension, located at the boundary of the accretion disk and hard surface of the compact object \cite{popham2001accretion}. Such a layer, due to the thermal compaction, produces a hard X-ray flux which dominates above 5 keV. Such a layer, however, may not be present in BHXB systems due to the absence of a hard surface.  

Such an idea can be verified with a candidate source, MAXI J1848-015, which shows HID distribution similar to NSXBs, while, other studies found evidence that the accretion disk extended to the hard surface of the compact object. Therefore, the proposed simple scheme may be used to provide model-independent preliminary ideas on the nature of a newly discovered source through proposed HID analysis and can further be followed up by spectro-timing studies from other observatories for further confirmation. 

We have also calculated the flare and outburst duration of each source by carefully selecting long flares and outbursts longer than 15 days. The analysis of 267 outbursts and flares from neutron star X-ray binaries (NSXBs) and black hole X-ray binaries (BHXBs) reveals a clear trend: short outbursts (lasting less than 60 days) are much more likely to come from NSXBs, with a probability of 60–80\%. In contrast, NSXBs are very unlikely (less than 20\% probability) to produce long outbursts lasting over 120 days. Black hole systems show the opposite behavior: long outbursts are statistically favoured over short ones.
Such an observation can be explained by the size of the accretion disc that participates in the outburst/flaring activities. Since accretion disc radius scales with compact object mass, more massive black hole systems tend to burn accreted material over a longer time and over a larger area than NS systems. Hence, longer and more massive outbursts are common in BHXB systems rather than NSBXs. More studies are required to establish such a hypothesis.

\section*{Acknowledgements}
We thank the referee for constructive suggestions that improve the quality of the manuscript.
This research has made use of MAXI data provided by RIKEN, JAXA, and the MAXI team \citep{maxi09}. \swift{}/BAT transient monitor results provided by the Swift/BAT team.

%%Appendix
\appendix

\section{Analysis of black hole sources}
\label{Analysis of black hole sources appendix}
\subsection{MAXI J1348-630}
\begin{figure*}
    \centering
    \includegraphics[scale=0.60]{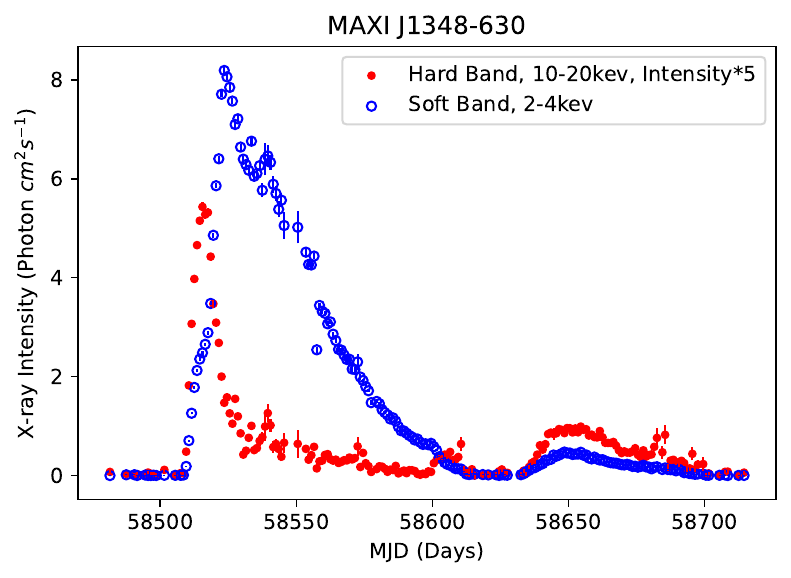}
    \includegraphics[scale=0.60]{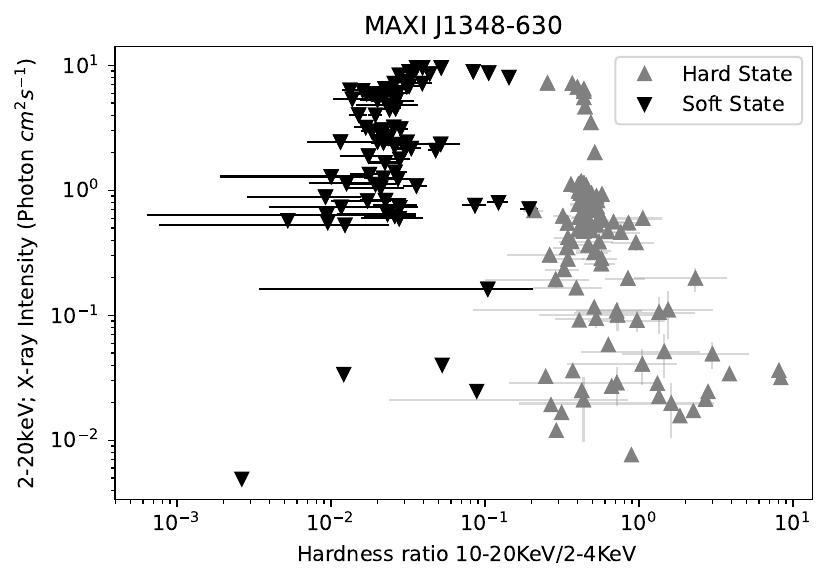}
    \captionsetup{labelformat=empty}  % Disable automatic "Figure X." prefix
    \caption{\textbf{Figure 15. }{\it Lightcurves and HID of MAXI J1348–630}: 
     Left panel: Zoomed view of the outburst observed with \maxi{} in the 2–4 keV (blue open circles) and 10–20 keV (red solid circles) bands. 
     Right panel: Corresponding Hardness Intensity Diagram (HID), where hardness is defined as the ratio of 10–20 keV to 2–4 keV intensity and intensity as the 2–20 keV count rate. Grey triangles denote the hard state and black inverted triangles denote the soft state.}

    \label{fig:MAXI J1348-630}
\end{figure*}

MAXI J1348-630 is categorized as a BHXB  with the mass of the compact object estimated within a range of 8.7--14.8 $M_\odot$ . In this observation, the \swift{} hard-band data are not available. The outburst is produced in the hard band and transitions into the soft band (left panel of Fig. \ref{fig:MAXI J1348-630}). The HID shows the conventional q-diagram (right panel of Fig. \ref{fig:MAXI J1348-630}).

\subsection{4U 1543-475}
\begin{figure*}
    \centering
    \includegraphics[scale=0.60]{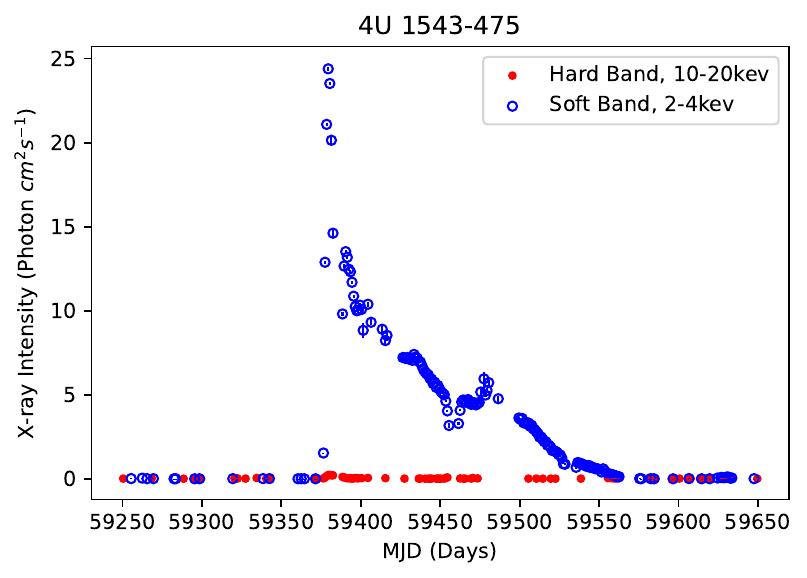}
    \includegraphics[scale=0.60]{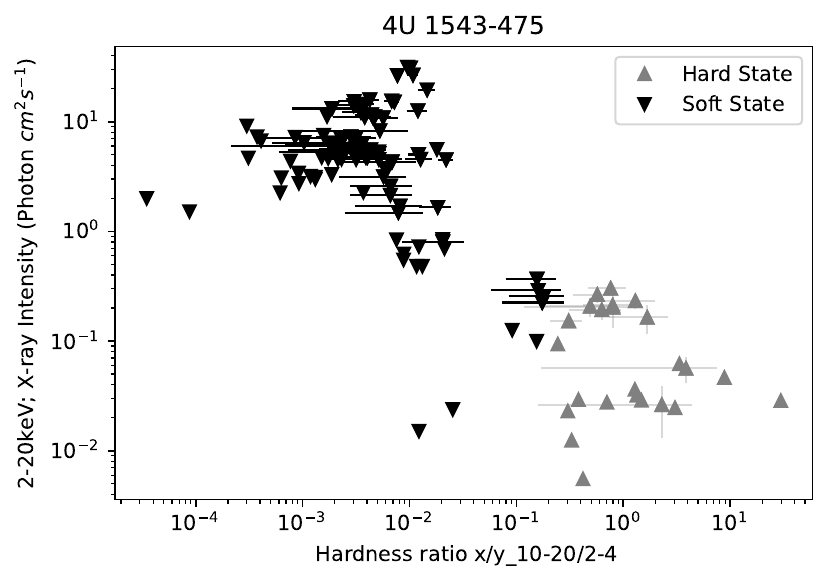}
    \captionsetup{labelformat=empty}  % Disable automatic "Figure X." prefix
    \caption{\textbf{Figure 16. }{\it Lightcurves and HID of 4U 1543–475}: 
    Left panel: Zoomed view of the outburst observed with \maxi{} in the 2–4 keV (blue open circles) and 10–20 keV (red solid circles) bands.
    Right panel: Corresponding Hardness Intensity Diagram (HID), where hardness is defined as the ratio of 10–20 keV to 2–4 keV intensity and intensity as the 2–20 keV count rate. Grey triangles denote the hard state and black inverted triangles denote the soft state.}

    \label{fig:4U 1543-475}
\end{figure*}

4U 1543-475 is a BHXB  with a BH mass of around $\sim$9.4$M_\odot$ . Only one outburst is observed in the soft band (left panel of Fig. \ref{fig:4U 1543-475}).

\subsection{GS 1354-64}
\begin{figure*}
    \centering
    \includegraphics[scale=0.60]{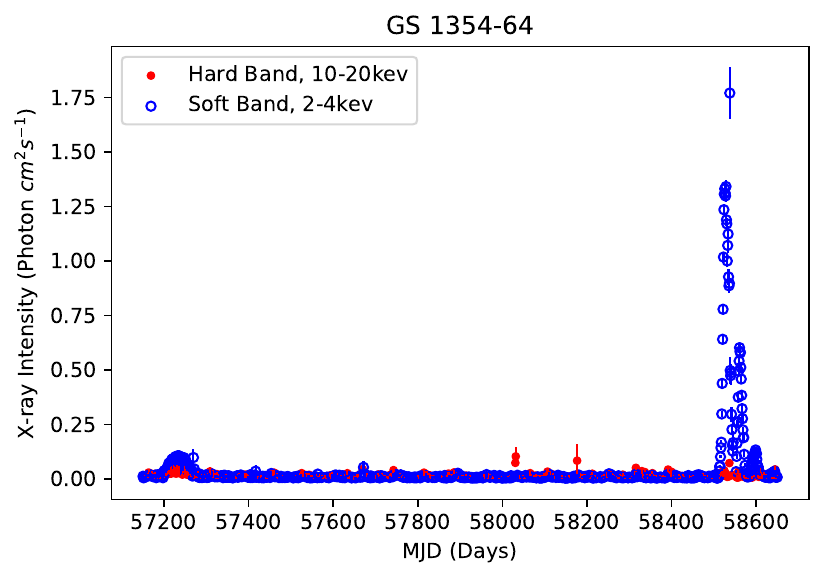}
    \includegraphics[scale=0.60]{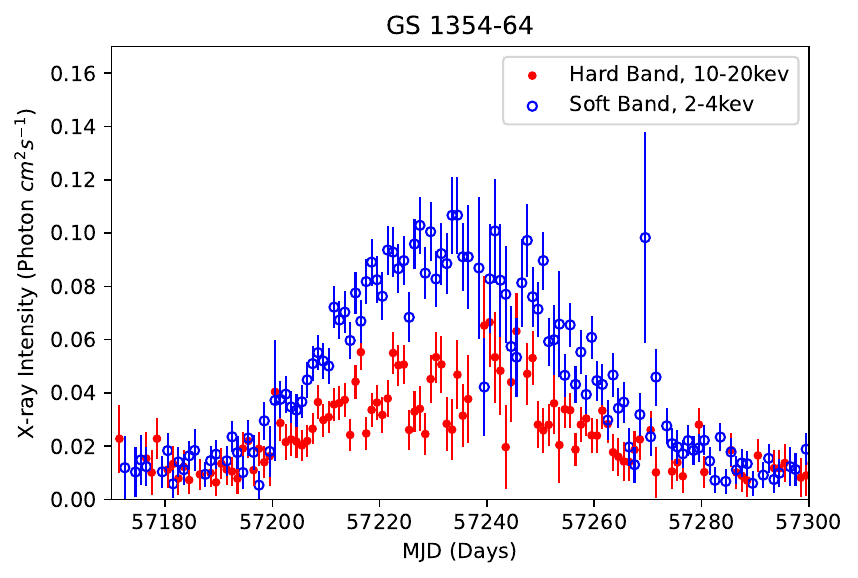}
    \includegraphics[scale=0.60]{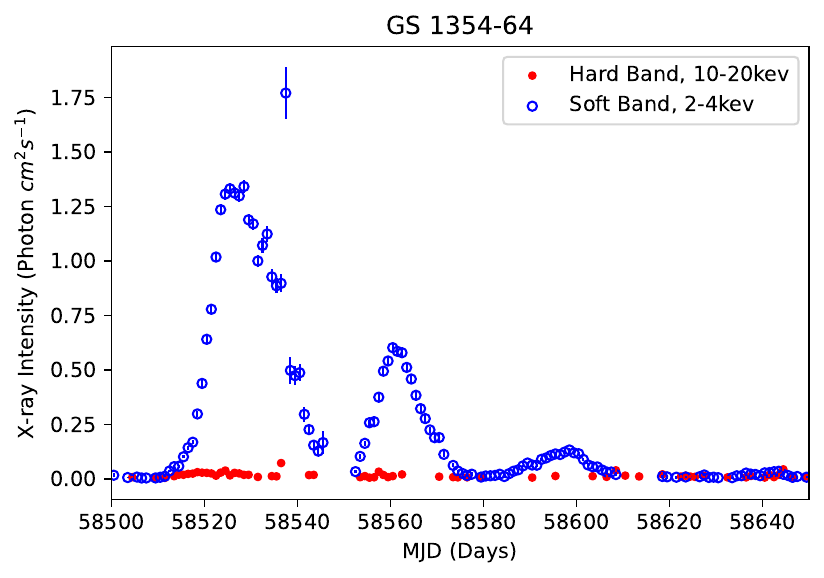}
    \includegraphics[scale=0.60]{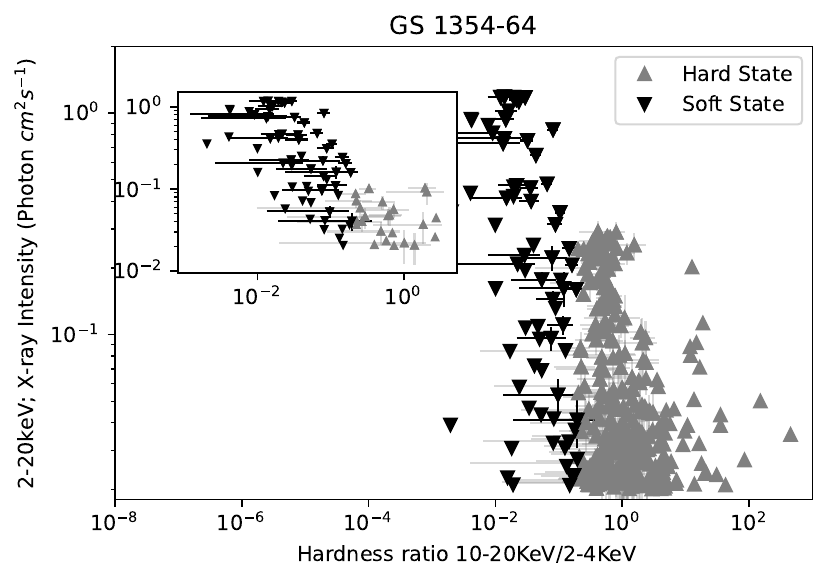}
    \captionsetup{labelformat=empty}  % Disable automatic "Figure X." prefix
    \caption{\textbf{Figure 17. }{\it Lightcurves and HID of GS 1354–64}: 
    Top left panel: Lightcurves obtained with \maxi{} in the 2–4 keV (blue open circles) and 10-20 keV (red solid circles) bands, during the full observation period.
    Top right panel: Zoomed view of the initial hard-band dominated (failed) outburst. 
    Bottom left panel: Zoomed view of a major outburst showing a gradual decline followed by several partial re-brightening events. 
    Bottom right panel: Corresponding Hardness Intensity Diagram (HID), where hardness is defined as the ratio of 10–20 keV to 2–4 keV intensity and intensity as the 2–20 keV count rate. The inset shows the HID during the main outburst phase.}

    \label{fig:GS 1354-64}
\end{figure*}

GS 1354-64 is classified as a low-mass BHXB. The mass of the compact object from the catalog of low-mass X-ray binary is $\sim$7.59$M_\odot$. Lightcurves spanning nearly $\sim$3.8 years are derived from \maxi{} (top left panel of Fig. \ref{fig:GS 1354-64}). On the left we observe a failed outburst (zoomed in view present in top right panel of Fig. \ref{fig:GS 1354-64}). An outburst is observed at the end of the light curve which actually reveals two outbursts on further zoom (bottom left panel of Fig. \ref{fig:GS 1354-64}). The second one is a failed outburst.

\subsection{MAXI J1910-057}
\begin{figure*}
    \centering
    \includegraphics[scale=0.60]{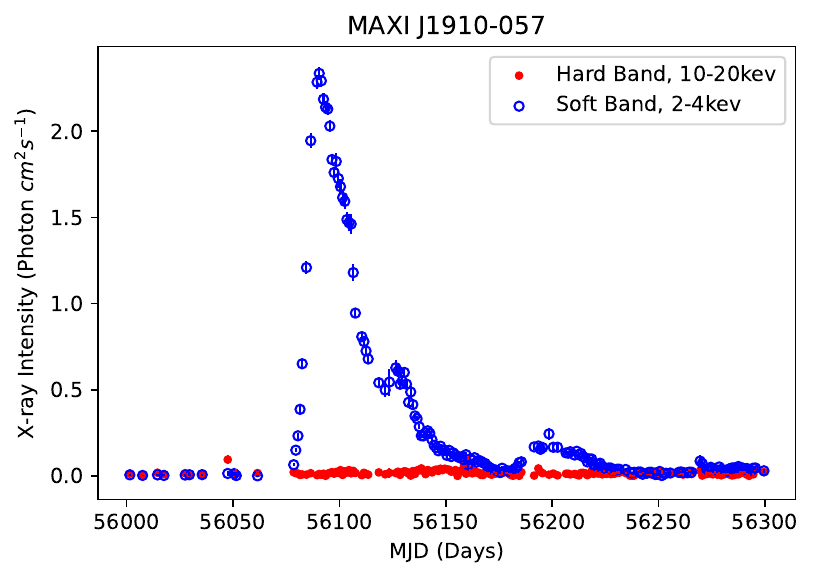}
    \includegraphics[scale=0.60]{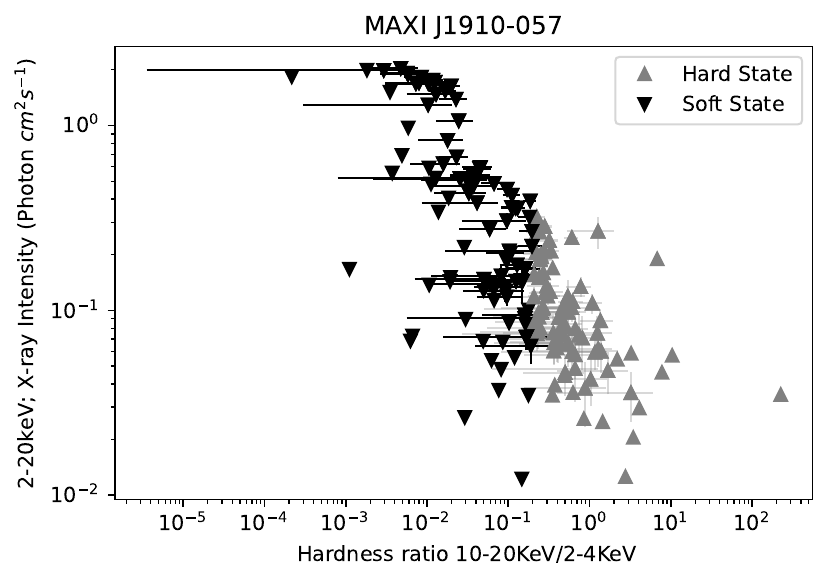}
    \captionsetup{labelformat=empty}  % Disable automatic "Figure X." prefix
    \caption{\textbf{Figure 18. }{\it Lightcurves and HID of MAXI J1910–057}: 
    Left panel: Zoomed view of the outburst observed with \maxi{} in the 2–4 keV (blue open circles) and 10–20 keV (red solid circles) bands. 
    Right panel: Corresponding Hardness Intensity Diagram (HID), where hardness is defined as the ratio of 10–20 keV to 2–4 keV intensity and intensity as the 2–20 keV count rate. Grey triangles denote the hard state and black inverted triangles denote the soft state.}

    \label{fig:MAXI J1910-057}
\end{figure*}

MAXI J1910-057 is estimated as a BHXB, where the mass of the central object lies in the range of 6.31-13.65$M_\odot$. An outburst is observed in the soft band (left panel of Fig. \ref{fig:MAXI J1910-057})

\subsection{GRS 1716-249}
\begin{figure*}
    \centering
    \includegraphics[scale=0.60]{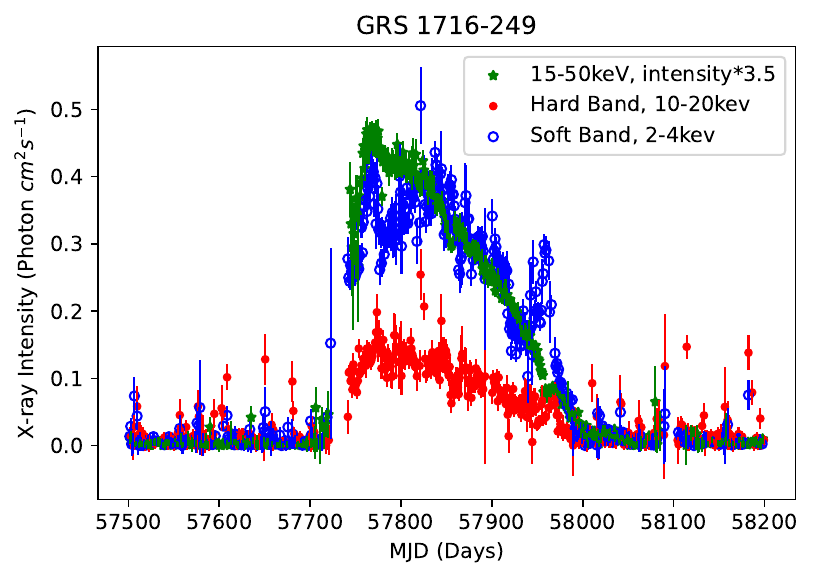}
    \includegraphics[scale=0.60]{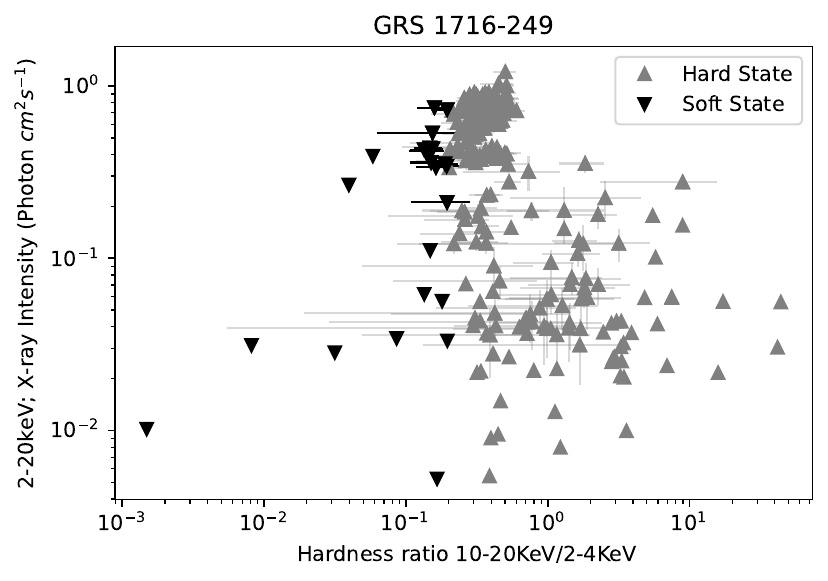}
    \captionsetup{labelformat=empty}  % Disable automatic "Figure X." prefix
    \caption{\textbf{Figure 19. }{\it Lightcurves and HID of GRS 1716–249}: 
     Left panel: Zoomed view of the outburst observed with \maxi{} in the 2–4 keV (blue open circles) and 10–20 keV (red solid circles) bands, along with \swift{}/BAT 15–50 keV data (green stars). The hard-band peak precedes the soft-band peak during the outburst. The \swift{} lightcurve is scaled by a factor of 3.5 for clarity. 
     Right panel: Corresponding Hardness Intensity Diagram (HID), where hardness is defined as the ratio of 10–20 keV to 2–4 keV intensity and intensity as the 2–20 keV count rate. Grey triangles denote the hard state and black inverted triangles denote the soft state.}

    \label{fig: GRS 1716−249}
\end{figure*}

With a central object mass of 4.4-9.6$M_\odot$, the source is categorized as a BHXB. An outburst is observed that transitions from the hard band to the soft band relatively quickly in a matter of $\sim$25-50 days. (left panel of Fig. \ref{fig: GRS 1716−249}). A somewhat failed outburst is also observed in the soft band during the gradual decay of the primary outburst.

\subsection{GRS 1739-278}
\begin{figure*}
    \centering
    \includegraphics[scale=0.60]{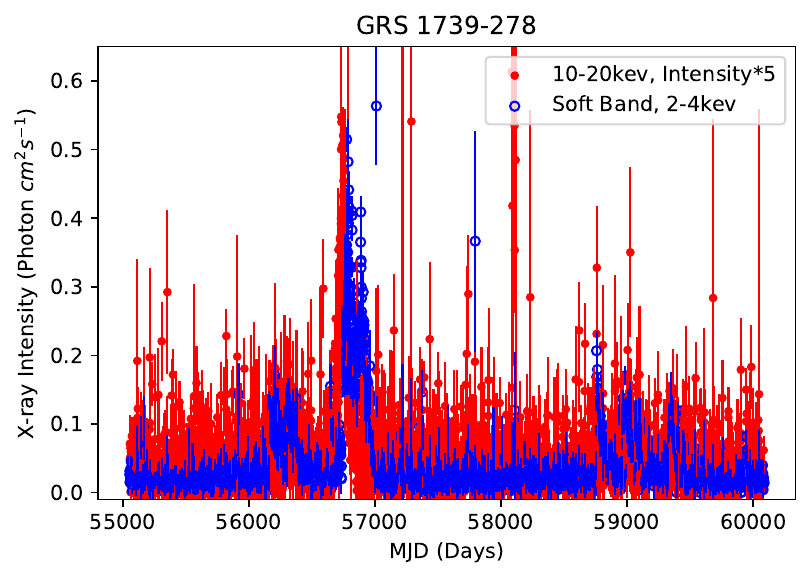}
    \includegraphics[scale=0.60]{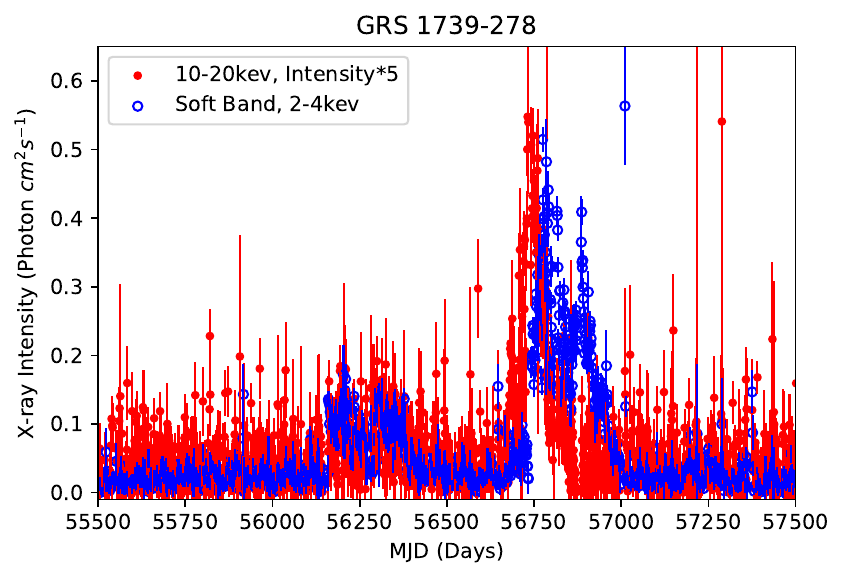}
    \includegraphics[scale=0.60]{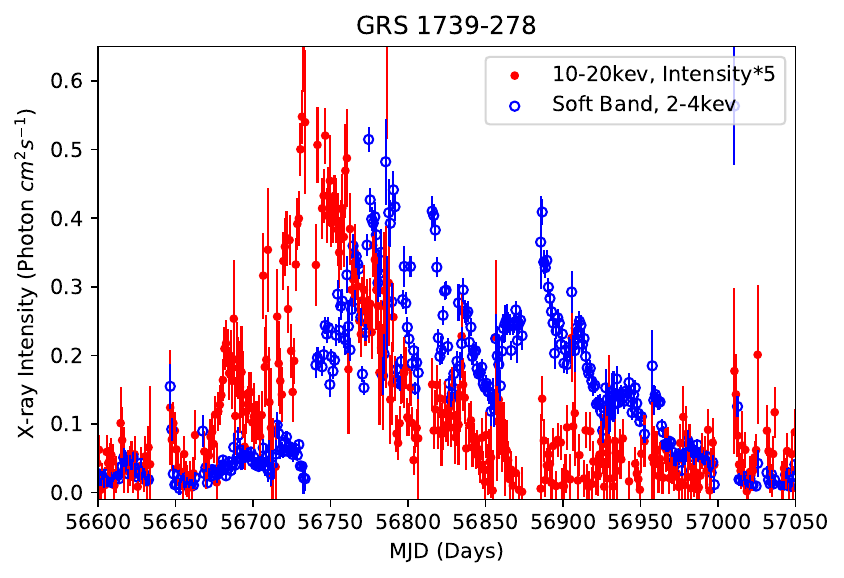}
    \includegraphics[scale=0.60]{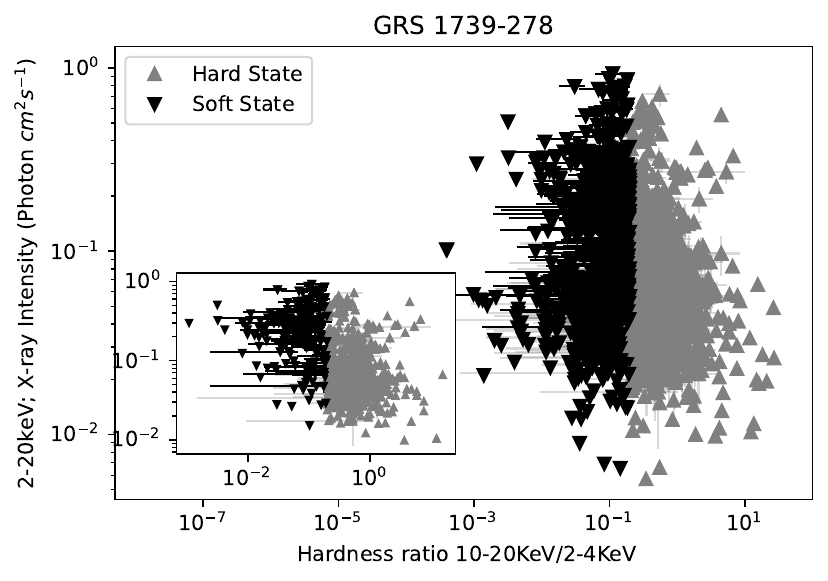}
    \captionsetup{labelformat=empty}  % Disable automatic "Figure X." prefix
    \caption{\textbf{Figure 20. }{\it Lightcurves and HID of GRS 1739–278}: 
    Top left panel: Lightcurves obtained with \maxi{} in the 2–4 keV (blue open circles) and 10–20 keV (red solid circles) bands, during the full observation period. The lightcurve shows a major outburst peak along with several hard-band dominated (failed) outbursts. 
    Top right panel: Zoomed view of the first failed outburst and the main intensity peak. 
    Bottom left panel: Zoomed view of the outburst phase highlighting variability in the intensity. 
    Bottom right panel: Corresponding Hardness Intensity Diagram (HID), where hardness is defined as the ratio of 10–20 keV to 2–4 keV intensity and intensity as the 2–20 keV count rate. The inset focuses on the HID during the main outburst as shown in the left bottom figure.}

    \label{fig: GRS 1739-278}
\end{figure*}

The GRS 1739-278 is an accreting transient black hole with mass in the range of 4-10$M_\odot$. One major outburst is observed along with a few failed outbursts (top left panel of Fig. \ref{fig: GRS 1739-278}). Once again, we see that the outburst transitions from a hard state to a soft state (bottom left panel of Fig. \ref{fig: GRS 1739-278}).

\subsection{MAXI J1543-564} 
\begin{figure*}
    \centering
    \centering
    \includegraphics[scale=0.60]{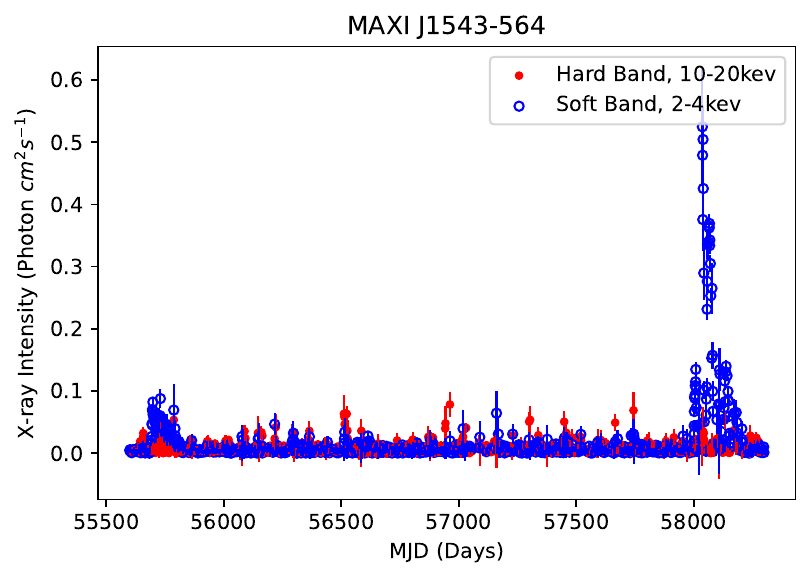}
    \includegraphics[scale=0.60]{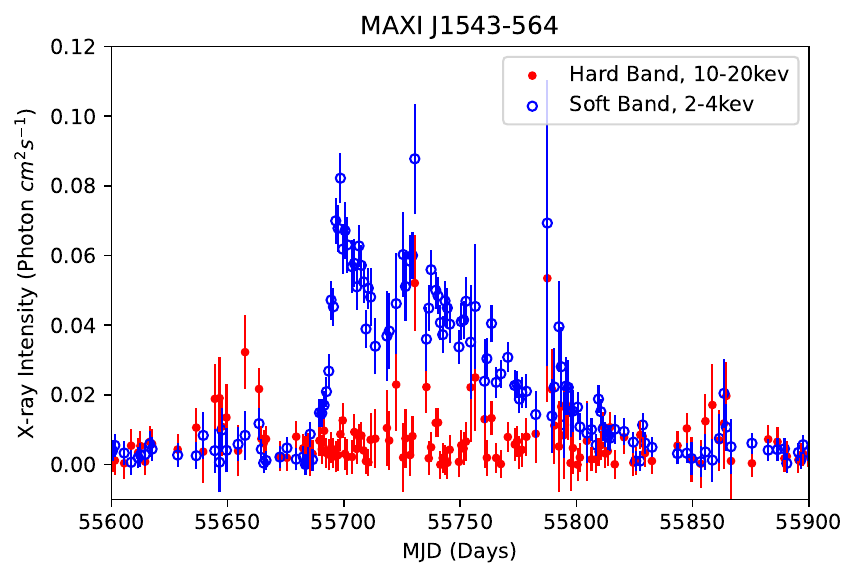}
    \includegraphics[scale=0.60]{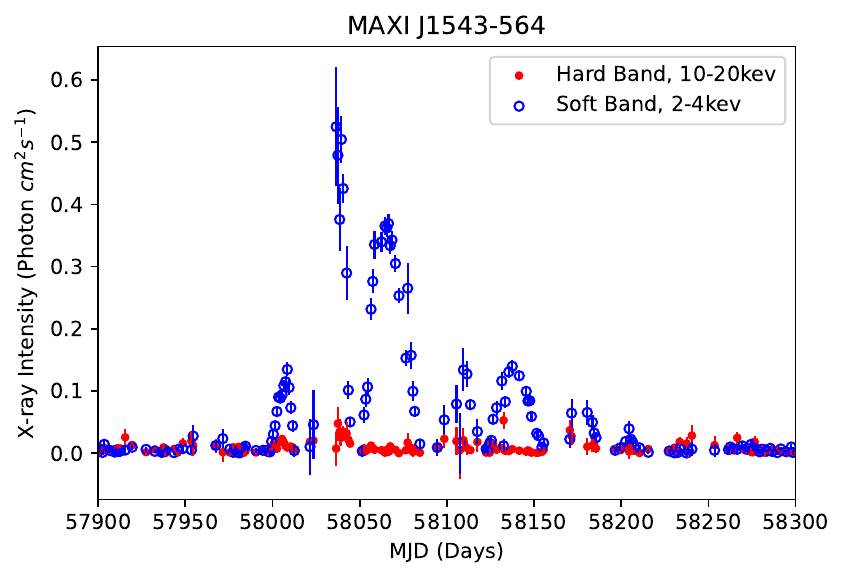}
    \includegraphics[scale=0.60]{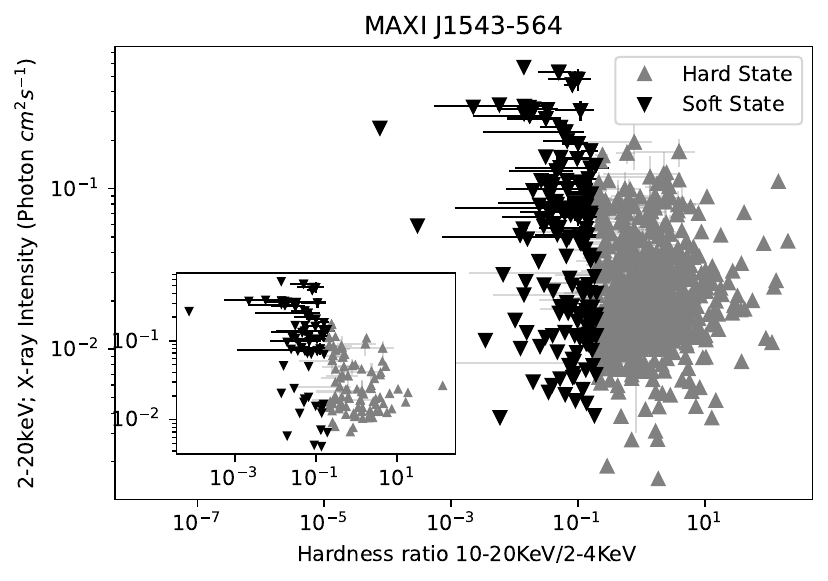}
    \captionsetup{labelformat=empty}  % Disable automatic "Figure X." prefix
    \caption{\textbf{Figure 21. }{\it Lightcurves and HID of MAXI J1543–564}: 
    Top left panel: Lightcurves obtained with \maxi{} in the 2–4 keV (blue open circles) and 10–20 keV (red solid circles) bands, over the full observation period. The lightcurve shows a major outburst along with an initial hard-band dominated (failed) outburst.
    Top right panel: Zoomed view of the failed outburst, showing the absence of a distinct peak. 
    Bottom left panel: Zoomed view of the main outburst, which decays through multiple re-brightening events. 
    Bottom right panel: Corresponding Hardness Intensity Diagram (HID) of the full observation period, where hardness is defined as the ratio of 10–20 keV to 2–4 keV intensity and intensity as the 2–20 keV count rate. The inset shows the HID during the main outburst in the bottom left panel.}

    \label{fig:MAXI J1543-564}
\end{figure*}

MAXI J1543-564 is a low-mass BHXB identified with black hole mass 12.6–14.0$M_\odot$. A major outburst can be observed in left panel of Fig. \ref{fig:MAXI J1543-564} along with a failed outbursts at the beginning of the observation (top right panel of Fig. \ref{fig:MAXI J1543-564} shows the zoomed in view). The view of the major outburst (bottom left panel of \ref{fig:MAXI J1543-564}) is zoomed in to show a series of outbursts as the major one decays.

\subsection{Swift J1728.9-3613}
\begin{figure*}
    \centering
    \includegraphics[scale=0.60]{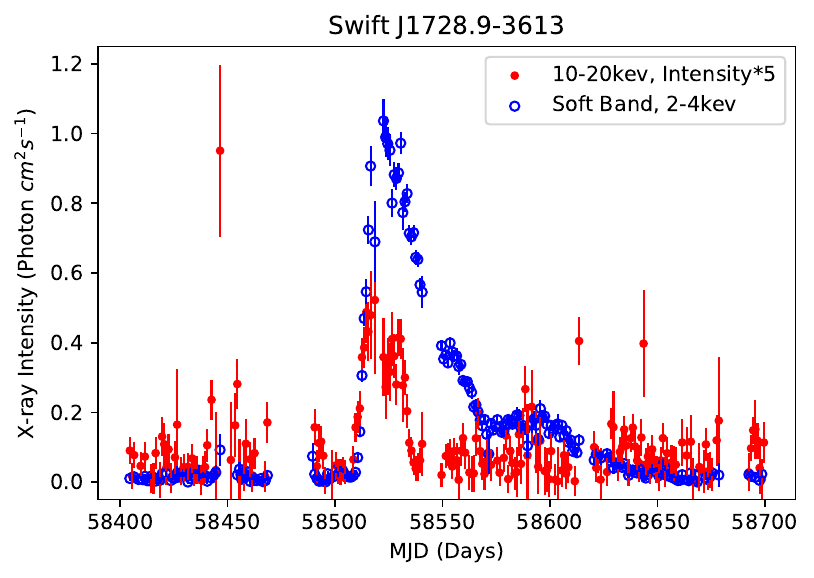}
    \includegraphics[scale=0.60]{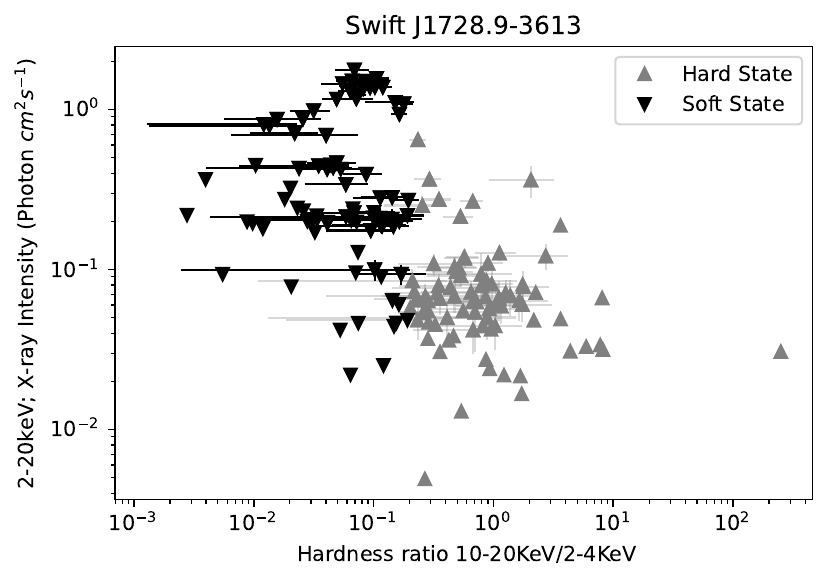}
    \captionsetup{labelformat=empty}  % Disable automatic "Figure X." prefix
    \caption{\textbf{Figure 22. }{\it Lightcurves and HID of Swift J1728.9–3613}: 
    Left panel: Zoomed view of the outburst observed with \maxi{} in the 2–4 keV (blue open circles) and 10–20 keV (red solid circles) bands.
    Right panel: Corresponding Hardness Intensity Diagram (HID), where hardness is defined as the ratio of 10–20 keV to 2–4 keV intensity and intensity as the 2–20 keV count rate. Grey triangles denote the hard state and black inverted triangles denote the soft state.}

    \label{fig:Swift J1728.9-3613}
\end{figure*}

Swift J1728.9-3613 has been identified as a BHXB candidate. An outburst is observed in the light curve (left panel of Fig. \ref{fig:Swift J1728.9-3613})

\subsection{MAXI J1535-571}
MAXI J1535-571 is classified as a BHXB candidate based on the X-ray emissions observed from the object. An outburst is observed with another failed outburst during the decay of the primary one.

\subsection{MAXI J1659-152}

MAXI J1659-152 is classified as a candidate for BHXB. The outburst initially occurs in the hard band and transitions to the soft band (left panel of Fig. \ref{fig:MAXI J1659-152}) in $\sim$25 days.

\subsection{EXO 1846-031}
EXO 1846-031 is identified as a black-hole candidate binary source based on the X-ray emissions observed from the object. Once again, the observed outburst transitions from the hard band to the soft band within a period of $\sim$50 days.

\subsection{MAXI J1803-298}
MAXI J1803-298 is categorized as a BHXB candidate. An outburst is visible during the observation period followed by a failed outburst during its decay.

\subsection{MAXI J1631-479}

MAXI J1631-479 is classified as a BHXB candidate. The lightcurve presents a prominent outburst peak (top left panel of Fig. \ref{fig:MAXI J1631-479}) and a few failed outbursts throughout the observation period (zoomed in view demonstrated in the top right panel of Fig. \ref{fig:MAXI J1631-479}).

\subsection{MAXI J1848-015}
MAXI J1848-015 is identified as a BHXB candidate. With a failed outburst (left panel of Fig. \ref{fig: MAXI J1848-015}).

\subsection{MAXI J1305-704}

MAXI J1305-704 is predicted to be a BHXB candidate. The left panel of Fig. \ref{fig: MAXI J1305-704} indicates a failed outburst. 

\subsection{MAXI J1836-194}

MAXI J1836-194 is categorized as a BHXB candidate. A failed outburst is observed in the light curve (left panel of Fig. \ref{fig: MAXI J1836-194}).

\begin{figure*}
    \centering
    \includegraphics[scale=0.60]{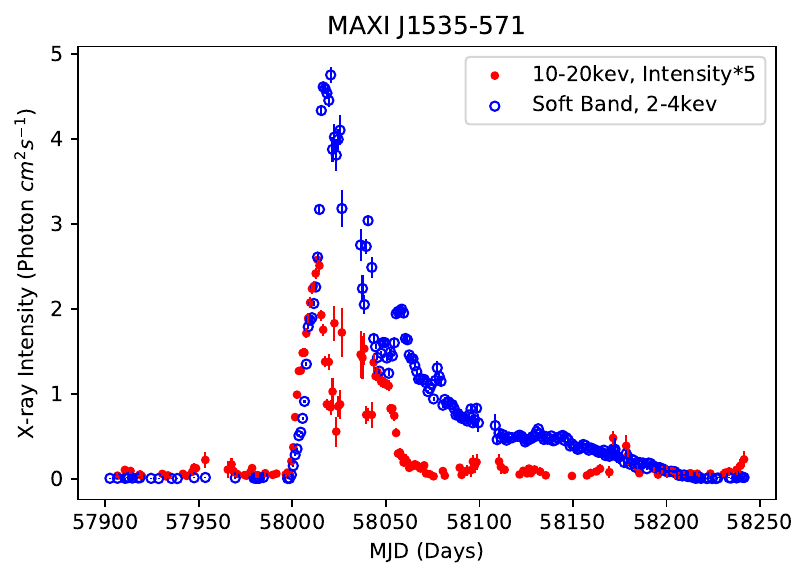}
    \includegraphics[scale=0.60]{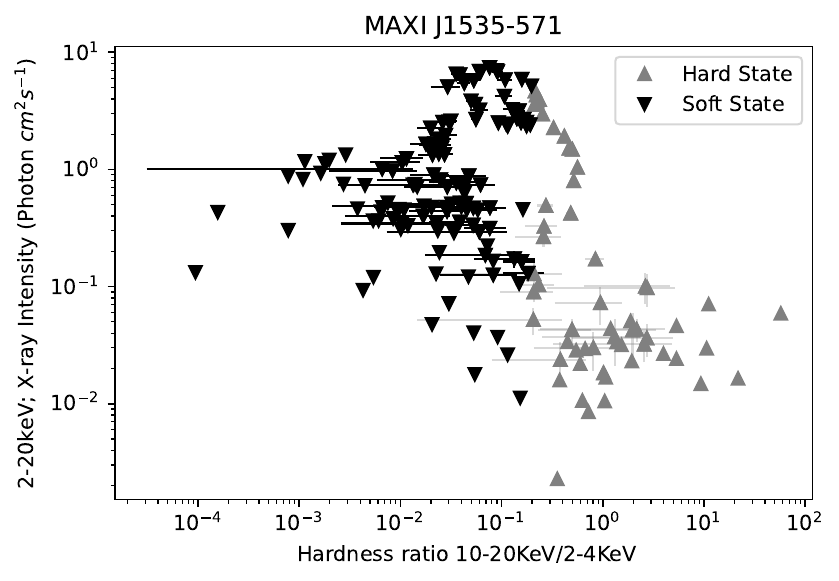}
    \captionsetup{labelformat=empty}  % Disable automatic "Figure X." prefix
    \caption{\textbf{Figure 23. }{\it Lightcurves and HID of MAXI J1535–571}: 
    Left panel: Zoomed view of the outburst observed with \maxi{} in the 2–4 keV (blue open circles) and 10–20 keV (red solid circles) bands.
   Right panel: Corresponding Hardness Intensity Diagram (HID), where hardness is defined as the ratio of 10–20 keV to 2–4 keV intensity and intensity as the 2–20 keV count rate. Grey triangles denote the hard state and black inverted triangles denote the soft state.}

    \label{fig:MAXI J1535-571}
\end{figure*}

\begin{figure*}
    \centering
    \includegraphics[scale=0.60]{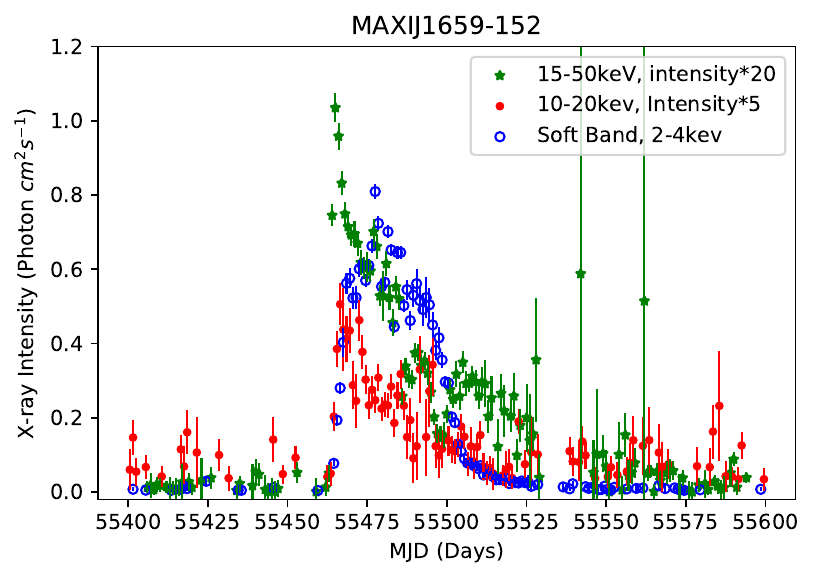}
    \includegraphics[scale=0.60]{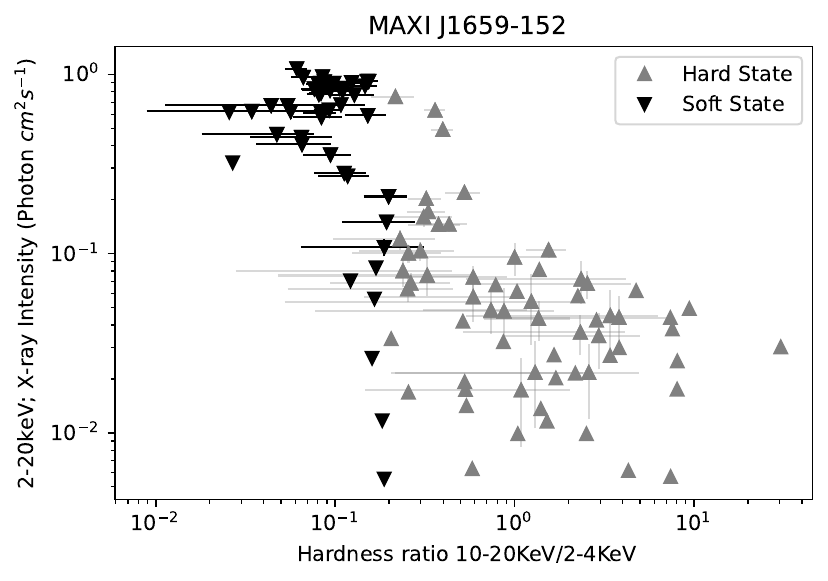}
    \captionsetup{labelformat=empty}  % Disable automatic "Figure X." prefix

    \caption{\textbf{Figure 24. }{\it Lightcurves and HID of MAXI J1659–152}: 
    Left panel: Zoomed view of the outburst observed with \maxi{} in the 2–4 keV (blue open circles) and 10–20 keV (red solid circles) bands, along with \swift{}/BAT 15–50 keV data (green stars). The hard-band peak precedes the soft-band peak during the outburst. The \swift{} lightcurve is scaled by a factor of 20 for clarity. 
    Right panel: Corresponding Hardness Intensity Diagram (HID), where hardness is defined as the ratio of 10–20 keV to 2–4 keV intensity and intensity as the 2–20 keV count rate. Grey triangles denote the hard state and black inverted triangles denote the soft state.}

\label{fig:MAXI J1659-152}
\end{figure*}

 \begin{figure*}
    \centering
    \includegraphics[scale=0.60]{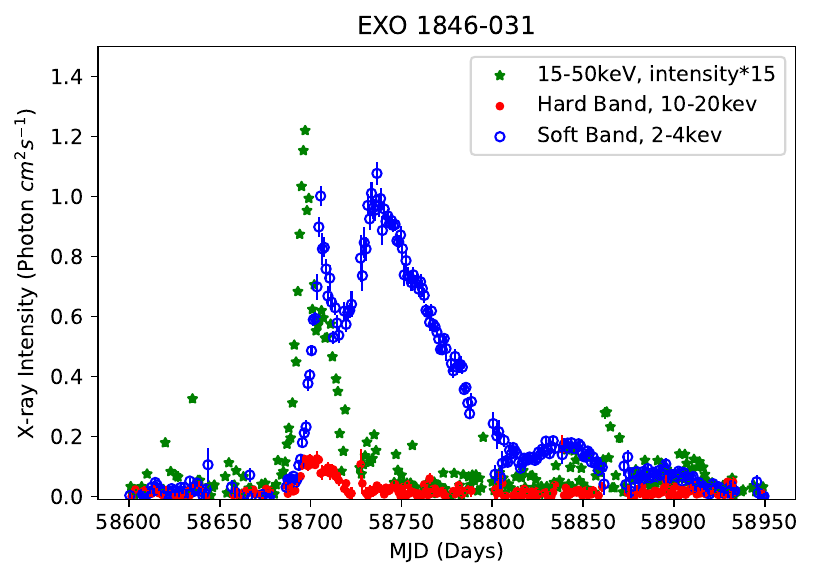}
    \includegraphics[scale=0.60]{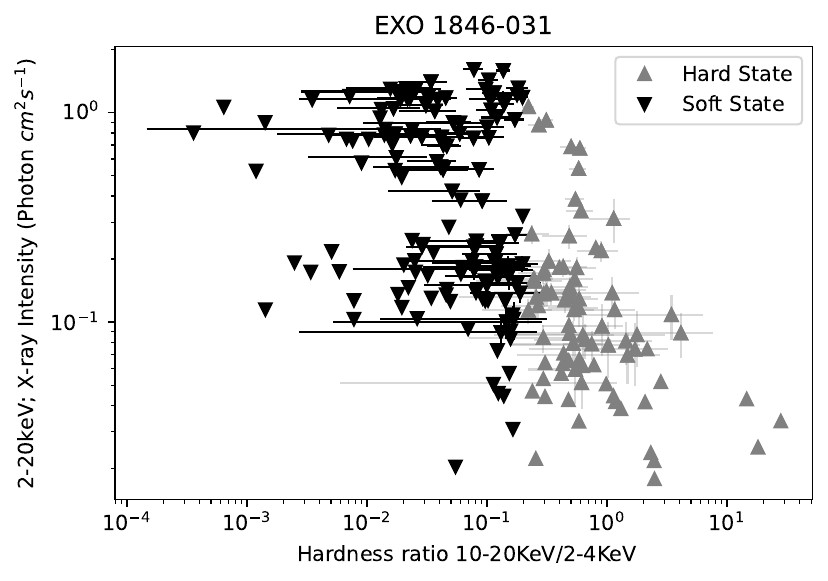}
    \captionsetup{labelformat=empty}  % Disable automatic "Figure X." prefix
    \caption{\textbf{Figure 25. }{\it Lightcurves and HID of EXO 1846–031}: 
    Left panel: Zoomed view of the outburst observed with \maxi{} in the 2–4 keV (blue open circles) and 10–20 keV (red solid circles) bands, along with \swift{}/BAT 15–50 keV data (green stars). The hard-band peak precedes the soft-band peak during the outburst, and a secondary soft-band enhancement is visible shortly after the main peak. The \swift{} lightcurve is scaled by a factor of 15 for clarity. 
    Right panel: Corresponding Hardness Intensity Diagram (HID), where hardness is defined as the ratio of 10–20 keV to 2–4 keV intensity and intensity as the 2–20 keV count rate. Grey triangles denote the hard state and black inverted triangles denote the soft state.}

    \label{fig: EXO 1846-031}
\end{figure*}

\begin{figure*}
    \centering
    \includegraphics[scale=0.60]{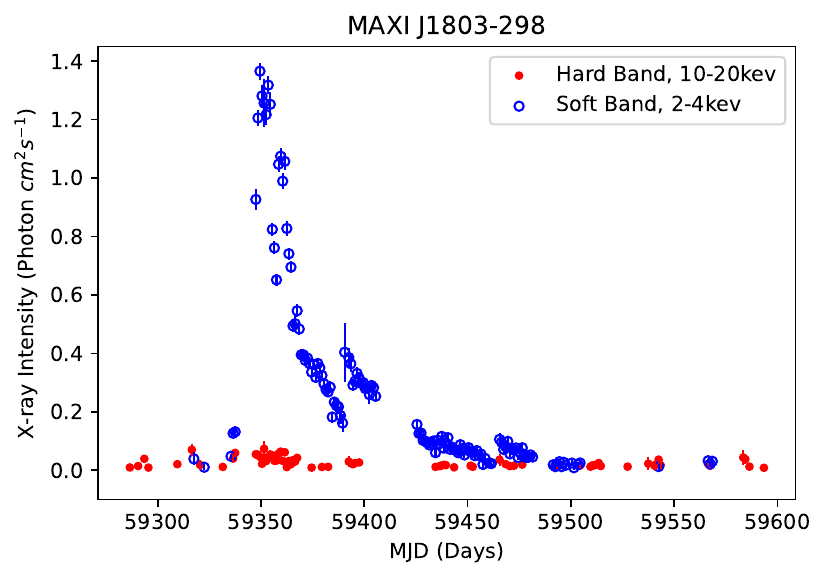}
    \includegraphics[scale=0.60]{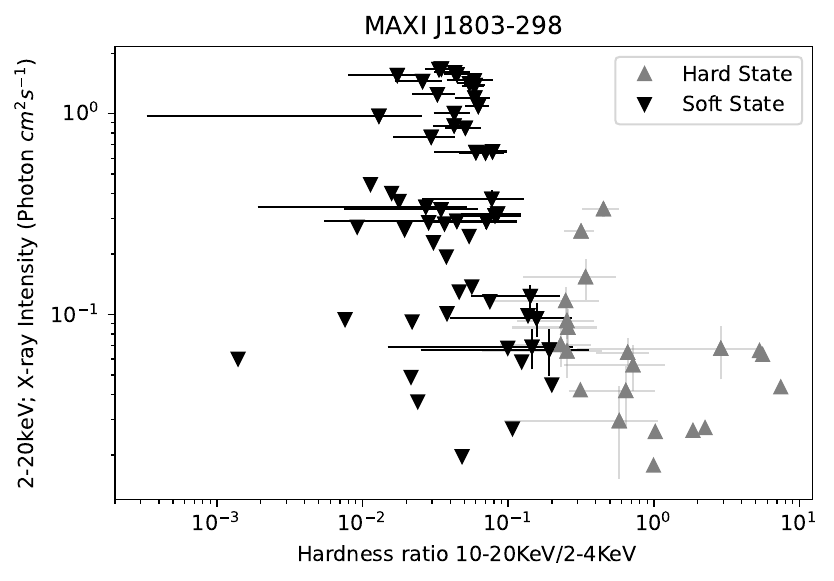}
    \captionsetup{labelformat=empty}  % Disable automatic "Figure X." prefix
    \caption{\textbf{Figure 26. }{\it Lightcurves and HID of MAXI J1803–298}: 
    Left panel: Zoomed view of the outburst observed with \maxi{} in the 2–4 keV (blue open circles) and 10–20 keV (red solid circles) bands. The hard-band peak precedes the soft-band peak during the outburst. A secondary soft-band enhancement is visible during the declining phase of the outburst.
    Right panel: Corresponding Hardness Intensity Diagram (HID), where hardness is defined as the ratio of 10–20 keV to 2–4 keV intensity and intensity as the 2–20 keV count rate. Grey triangles denote the hard state and black inverted triangles denote the soft state.}

    \label{fig: MAXI J1803-298}
\end{figure*}

\begin{figure*}
    \centering
    \includegraphics[scale=0.60]{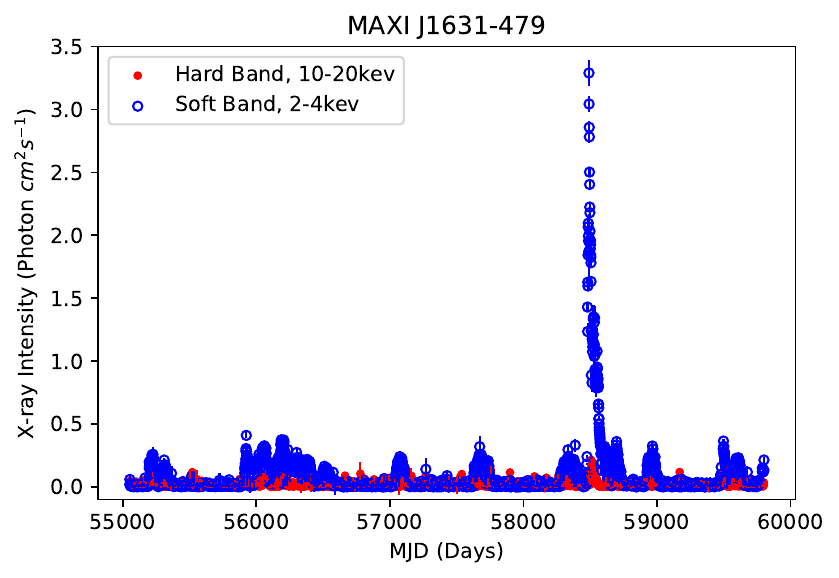}
    \includegraphics[scale=0.60]{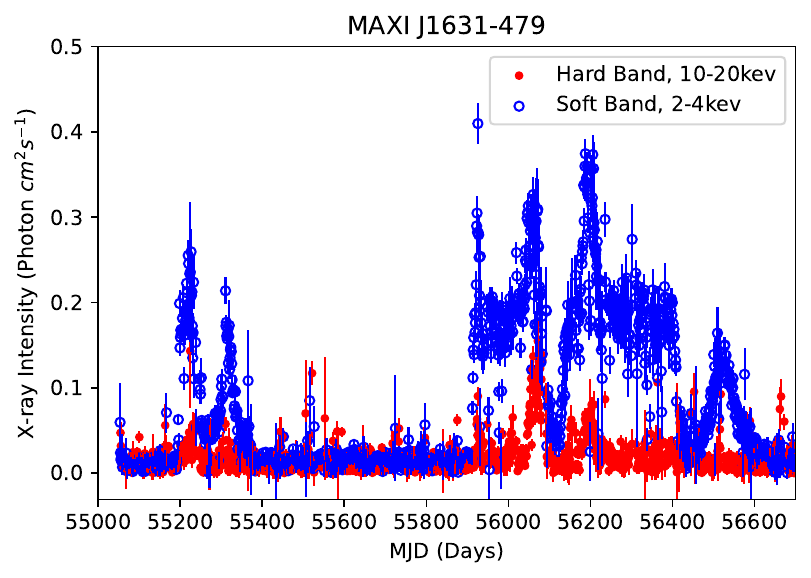}
    \includegraphics[scale=0.60]{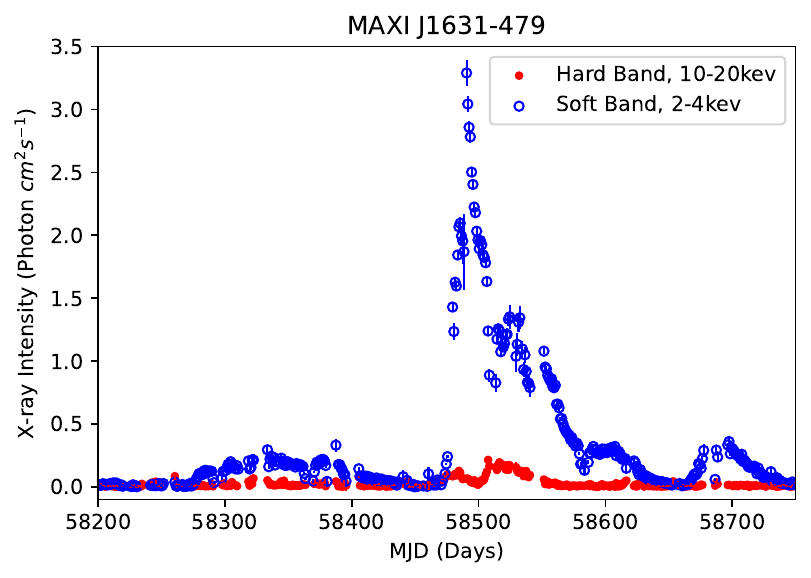}
    \includegraphics[scale=0.60]{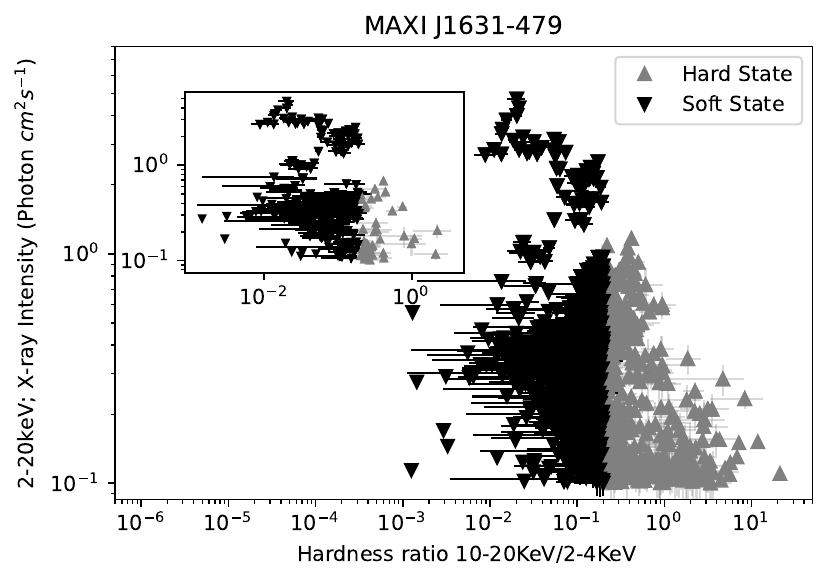}
    \captionsetup{labelformat=empty}  % Disable automatic "Figure X." prefix

    \caption{\textbf{Figure 27. }{\it Lightcurves and HID of MAXI J1631–479}: 
    Top left panel: Lightcurves obtained with \maxi{} in the 2–4 keV (blue open circles) and 10–20 keV (red solid circles) bands, showing a prominent outburst and several hard-band dominated (failed) outbursts during the full observation period. 
    Top right panel: Zoomed view of a representative failed outburst. 
    Bottom left panel: Zoomed view of the main outburst highlighting its temporal evolution. 
    Bottom right panel: Corresponding Hardness Intensity Diagram (HID) of the entire lightcurve, where hardness is defined as the ratio of 10–20 keV to 2–4 keV intensity and intensity as the 2–20 keV count rate. The inset shows the HID during the principal outburst in the bottom left panel.}

    \label{fig:MAXI J1631-479}
\end{figure*}

\begin{figure*}
    \centering
    \includegraphics[scale=0.60]{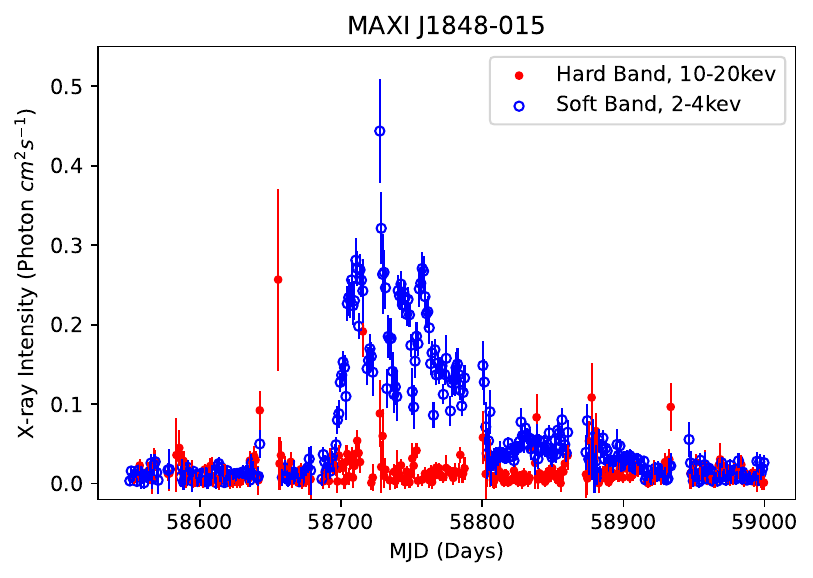}
    \includegraphics[scale=0.60]{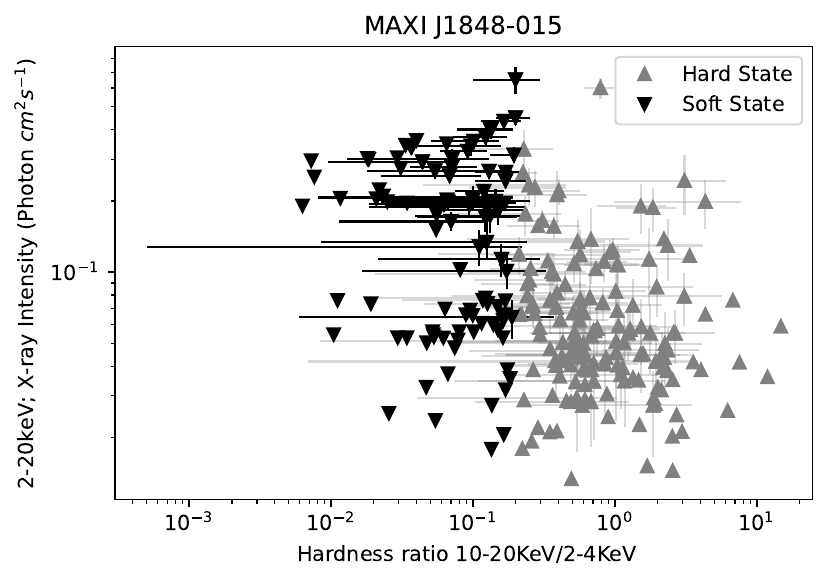}
    \captionsetup{labelformat=empty}  % Disable automatic "Figure X." prefix

    \caption{\textbf{Figure 28. }{\it Lightcurves and HID of MAXI J1848–015}: 
    Left panel: Zoomed view of the outburst observed with \maxi{} in the 2–4 keV (blue open circles) and 10–20 keV (red solid circles) bands.
    Right panel: Corresponding Hardness Intensity Diagram (HID), where hardness is defined as the ratio of 10–20 keV to 2–4 keV intensity and intensity as the 2–20 keV count rate. Grey triangles denote the hard state and black inverted triangles denote the soft state.}

    \label{fig: MAXI J1848-015}
\end{figure*}

\begin{figure*}
    \centering
    \includegraphics[scale=0.60]{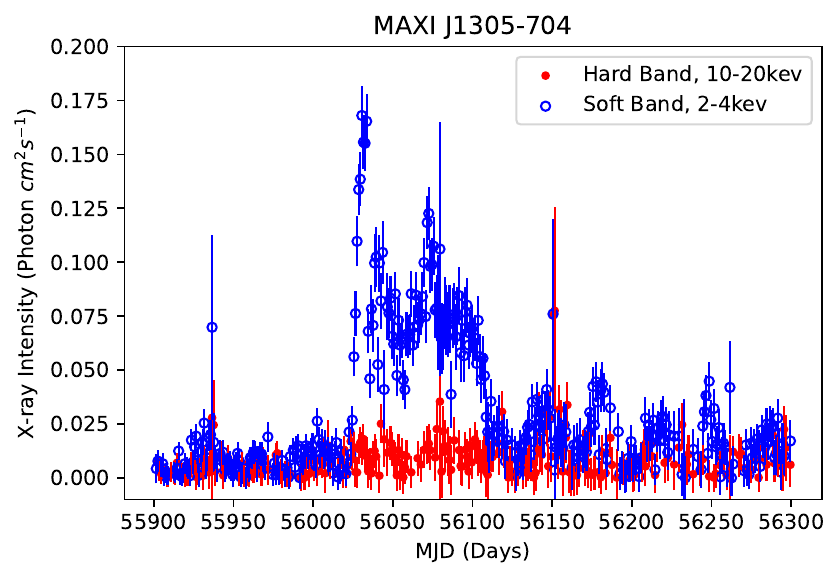}
    \includegraphics[scale=0.60]{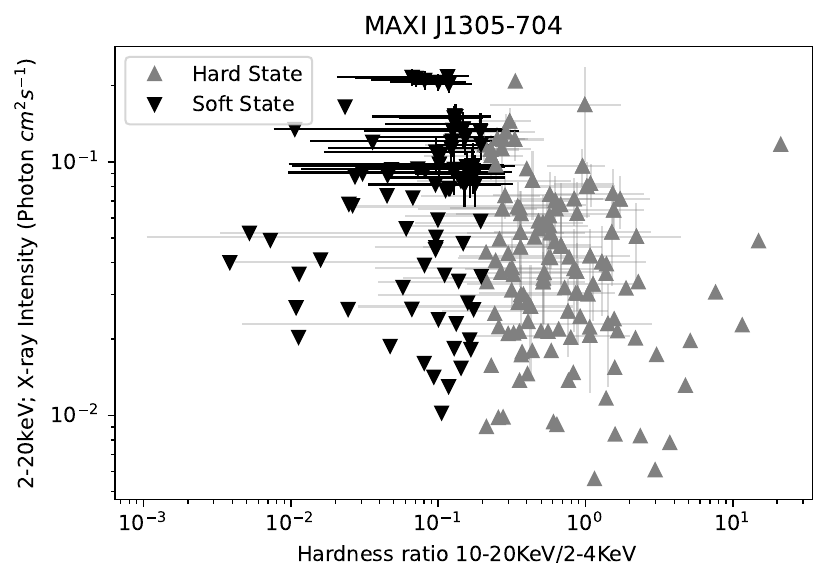}
    \captionsetup{labelformat=empty}  % Disable automatic "Figure X." prefix
    \caption{\textbf{Figure 29. }{\it Lightcurves and HID of MAXI J1305–704}: 
    Left panel: Zoomed view of the hard-band dominated (failed) outburst observed with \maxi{} in the 2–4 keV (blue open circles) and 10–20 keV (red solid circles) bands. 
    Right panel: Corresponding Hardness Intensity Diagram (HID), where hardness is defined as the ratio of 10–20 keV to 2–4 keV intensity and intensity as the 2–20 keV count rate. Grey triangles denote the hard state and black inverted triangles denote the soft state.}

    \label{fig: MAXI J1305-704}
\end{figure*}

\begin{figure*}
    \centering
    \includegraphics[scale=0.60]{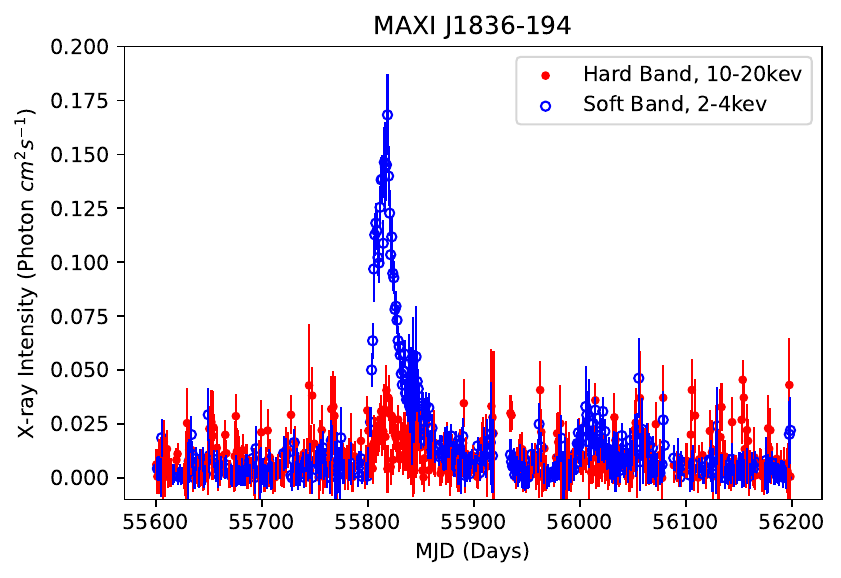}
    \includegraphics[scale=0.60]{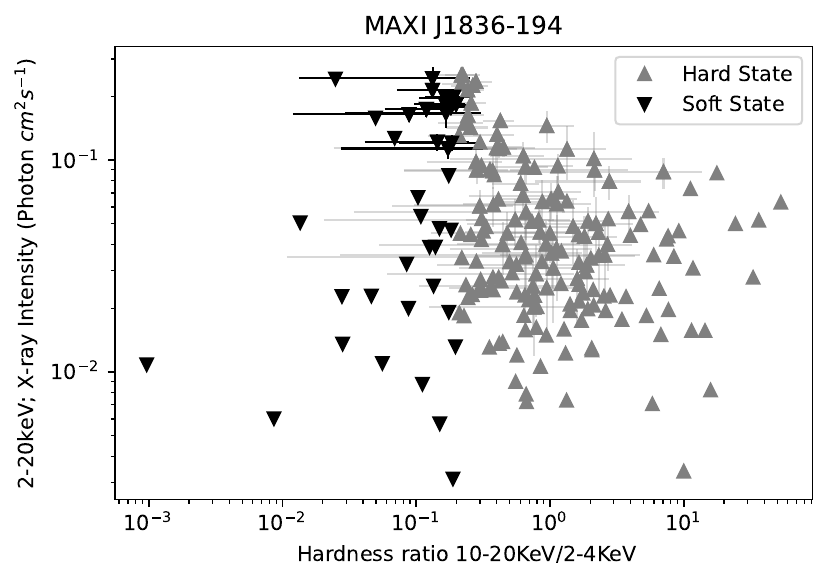}
    \captionsetup{labelformat=empty}  % Disable automatic "Figure X." prefix

    \caption{\textbf{Figure 30. }{\it Lightcurves and HID of MAXI J1836–194}: 
    Left panel: Lightcurves obtained with \maxi{} in the 2–4 keV (blue open circles) and 10–20 keV (red solid circles) bands, along with \swift{}/BAT 15–50 keV data (green stars), showing a representative outburst during the observation period. The \swift{} lightcurve is scaled appropriately for clarity. 
    Right panel: Corresponding Hardness Intensity Diagram (HID), where hardness is defined as the ratio of 10–20 keV to 2–4 keV intensity and intensity as the 2–20 keV count rate. Grey triangles denote the hard state and black inverted triangles denote the soft state.}

    \label{fig: MAXI J1836-194}
\end{figure*}

\section{Analysis of Neutron star sources}
\label{Analysis of Neutron star sources appendix}
\subsection{1A 0535+262}

\begin{figure*}
    \centering
    \includegraphics[scale=0.60]{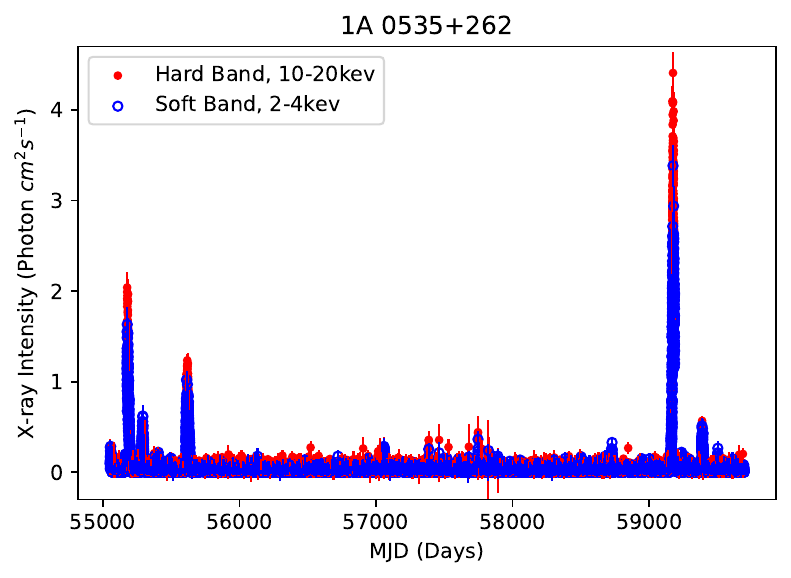}
     \includegraphics[scale=0.60]{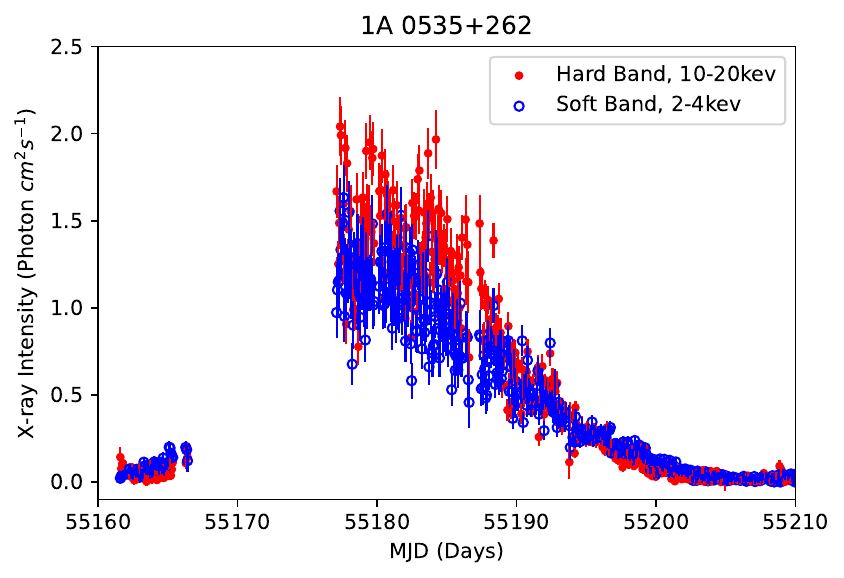}
    \includegraphics[scale=0.60]{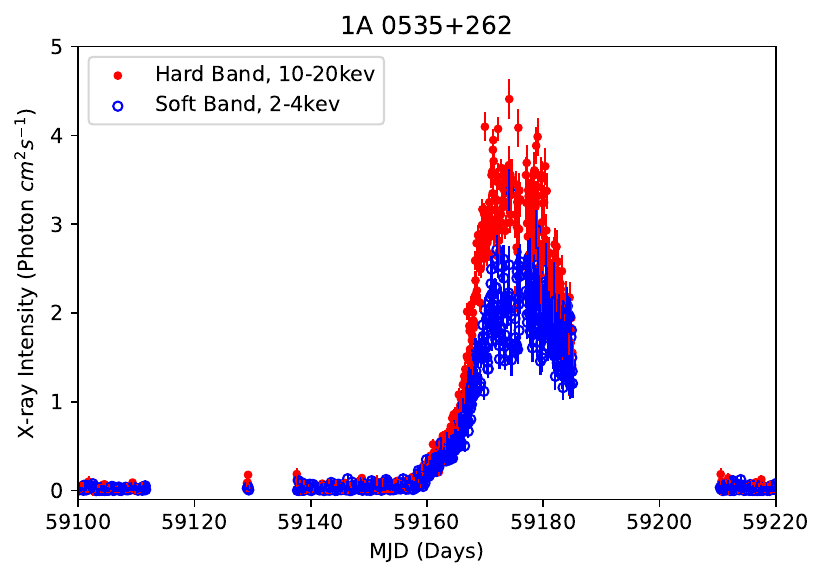}
    \includegraphics[scale=0.60]{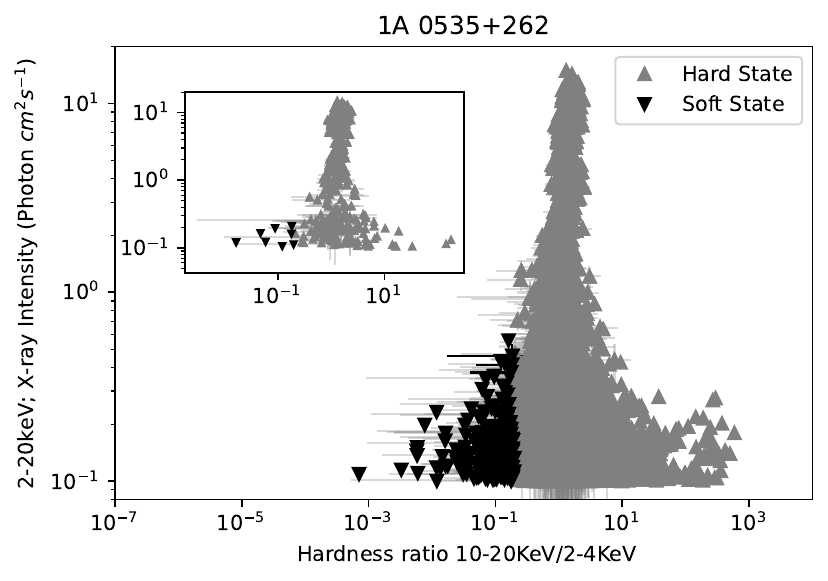}
    
    \captionsetup{labelformat=empty}  % Disable automatic "Figure X." prefix
    \caption{\textbf{Figure 31. }{\it Lightcurves and HID of 1A 0535+262}: 
    Top left panel: Lightcurves obtained with \maxi{} in the 2–4 keV (blue open circles) and 10–20 keV (red solid circles) bands, showing multiple outbursts during the full observation period. 
    Top right panel: Zoomed view of the first outburst. 
    Bottom left panel: Zoomed view of the last outburst, where intensity enhancements are observed over the same time interval in both energy bands. 
     Bottom right panel: Corresponding Hardness Intensity Diagram (HID) of the entire lightcurve, where hardness is defined as the ratio of 10–20 keV to 2–4 keV intensity and intensity as the 2–20 keV count rate. The distribution of points extends toward higher hardness values during peak intensity. The inset focuses on the HID of the outburst in the left bottom panel.}

    \label{fig:1A 0535+262}
\end{figure*}

1A 0535+262 is categorized as a NSXB. During the observation period, there are a total of three outbursts (top left panel of Fig. \ref{fig:1A 0535+262}). Zoomed in view of the outbursts show that the intensity peaks occur in both the hard and soft bands simultaneously (top right and bottom left panel of \ref{fig:1A 0535+262}). An intensity peak can also be seen in the HID (bottom right panel of \ref{fig:1A 0535+262}) show that the spectrum of the source is dominated by the hard band. 

\subsection{GRO J1008-57}

\begin{figure*}
    \centering
    \includegraphics[scale=0.60]{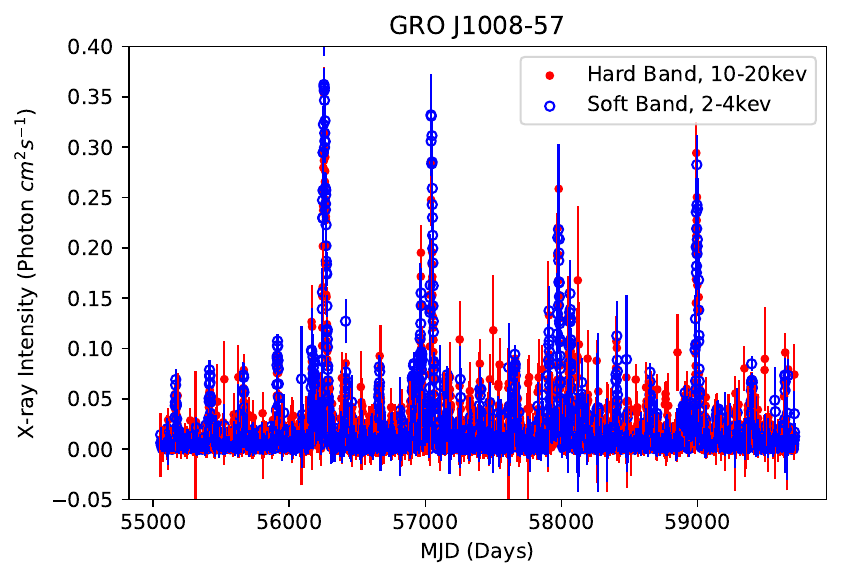}
    \includegraphics[scale=0.60]{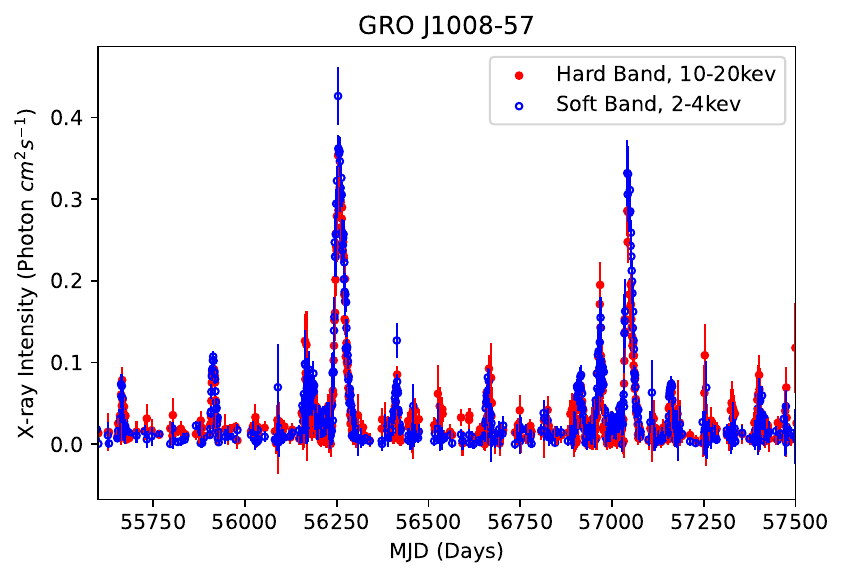}
    \includegraphics[scale=0.60]{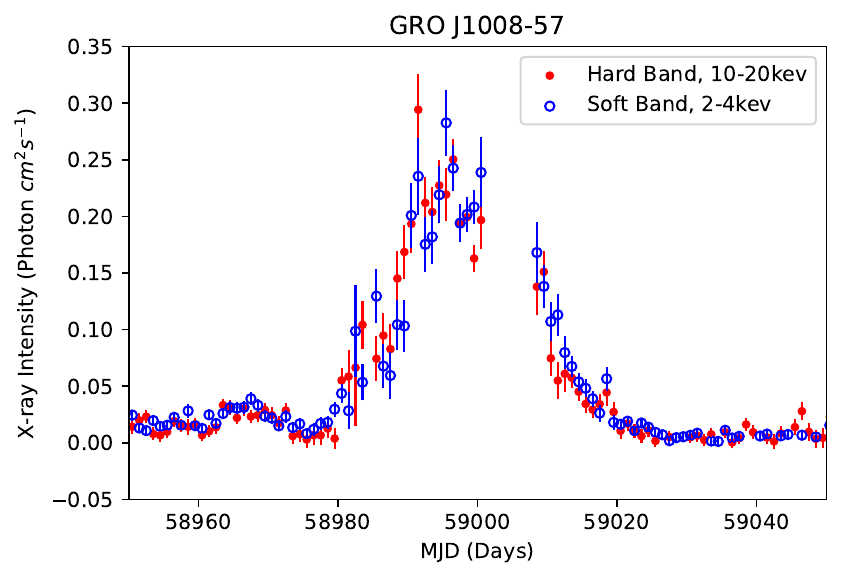}
    \includegraphics[scale=0.60]{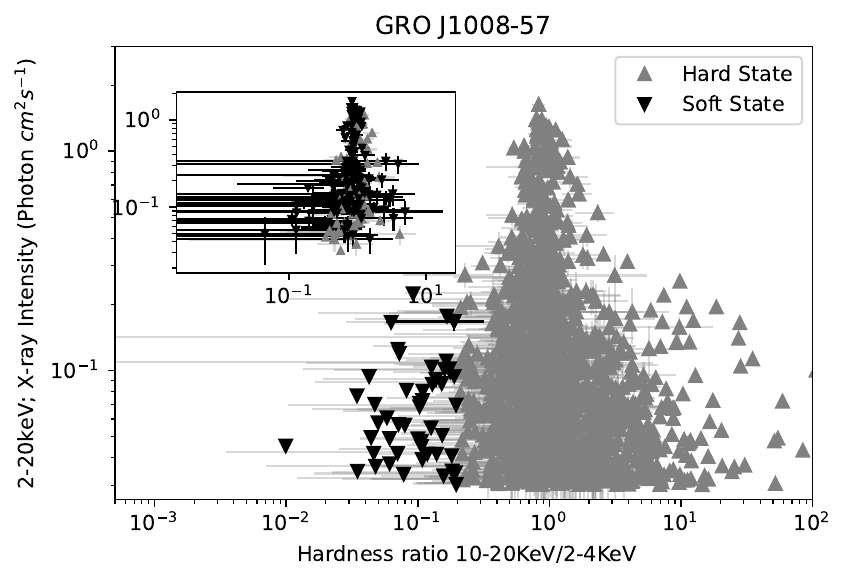}
    \captionsetup{labelformat=empty}  % Disable automatic "Figure X." prefix

    \caption{\textbf{Figure 32. }{\it Lightcurves and HID of GRO J1008–57}: 
    Top left panel: Lightcurves obtained with \maxi{} in the 2–4 keV (blue open circles) and 10–20 keV (red solid circles) bands, showing multiple outbursts and several hard-band dominated (failed) outbursts during the full observation period.
    Top right panel: Partially zoomed view of the lightcurve highlighting the sequence of outbursts. 
    Bottom left panel: Zoomed view of a representative outburst, where intensity enhancements are observed over the same time interval in both energy bands. 
    Bottom right panel: Corresponding Hardness Intensity Diagram (HID) of the entire lightcurve, where hardness is defined as the ratio of 10–20 keV to 2–4 keV intensity and intensity as the 2–20 keV count rate. The distribution of points extends toward higher hardness values near peak intensity. The inset focuses on the HID of the outburst in the left bottom panel.}

    \label{fig:GRO J1008-57}
\end{figure*}

GRO J1008-57 is categorized a an NSXB. A series is outbursts can be seen during the observation period along with multiple failed outbursts (top left panel of Fig. \ref{fig:GRO J1008-57}). The zoomed in view of a particular outburst (bottom left panel of Fig. \ref{fig:GRO J1008-57}) shows that the event occurs simultaneously in both bands. The HID (bottom right panel of Fig. \ref{fig:GRO J1008-57}) once again indicates that the spectrum is dominant in the hard band.

\subsection{2S 1417-624}

\begin{figure*}
    \centering
    \includegraphics[scale=0.60]{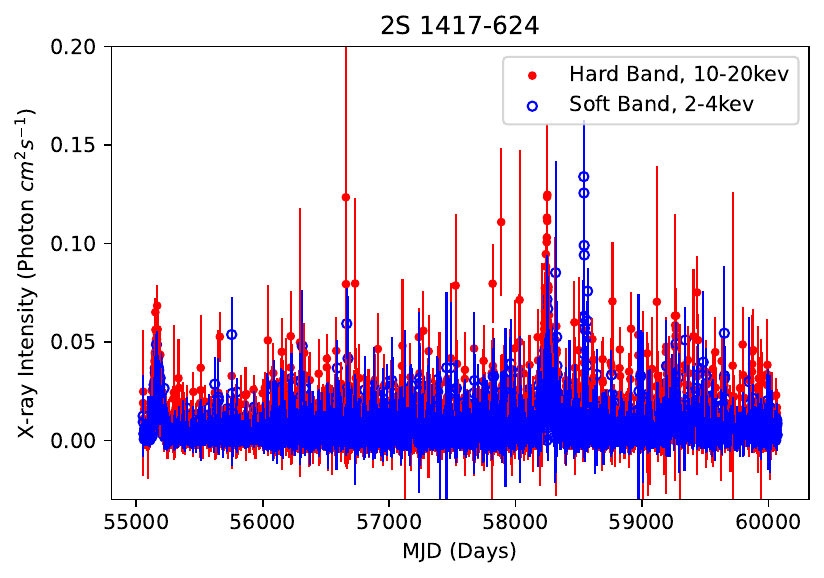}
    \includegraphics[scale=0.60]{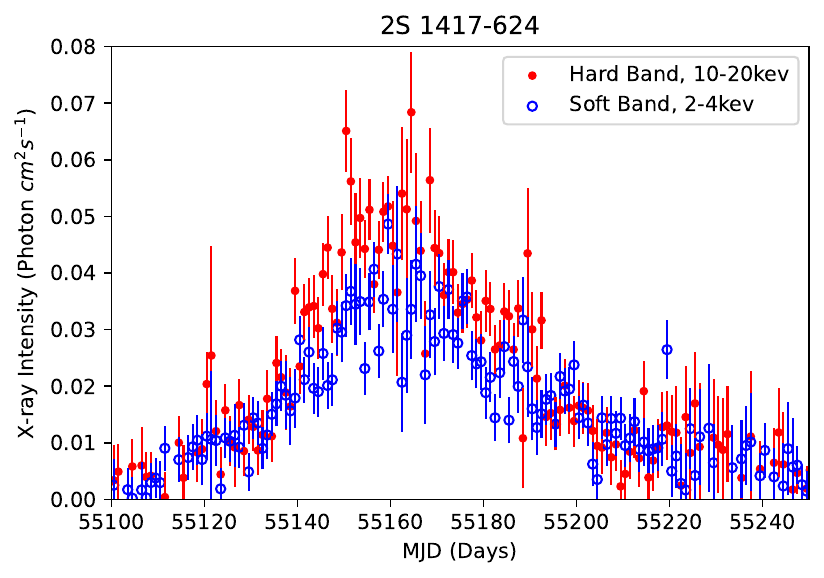}
    \includegraphics[scale=0.60]{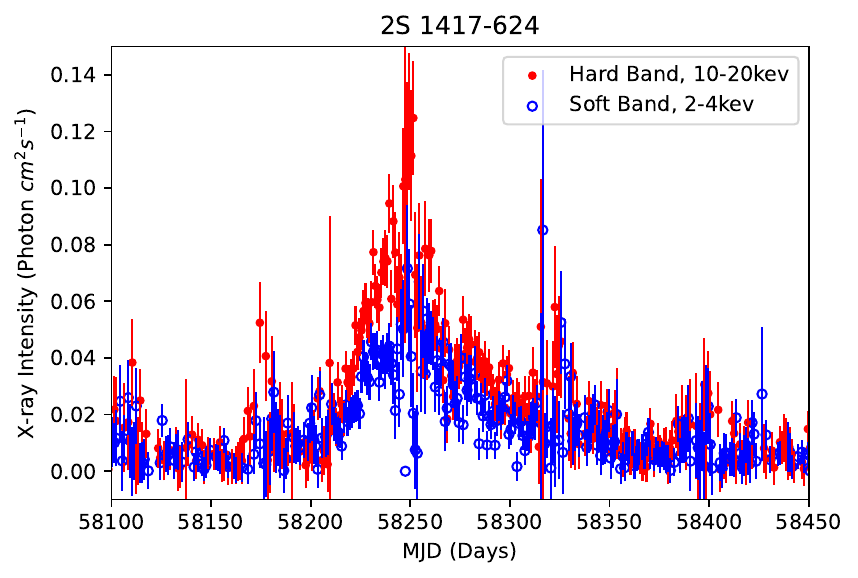}
    \includegraphics[scale=0.60]{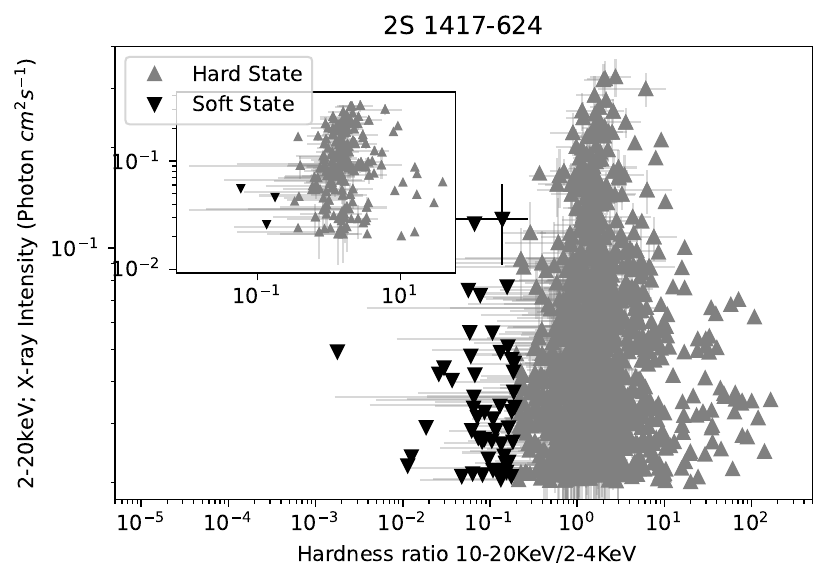}
    \captionsetup{labelformat=empty}  % Disable automatic "Figure X." prefix

    \caption{\textbf{Figure 33. }{\it Lightcurves and HID of 2S 1417–624}: 
    Top left panel: Lightcurves obtained with \maxi{} in the 2–4 keV (blue open circles) and 10–20 keV (red solid circles) bands, during the full observation period.
    Top right panel: Zoomed view of a hard-band dominated (failed) outburst. 
    Bottom left panel: Zoomed view of a short-duration intensity enhancement. 
    Bottom right panel: Corresponding Hardness Intensity Diagram (HID) of the entire lightcurve, where hardness is defined as the ratio of 10–20 keV to 2–4 keV intensity and intensity as the 2–20 keV count rate. The inset shows the HID during the failed outburst as shown in the left bottom panel.}

    \label{fig: 2S 1417-624}
\end{figure*}

2S 1417-624 is identified as a NSXB. The observation shows that there are a few intensity peaks, probably due to failed outbursts (Top left and top right panel of Fig. \ref{fig: 2S 1417-624}). Once again, the HID is dominated by the spectra in the hard band (bottom right panel of Fig. \ref{fig: 2S 1417-624}).

\subsection{GX 304-1}

\begin{figure*}
    \centering
    \includegraphics[scale=0.60]{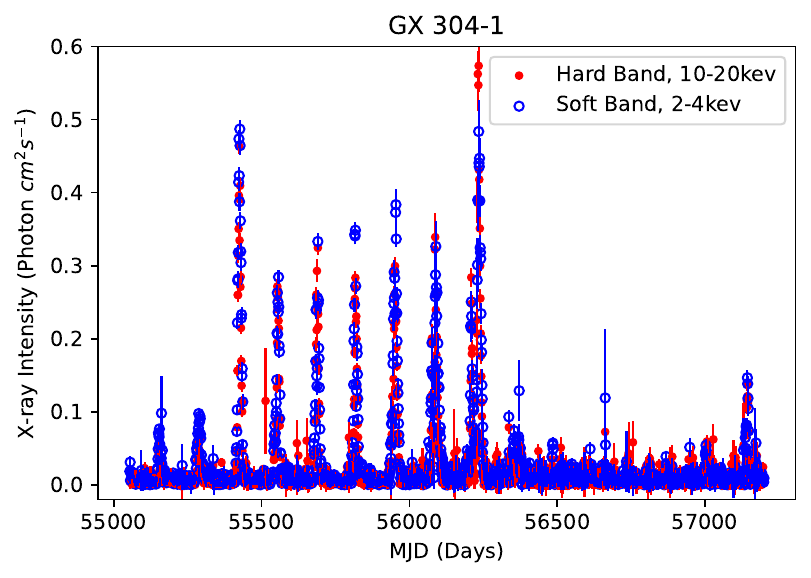}
    \includegraphics[scale=0.60]{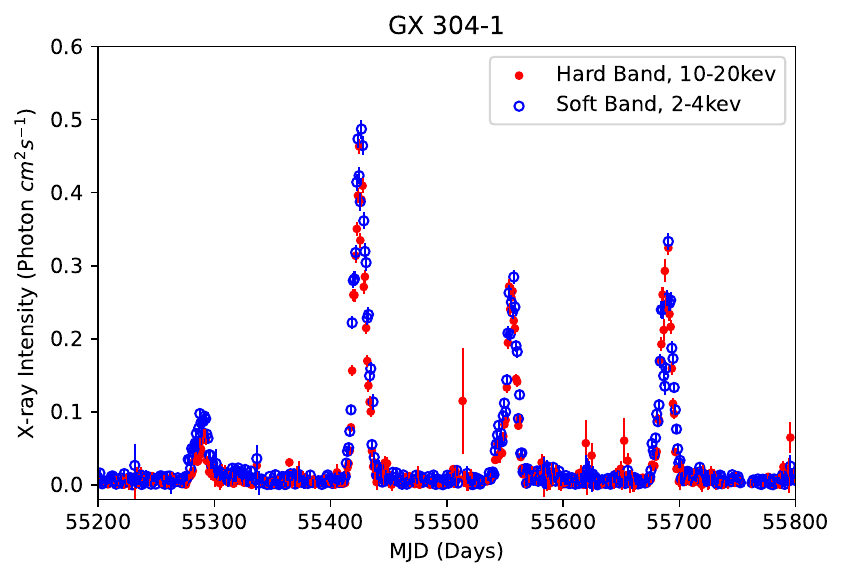}
    \includegraphics[scale=0.60]{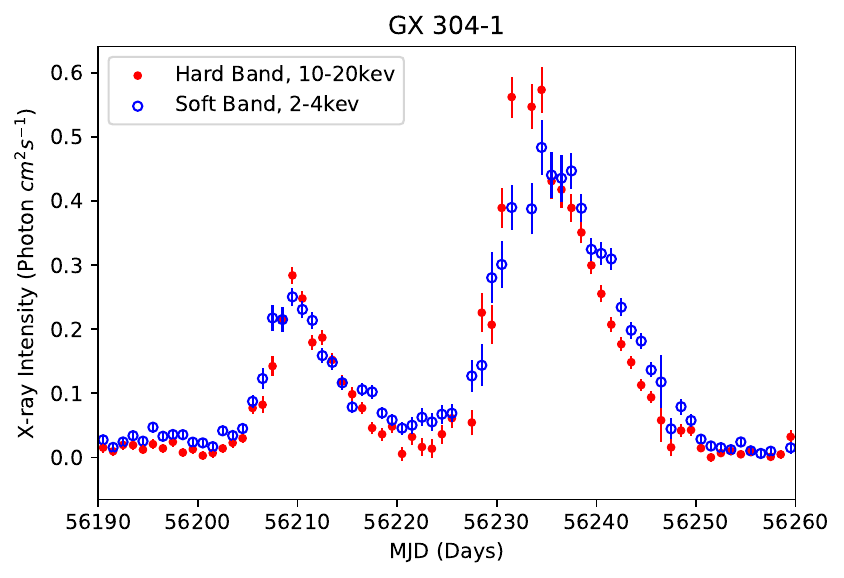}
    \includegraphics[scale=0.60]{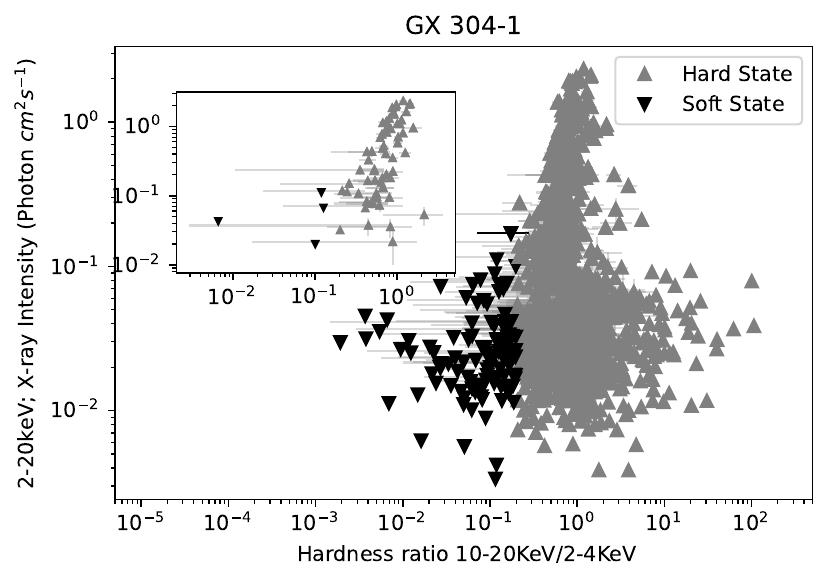}
    \captionsetup{labelformat=empty}  % Disable automatic "Figure X." prefix

    \caption{\textbf{Figure 34. }{\it Lightcurves and HID of GX 304–1}: 
    Top left panel: Lightcurves obtained with \maxi{} in the 2–4 keV (blue open circles) and 10–20 keV (red solid circles) bands, showing multiple outburst peaks during the full observation period.
    Top right panel: Zoomed view of representative outbursts and hard-band dominated (failed) outbursts. 
    Bottom left panel: Further zoomed view of selected outbursts, where intensity enhancements are observed over the same time interval in both energy bands. 
    Bottom right panel: Corresponding Hardness Intensity Diagram (HID) of the entire lightcurve, where hardness is defined as the ratio of 10–20 keV to 2–4 keV intensity and intensity as the 2–20 keV count rate. The distribution of points is concentrated toward higher hardness values. The inset shows the HID during a representative outburst as shown in the bottom left panel.}

    \label{fig: GX 304-1}
\end{figure*}

GX 304-1 is categorized as a NSXB. Multiple outbursts can be observed during the observation period (Top left panel of Fig. \ref{fig: GX 304-1}). The outburst occurs simultaneously in both the hard and soft bands (Bottom left panel of Fig .\ref{fig: GX 304-1}). The HID is dominated by the hard band (Bottom right panel of Fig. \ref{fig: GX 304-1}).

\subsection{4U 0115+63} 

\begin{figure*}
    \centering
    \includegraphics[scale=0.60]{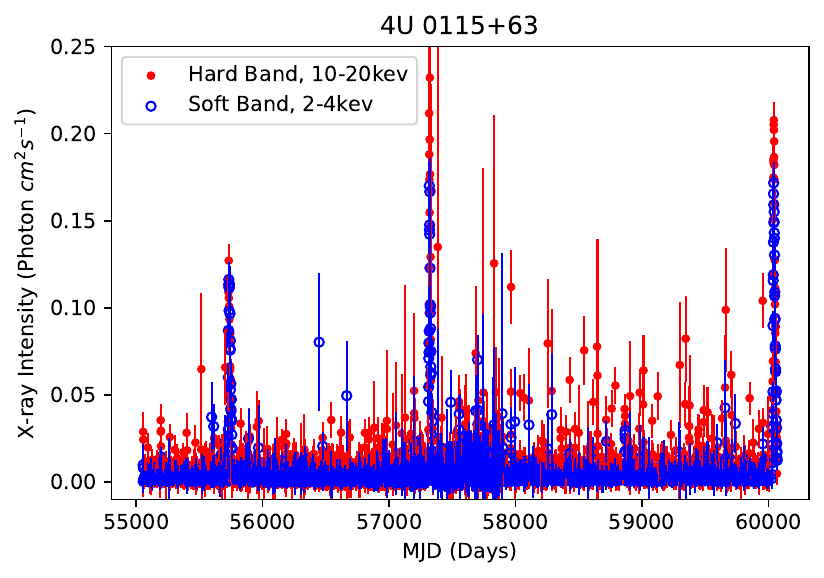}
    \includegraphics[scale=0.60]{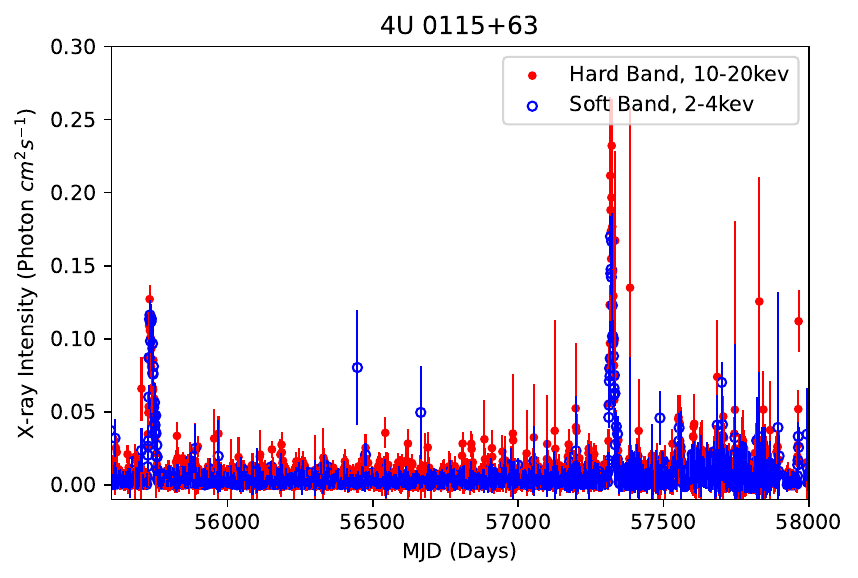}
    \includegraphics[scale=0.60]{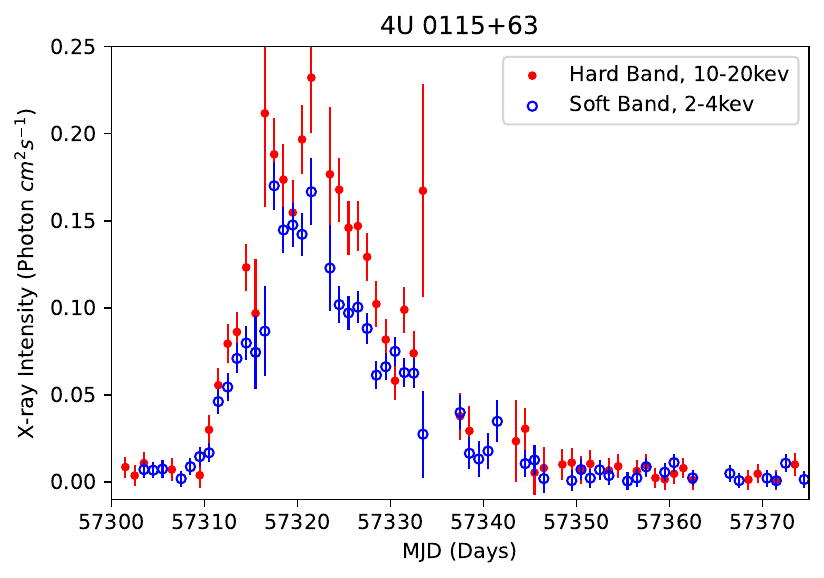}
    \includegraphics[scale=0.60]{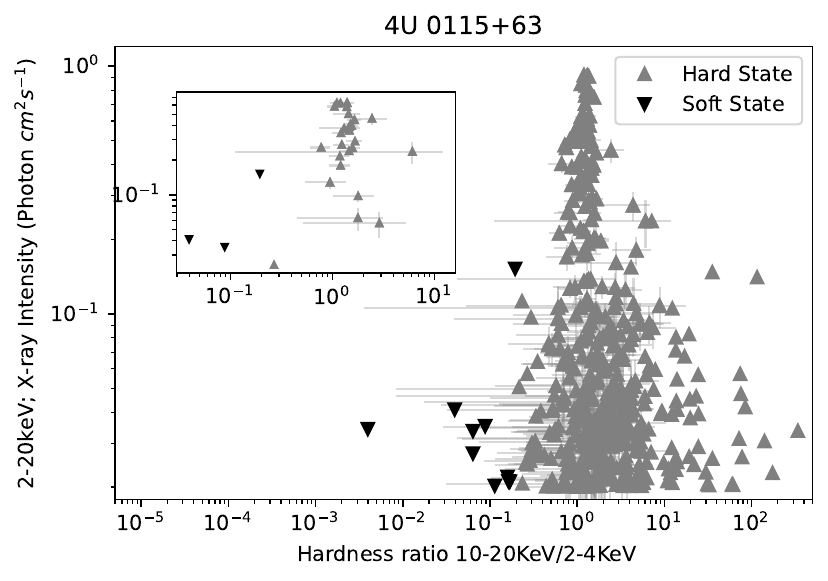}
    \captionsetup{labelformat=empty}  % Disable automatic "Figure X." prefix
    
    \caption{\textbf{Figure 35. }{\it Lightcurves and HID of 4U 0115+63}: 
    Top left panel: Lightcurves obtained with \maxi{} in the 2–4 keV (blue open circles) and 10–20 keV (red solid circles) bands, showing multiple outbursts during the full observation period. 
    Top right panel: Zoomed view of the lightcurve highlighting a prominent outburst and a hard-band dominated (failed) outburst. 
    Bottom left panel: Zoomed view of a representative outburst illustrating its temporal evolution. 
    Bottom right panel: Corresponding Hardness Intensity Diagram (HID) of the entire lightcurve, where hardness is defined as the ratio of 10–20 keV to 2–4 keV intensity and intensity as the 2–20 keV count rate. The inset shows the HID during the selected outburst as shown in the left bottom panel.}

    \label{fig: 4U 0115+63}
\end{figure*}

4U 0115+63 is classified as an NSXB. The light curve shows multiple outburst peaks during the observation period (Top left panel of Fig. \ref{fig: 4U 0115+63}). Zoomed in view of the outburst shows that the outburst peak is slightly asymmetric (Top right panel of Fig/. \ref{fig: 4U 0115+63}) but the ascent is not as steep as observed in the BH lightcurves. Peaks in both bands occur at the same time and the spectrum is dominated by emissions from the hard band (Bottom right panel of Fig. \ref{fig: 4U 0115+63}). 

\subsection{NGC 6440}

\begin{figure*}
    \centering
    \includegraphics[scale=0.60]{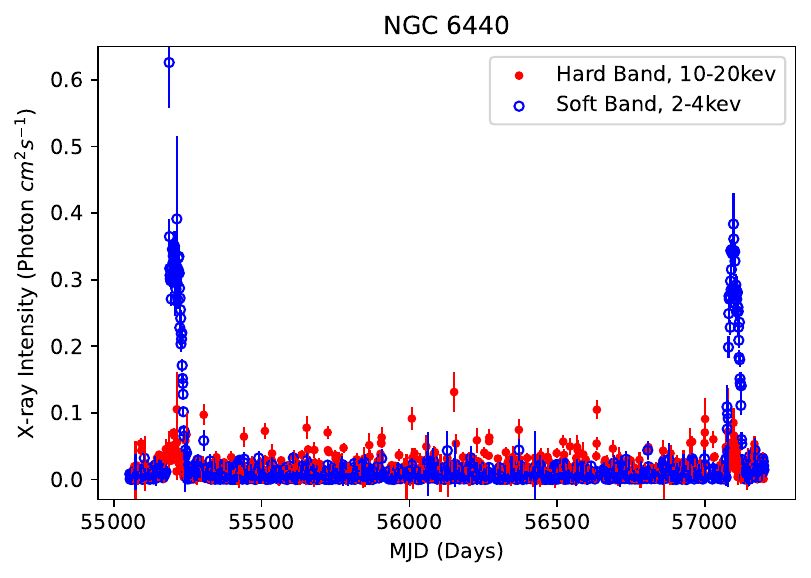}
    \includegraphics[scale=0.60]{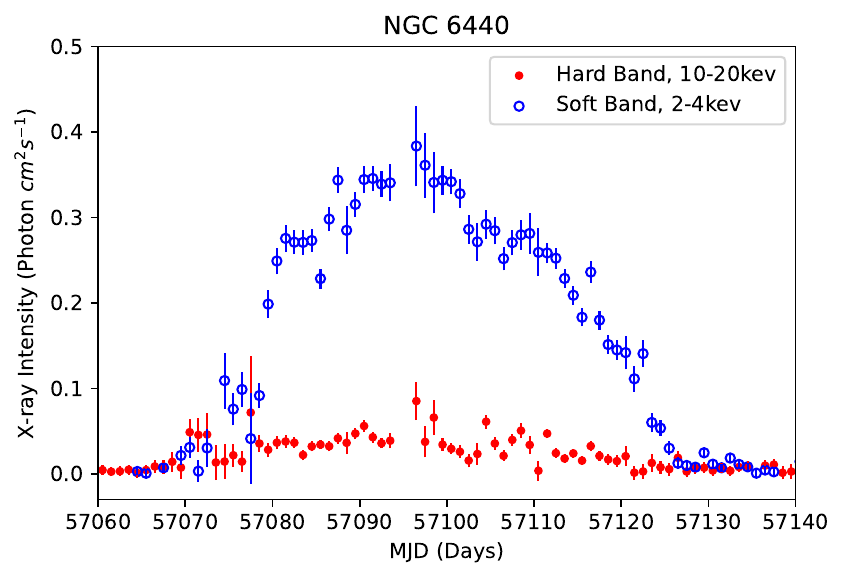}
    \includegraphics[scale=0.60]{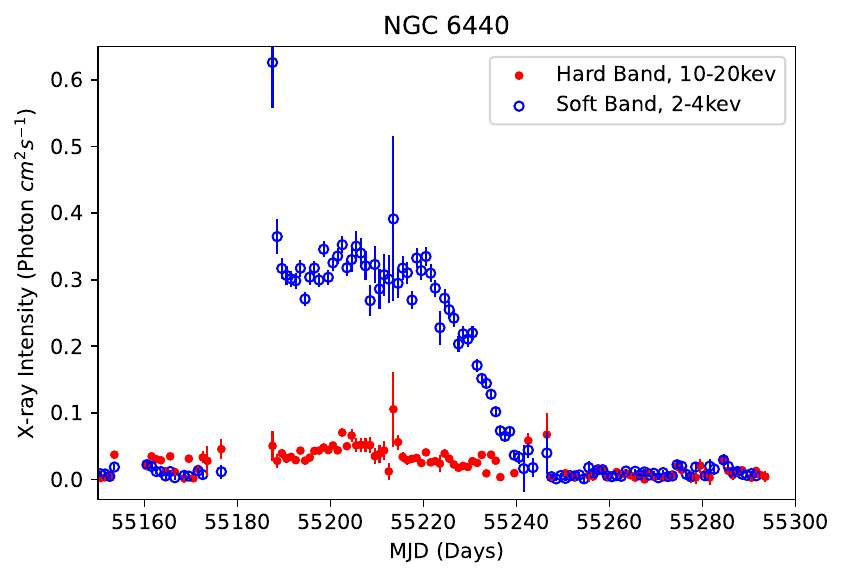}
    \includegraphics[scale=0.60]{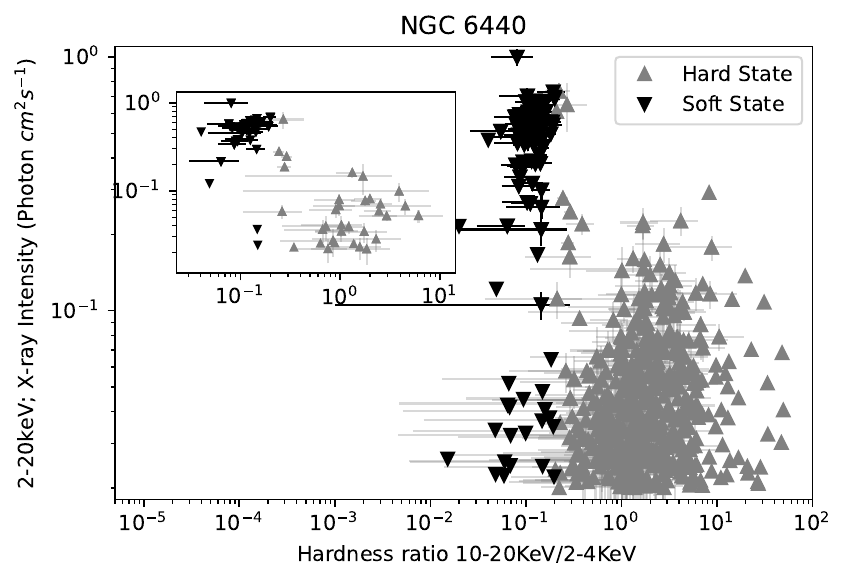}
    \captionsetup{labelformat=empty}  % Disable automatic "Figure X." prefix

    \caption{\textbf{Figure 36. }{\it Lightcurves and HID of NGC 6440}: 
    Top left panel: Lightcurves obtained with \maxi{} in the 2–4 keV (blue open circles) and 10–20 keV (red solid circles) bands, showing two distinct outbursts during the full observation period. 
    Top right panel: Zoomed view of the second outburst. 
    Bottom left panel: Zoomed view of the first outburst. 
    Bottom right panel: Corresponding Hardness Intensity Diagram (HID) of the entire lightcurve, where hardness is defined as the ratio of 10–20 keV to 2–4 keV intensity and intensity as the 2–20 keV count rate. The inset shows the HID during a representative outburst as shown in the left bottom panel.}

    \label{fig: NGC 6440}
\end{figure*}

NGC 6440 is classified as a NSXB. The light curve during the observation period shows two outbursts (top left panel of Fig. \ref{fig: NGC 6440}). Both outbursts occur simultaneously in both bands (top right panel and bottom left panel of Fig. \ref{fig: NGC 6440}). From the HID (bottom right panel of Fig .\ref{fig: NGC 6440}), the soft X-rays are detected for a longer time, but the intensity of the soft band is more compared to the hard band.

\subsection{SMC X-3}
\begin{figure*}
    \centering
    \includegraphics[scale=0.60]{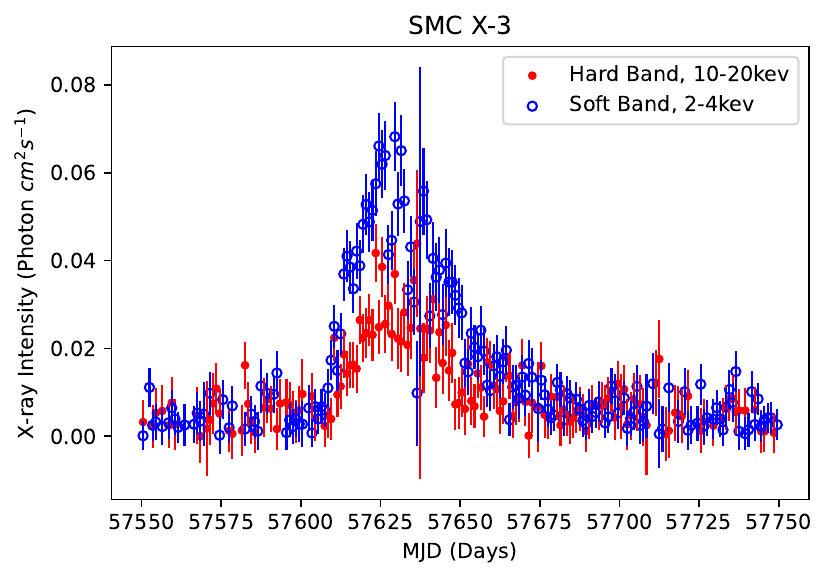}
    \includegraphics[scale=0.60]{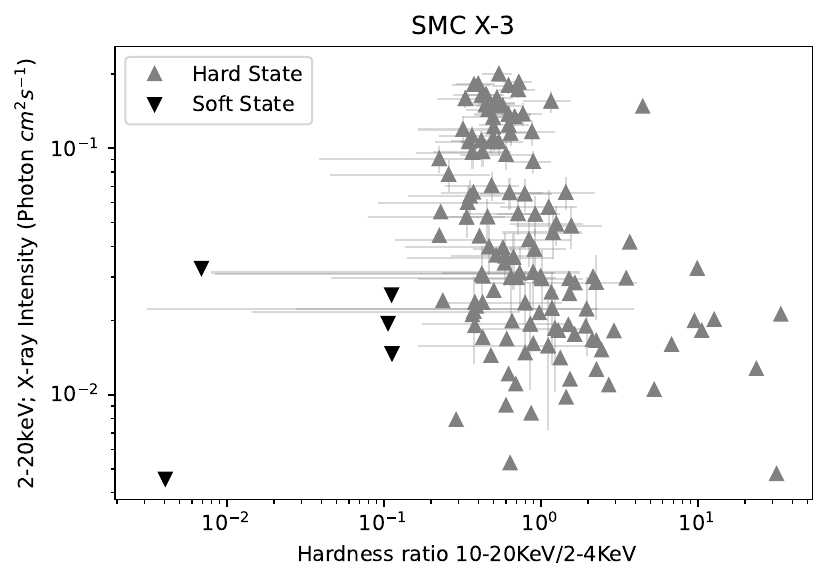}
    \captionsetup{labelformat=empty}  % Disable automatic "Figure X." prefix
    \caption{\textbf{Figure 37. }{\it Lightcurves and HID of SMC X–3}: 
    Left panel: Lightcurves obtained with \maxi{} in the 2–4 keV (blue open circles) and 10–20 keV (red solid circles) bands, showing a hard-band dominated (failed) outburst during the observation period. 
    Right panel: Corresponding Hardness Intensity Diagram (HID), where hardness is defined as the ratio of 10–20 keV to 2–4 keV intensity and intensity as the 2–20 keV count rate. The distribution of points is concentrated predominantly in the high-hardness region.}

    \label{fig: SMC X-3}
\end{figure*}

SMC X-3 is classified as a NSXB. The lightcurve shows a failed outburst (left panel of Fig. \ref{fig: SMC X-3}). The spectrum is primarily dominated by the hard band (right panel of Fig. \ref{fig: SMC X-3})

\subsection{XTE J1739-285}
\begin{figure*}
    \centering
    \includegraphics[scale=0.60]{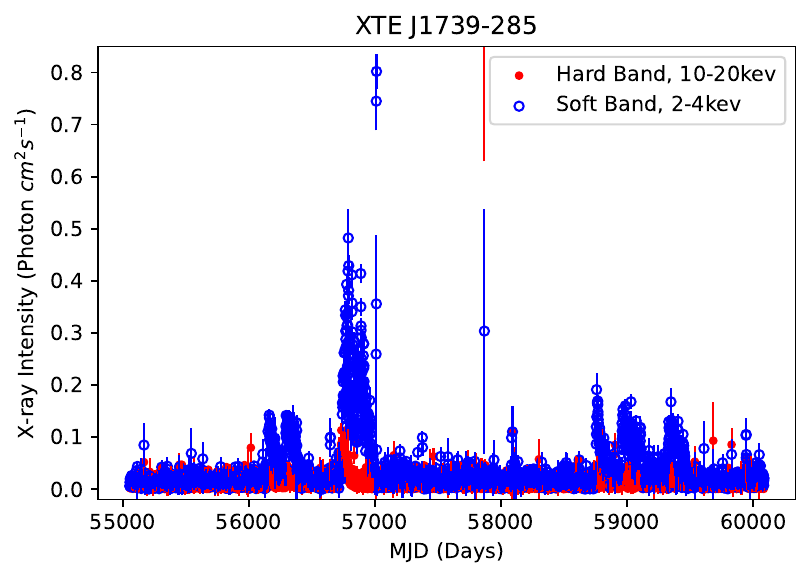}
    \includegraphics[scale=0.60]{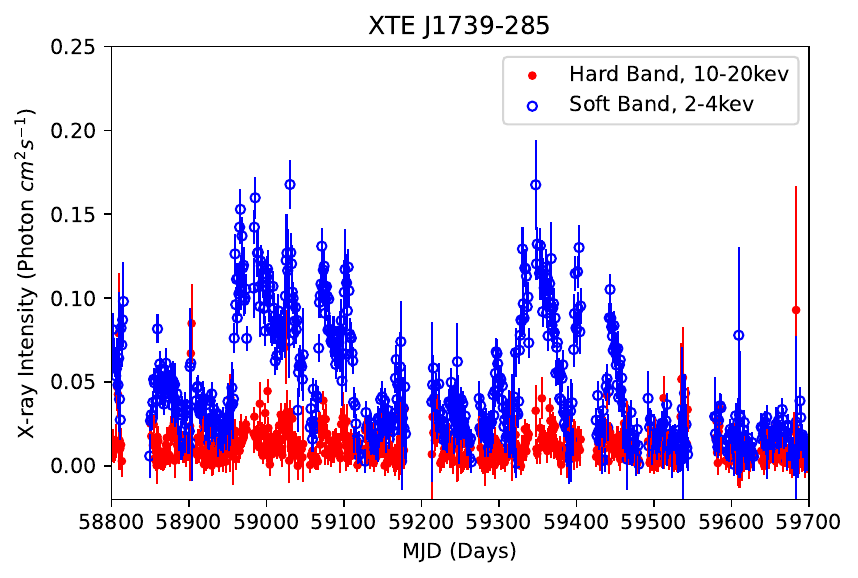}
    \includegraphics[scale=0.60]{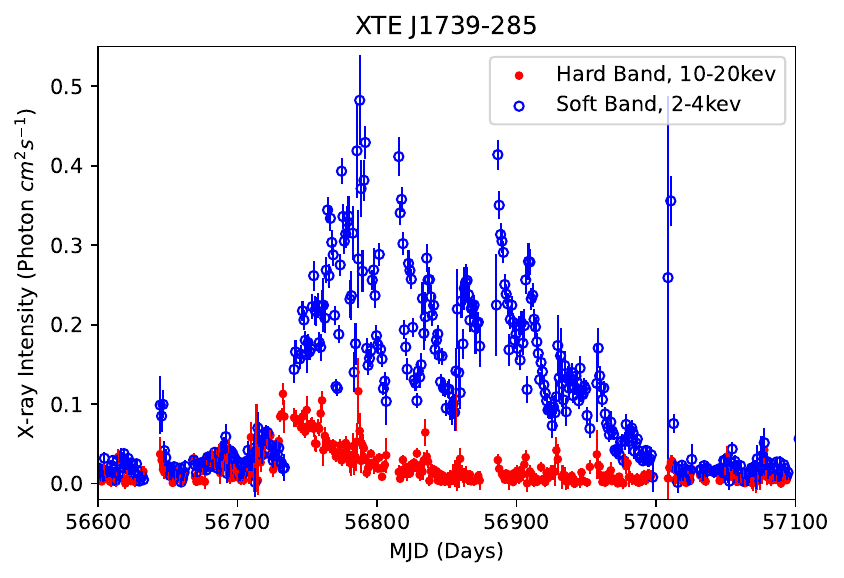}
    \includegraphics[scale=0.60]{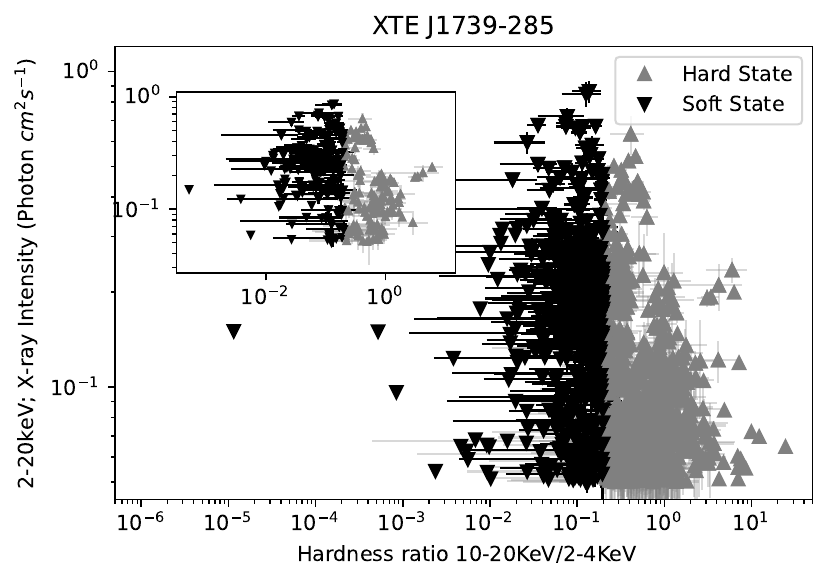}
    \captionsetup{labelformat=empty}  % Disable automatic "Figure X." prefix
    \caption{\textbf{Figure 38. }{\it Lightcurves and HID of XTE J1739–285}: 
    Top left panel: Lightcurves obtained with \maxi{} in the 2–4 keV (blue open circles) and 10–20 keV (red solid circles) bands, showing multiple outbursts during the full observation period. 
    Top right panel: Zoomed view of the hard-band dominated (failed) outbursts at the end of the observation period. 
    Bottom left panel: Zoomed view of representative outbursts highlighting their temporal evolution. 
    Bottom right panel: Corresponding Hardness Intensity Diagram (HID) of the entire lightcurve, where hardness is defined as the ratio of 10–20 keV to 2–4 keV intensity and intensity as the 2–20 keV count rate. The inset shows the HID during the outburst phases as shown in the left bottom panel.}

    \label{fig: XTE J1739-285}
\end{figure*}

XTE J1739-285 is a NSXB with a mass estimation of $\sim$1.51$M_\odot$. The light curve boasts multiple outburst peaks (Top left panel of Fig .\ref{fig: XTE J1739-285}). Although the outburst peaks are not clearly resolvable (bottom left panel of Fig. \ref{fig: XTE J1739-285}), it spans over $\sim$300 days. The HID shows nearly equal intensity distributions in both the hard and soft bands.

\subsection{1RXS J180408.9-342058} 
\begin{figure*}
    \centering
    \includegraphics[scale=0.60]{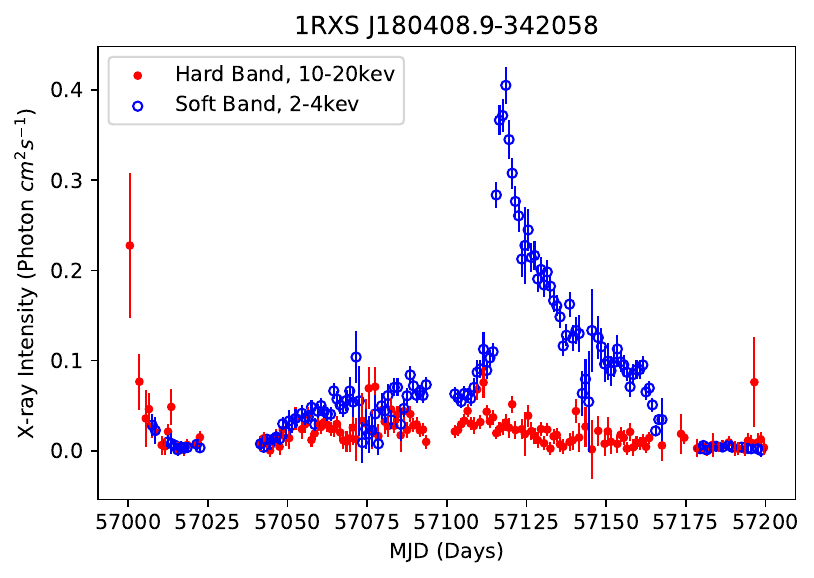}
    \includegraphics[scale=0.60]{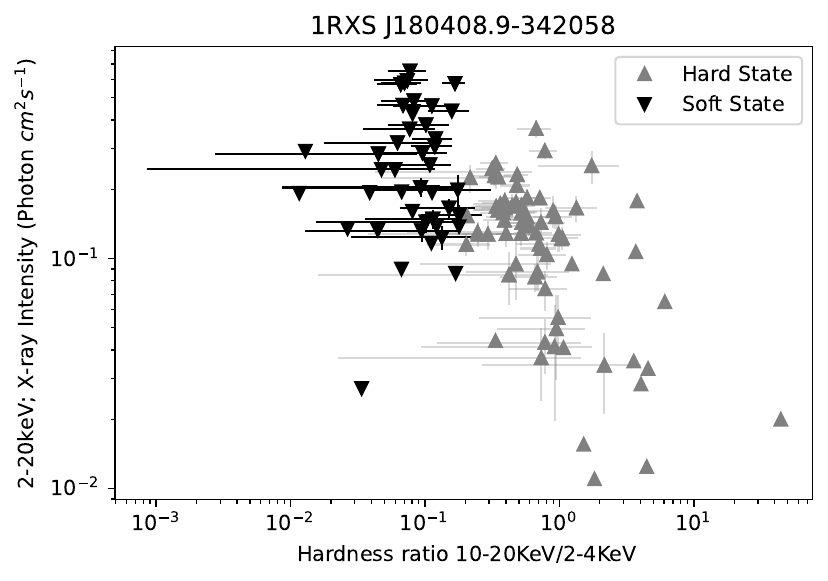}
    \captionsetup{labelformat=empty}  % Disable automatic "Figure X." prefix
    \caption{\textbf{Figure 39. }{\it Lightcurves and HID of 1RXS J180408.9–342058}: 
    Left panel: Lightcurves obtained with \maxi{} in the 2–4 keV (blue open circles) and 10–20 keV (red solid circles) bands.
    Right panel: Corresponding Hardness Intensity Diagram (HID), where hardness is defined as the ratio of 10–20 keV to 2–4 keV intensity and intensity as the 2–20 keV count rate. Grey triangles denote the hard state and black inverted triangles denote the soft state.}

    \label{fig:1RXS J180408.9-342058}
\end{figure*}

1RXS J180408.9-342058 is a low-mass NSXB. A typical outburst peak is observed (left panel of Fig. \ref{fig:1RXS J180408.9-342058}) with a steep ascent and gradual descent.

\subsection{4U 1624-490} 
\begin{figure*}
    \centering
    \includegraphics[scale=0.60]{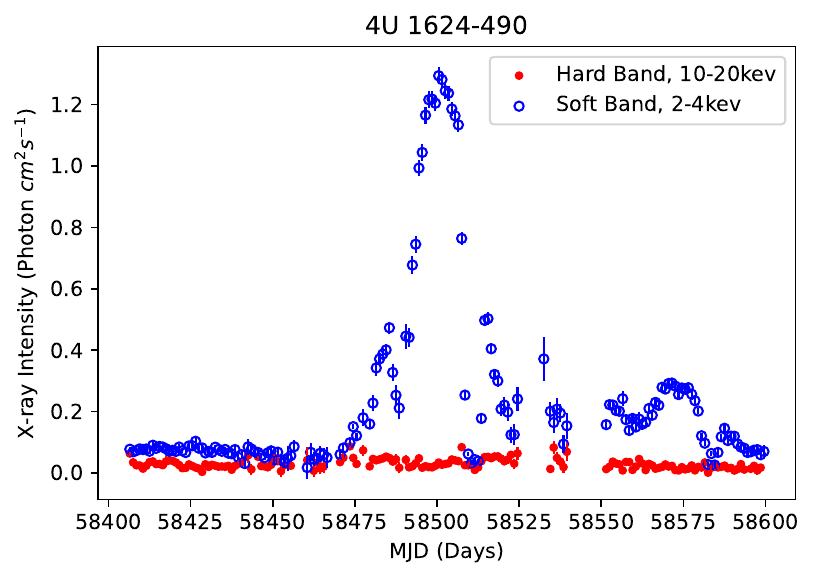}
    \includegraphics[scale=0.60]{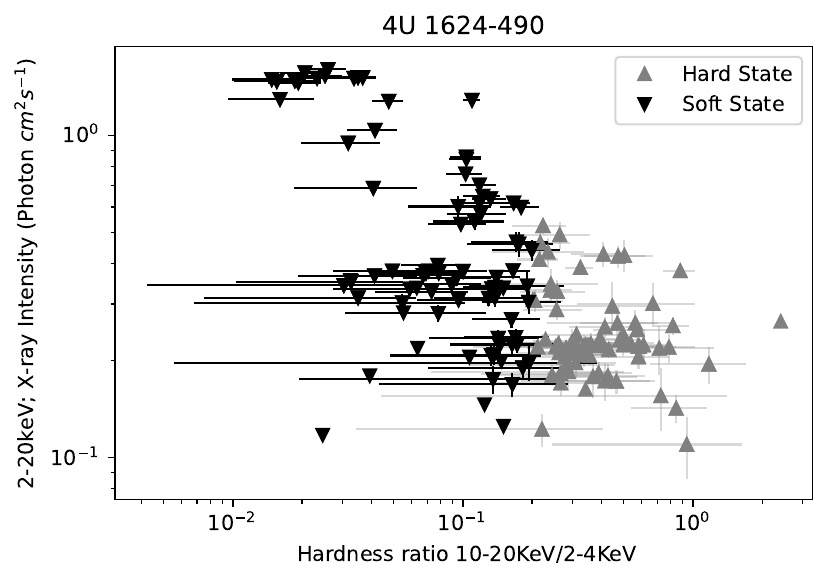}
    \captionsetup{labelformat=empty}  % Disable automatic "Figure X." prefix
    \caption{\textbf{Figure 40. }{\it Lightcurves and HID of 4U 1624–490}: 
    Left panel: Lightcurves obtained with \maxi{} in the 2–4 keV (blue open circles) and 10–20 keV (red solid circles) bands.
    Right panel: Corresponding Hardness Intensity Diagram (HID), where hardness is defined as the ratio of 10–20 keV to 2–4 keV intensity and intensity as the 2–20 keV count rate. Grey triangles denote the hard state and black inverted triangles denote the soft state.}

    \label{fig:4U 1624-490}
\end{figure*}

4U 1624-490 is classified as a low mass NSXB. An outburst peak can be observed during the observation period followed by a few failed outbursts (Left panel of Fig. \ref{fig:4U 1624-490}.)

\subsection{EXO 1722-363} 
\begin{figure*}
    \centering
    \includegraphics[scale=0.60]{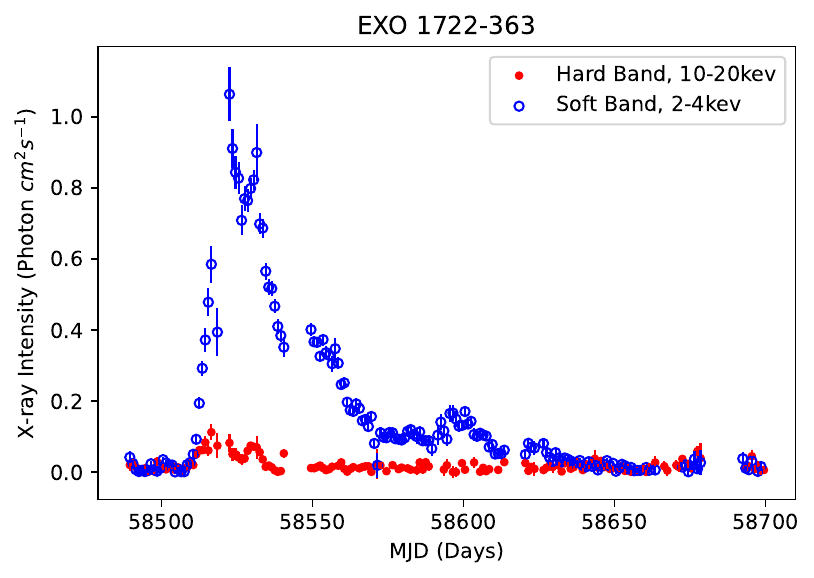}
    \includegraphics[scale=0.60]{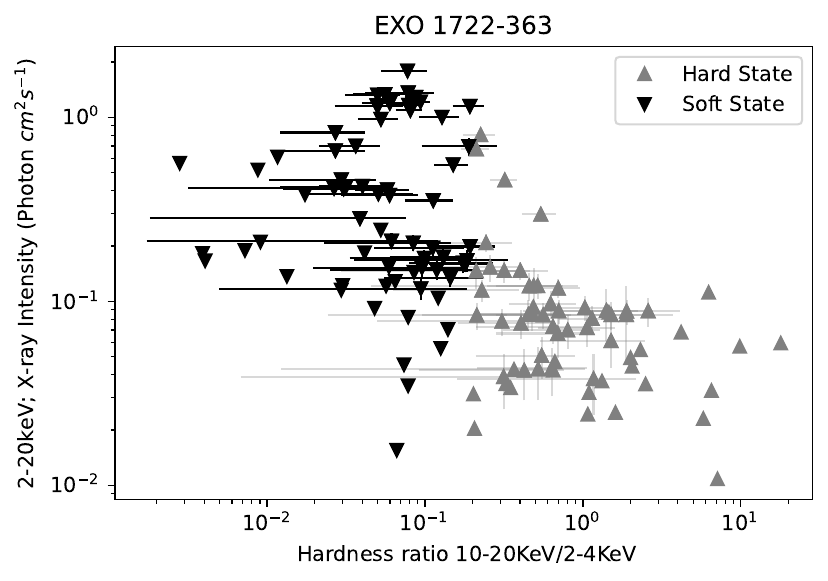}
    \captionsetup{labelformat=empty}  % Disable automatic "Figure X." prefix
    \caption{\textbf{Figure 41. }{\it Lightcurves and HID of EXO 1722–363}: 
    Left panel: Lightcurves obtained with \maxi{} in the 2–4 keV (blue open circles) and 10–20 keV (red solid circles) bands.
    Right panel: Corresponding Hardness Intensity Diagram (HID), where hardness is defined as the ratio of 10–20 keV to 2–4 keV intensity and intensity as the 2–20 keV count rate. The HID exhibits the characteristic q-shaped evolution during the outburst.
}

    \label{fig:EXO 1722-363}
\end{figure*}

The mass of the NSXB lies in the range of $\sim$1.5-1.6$M_\odot$. A typical outburst is observed during the observation period with a steep ascent and a gradual descent (Left panel of Fig. \ref{fig:EXO 1722-363}). A q-diagram can be observed in the HID (Right panel of Fig. \ref{fig:EXO 1722-363}).

\subsection{IGR J18483-0311}
\begin{figure*}
    \centering
    \includegraphics[scale=0.60]{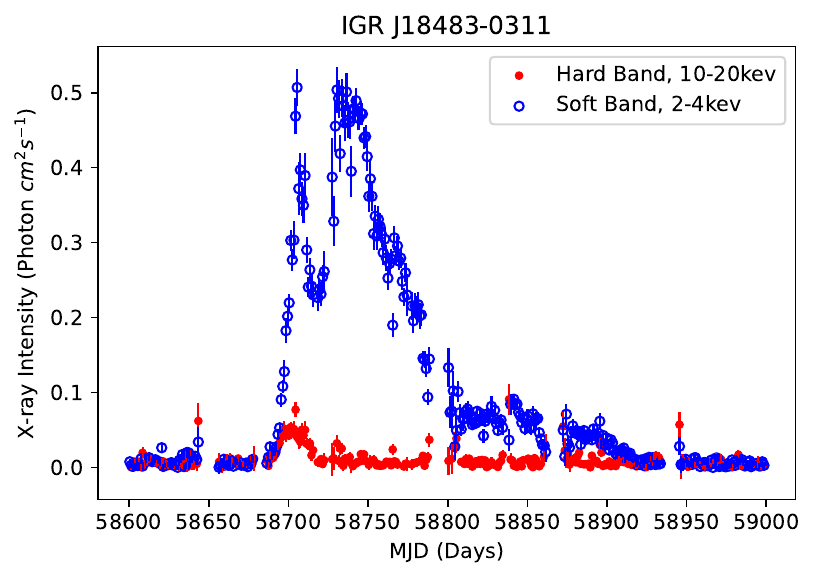}
    \includegraphics[scale=0.60]{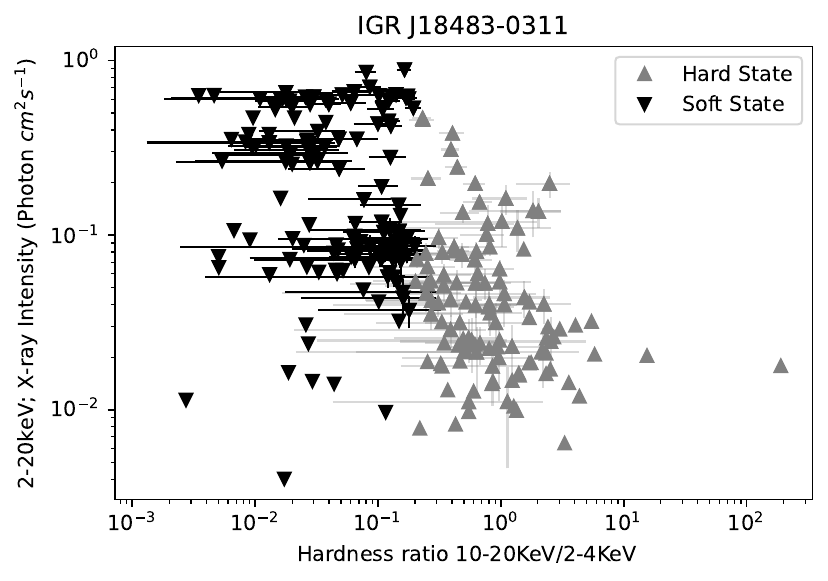}
    \captionsetup{labelformat=empty}  % Disable automatic "Figure X." prefix
    \caption{\textbf{Figure 42. }{\it Lightcurves and HID of IGR J18483–0311}: 
    Left panel: Lightcurves obtained with \maxi{} in the 2–4 keV (blue open circles) and 10–20 keV (red solid circles) bands.
    Right panel: Corresponding Hardness Intensity Diagram (HID), where hardness is defined as the ratio of 10–20 keV to 2–4 keV intensity and intensity as the 2–20 keV count rate. Grey triangles denote the hard state and black inverted triangles denote the soft state.}

    \label{fig: IGR J18483-0311}
\end{figure*}

IGR J18483-031 is classified as a NSXB. Two outburst intensity peaks can be seen in the lightcurve. The second peak can be observed midway through the descent of the first peak (Left panel of Fig. \ref{fig: IGR J18483-0311}). The HID demonstrates a q-diagram (Right panel of Fig. \ref{fig: IGR J18483-0311}).

\subsection{XTE J1807-294}
\begin{figure*}
    \centering
    \includegraphics[scale=0.60]{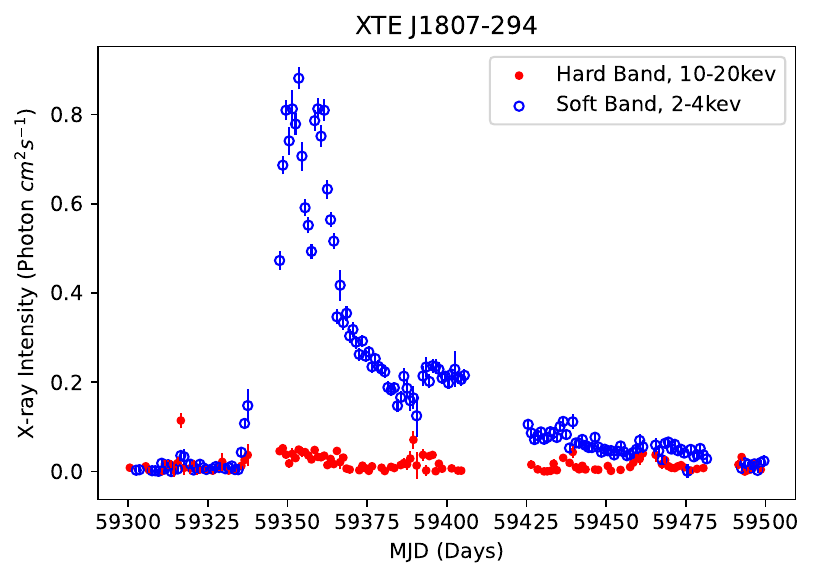}
    \includegraphics[scale=0.60]{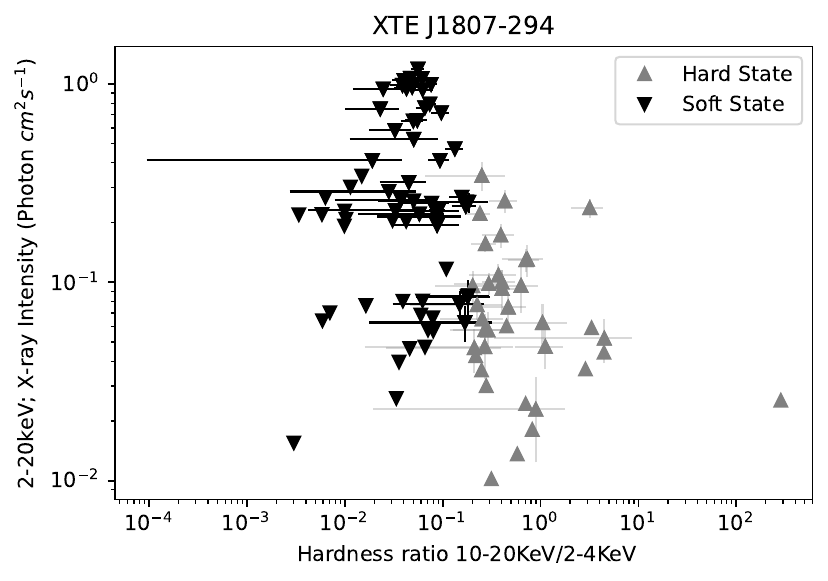}
    \captionsetup{labelformat=empty}  % Disable automatic "Figure X." prefix
    \caption{\textbf{Figure 43. }{\it Lightcurves and HID of XTE J1807–294}: 
    Left panel: Lightcurves obtained with \maxi{} in the 2–4 keV (blue open circles) and 10–20 keV (red solid circles) bands.
    Right panel: Corresponding Hardness Intensity Diagram (HID), where hardness is defined as the ratio of 10–20 keV to 2–4 keV intensity and intensity as the 2–20 keV count rate. Grey triangles denote the hard state and black inverted triangles denote the soft state.}

    \label{fig: XTE J1807−294}
\end{figure*}

XTE J1807-294 is an accreting millisecond pulsar with a mass range $\sim$1-2.5$M_\odot$. A typical outburst is visible during observation (Left panel of Fig. \ref{fig: XTE J1807−294}).

\subsection{XTE J1543-568}
\begin{figure*}
    \centering
    \includegraphics[scale=0.60]{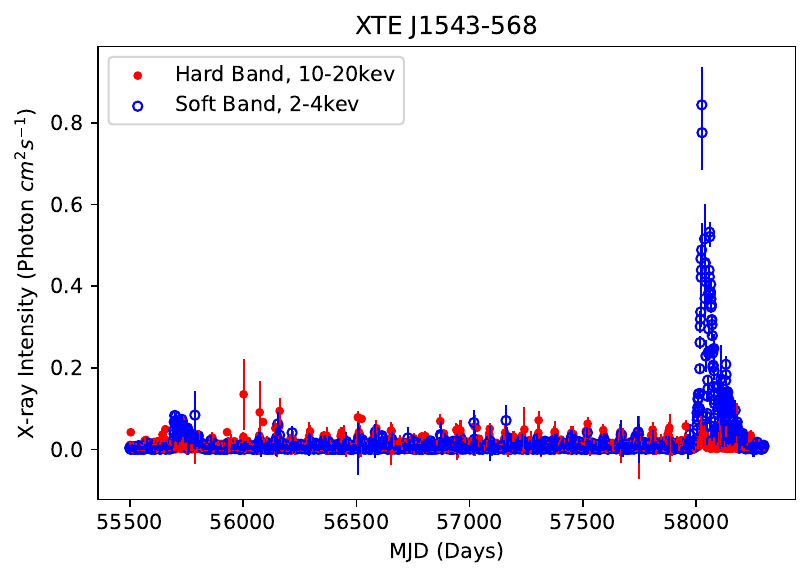}
    \includegraphics[scale=0.60]{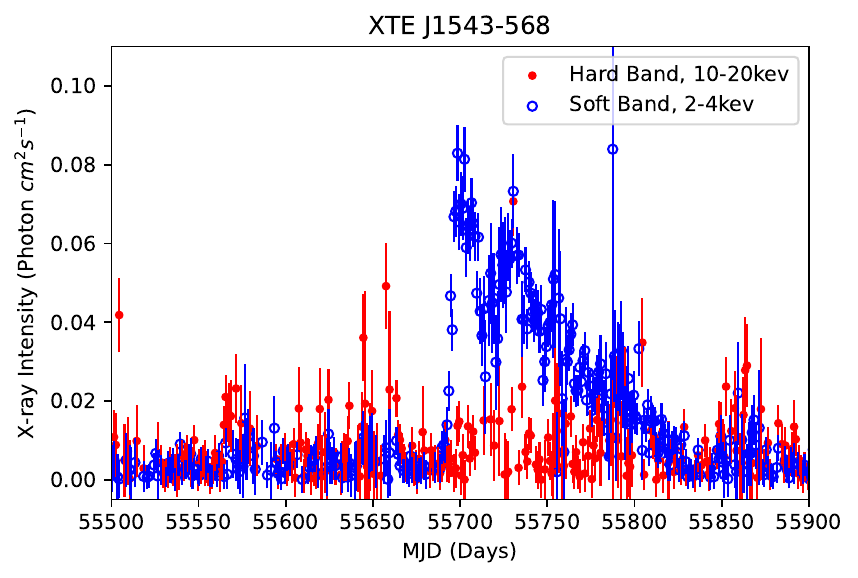}
    \includegraphics[scale=0.60]{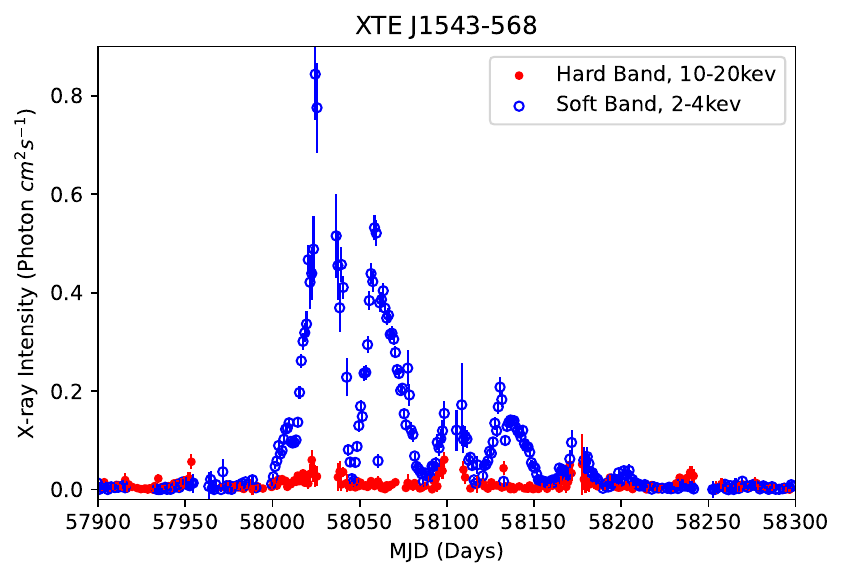}
    \includegraphics[scale=0.60]{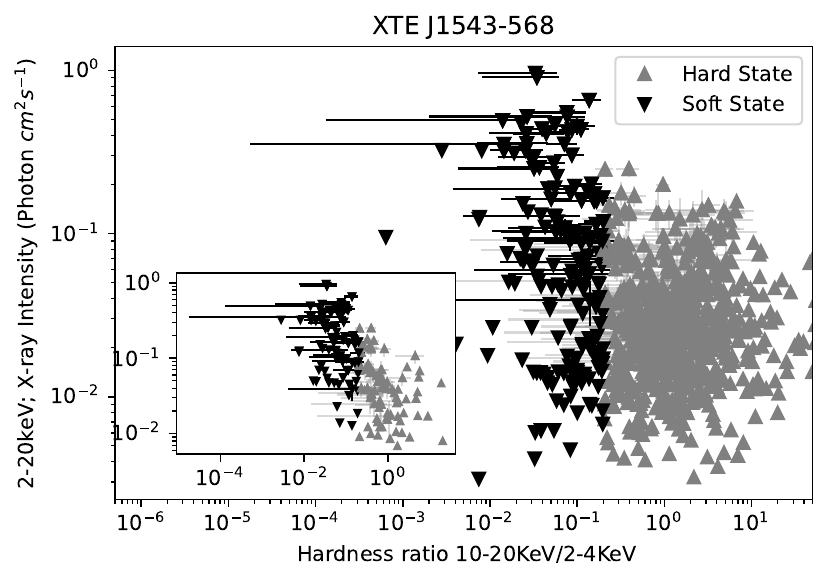}

    \captionsetup{labelformat=empty}  % Disable automatic "Figure X." prefix
\caption{\textbf{Figure 44. }{\it Lightcurves and HID of XTE J1543–568}: 
Top left panel: Lightcurves obtained with \maxi{} in the 2–4 keV (blue open circles) and 10–20 keV (red solid circles) bands.
Top right panel: Zoomed view of the initial hard-band dominated (failed) outburst. 
Bottom left panel: Zoomed view of the main outburst phase showing a sequence of re-brightening events with progressively decreasing peak intensities. 
Bottom right panel: Corresponding Hardness Intensity Diagram (HID) of the entire lightcurve, where hardness is defined as the ratio of 10–20 keV to 2–4 keV intensity and intensity as the 2–20 keV count rate. The inset shows the HID during the principal outburst.}

    \label{fig: XTE J1543-568}
\end{figure*}

XTE J15430-568 is classified as an NSXB. A major outburst and a failed outburst can be seen during the observation (Top left panel of Fig. \ref{fig: XTE J1543-568}). A series of outbursts with gradually decreasing intensity peaks can be observed from the zoomed in view of the outburst (Bottom left panel of Fig. \ref{fig: XTE J1543-568}).

\subsection{Terzan 5}
\begin{figure*}
    \centering
    \includegraphics[scale=0.60]{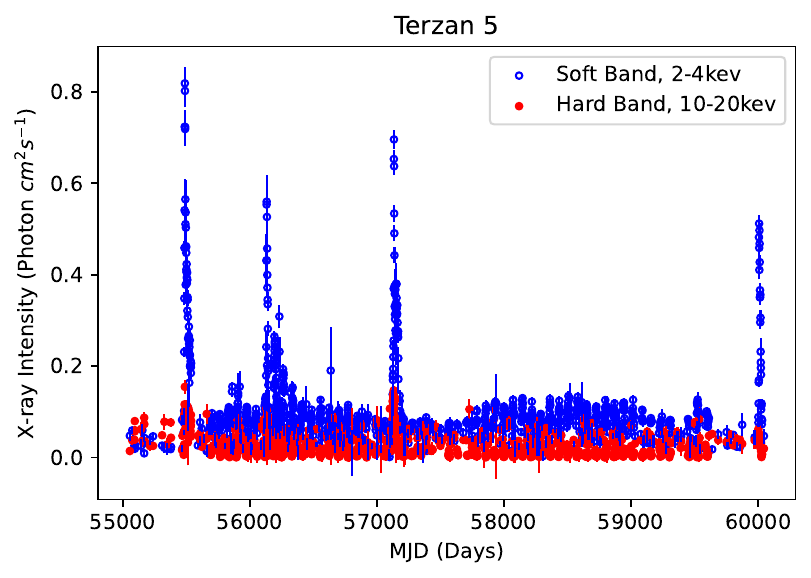}
    \includegraphics[scale=0.60]{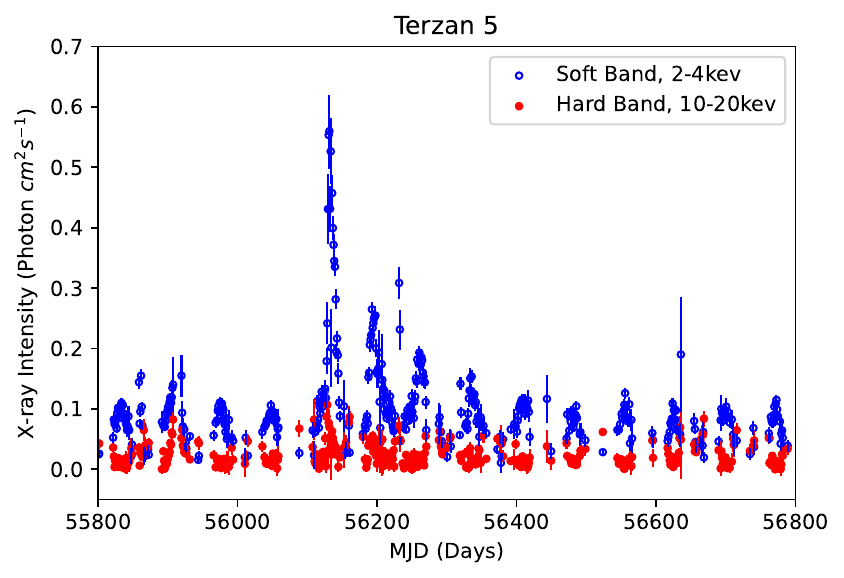}
    \includegraphics[scale=0.60]{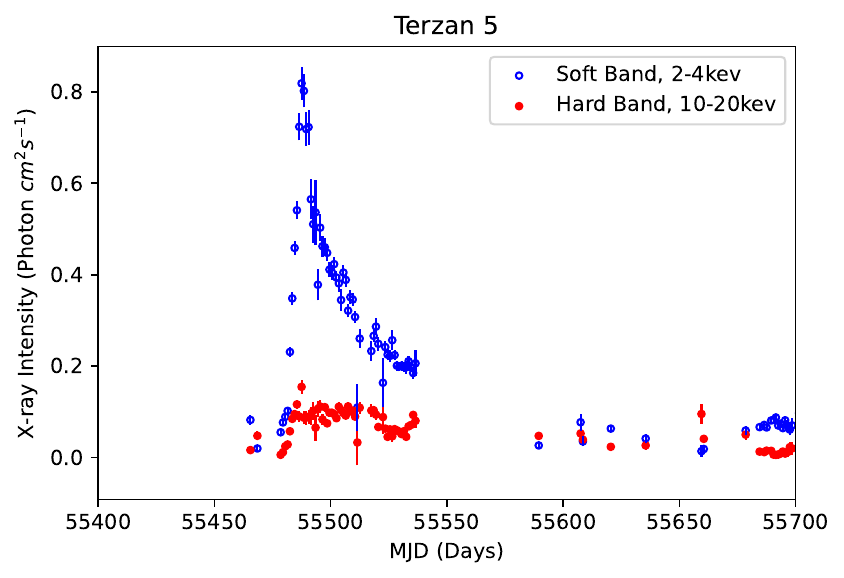}
    \includegraphics[scale=0.60]{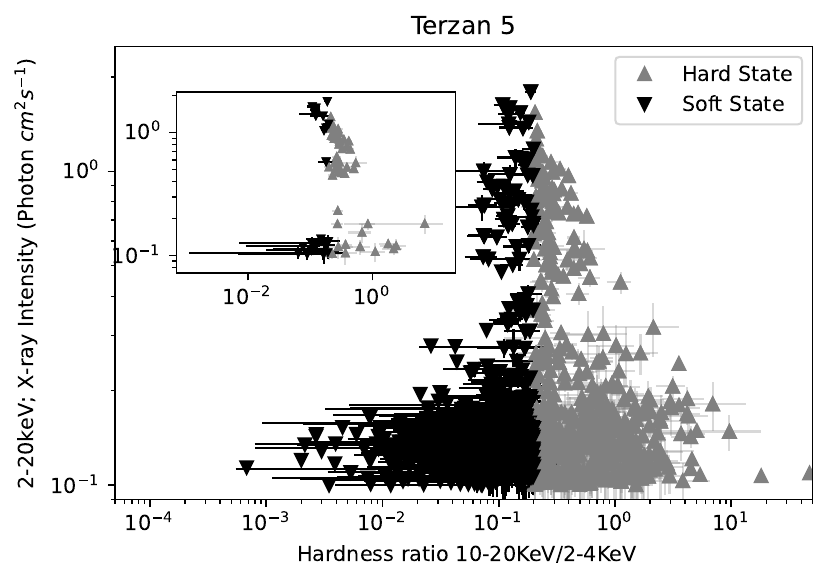}
    \captionsetup{labelformat=empty}  % Disable automatic "Figure X." prefix
    \caption{\textbf{Figure 45. }{\it Lightcurves and HID of Terzan 5}: 
    Top left panel: Lightcurves obtained with \maxi{} in the 2–4 keV (blue open circles) and 10–20 keV (red solid circles) bands.
    Top right panel: Zoomed view of the first outburst. 
    Bottom left panel: Further zoomed view of the same outburst highlighting its temporal evolution. 
    Bottom right panel: Corresponding Hardness Intensity Diagram (HID) of the entire lightcurve, where hardness is defined as the ratio of 10–20 keV to 2–4 keV intensity and intensity as the 2–20 keV count rate. The inset shows the HID during the first outburst.}

    \label{fig: Terzan 5}
\end{figure*}

Terzan 5 is one of three newly discovered millisecond pulsars reported. Four distinct intensity peaks can be observed during the observation (Top left panel of Fig. \ref{fig: Terzan 5}). Zooming onto a particular intensity peak, we can clearly see the outburst with a steep ascent followed by a gradual descent in intensity.

\subsection{XTE J1709–267}
\begin{figure*}
    \centering
    \includegraphics[scale=0.60]{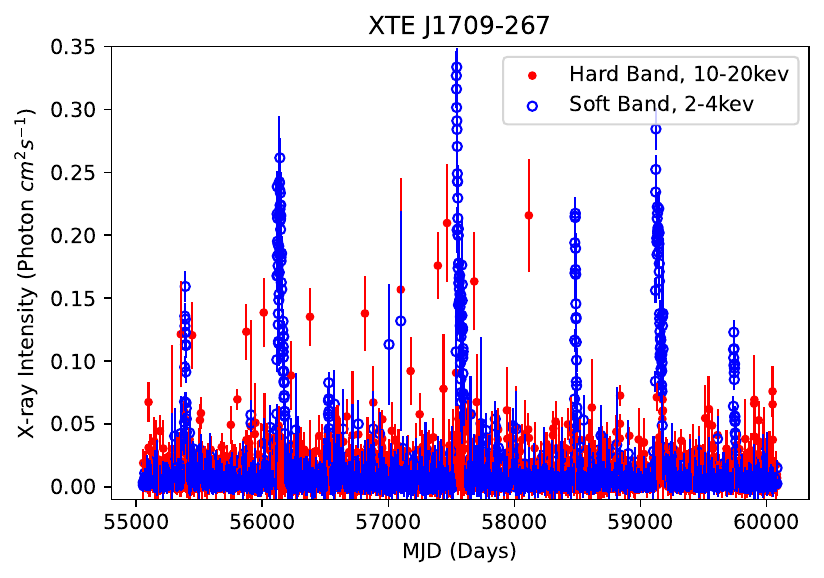}
    \includegraphics[scale=0.60]{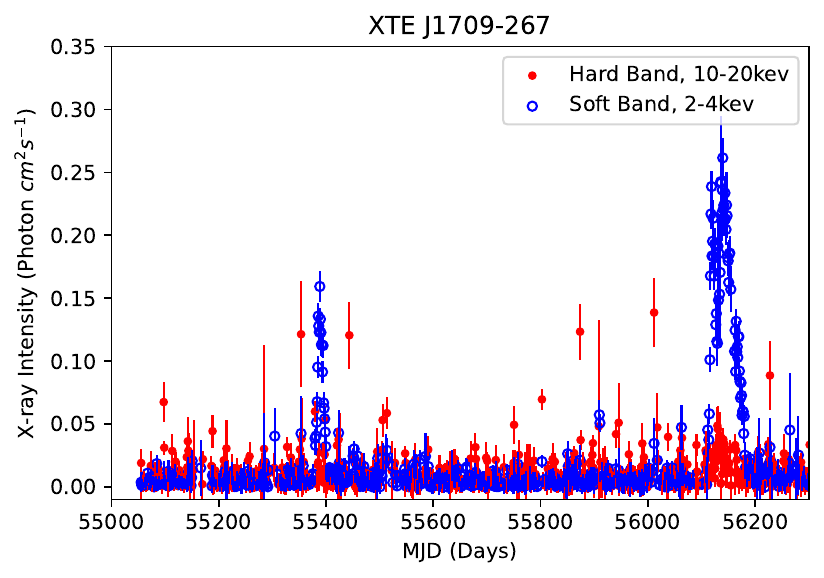}
    \includegraphics[scale=0.60]{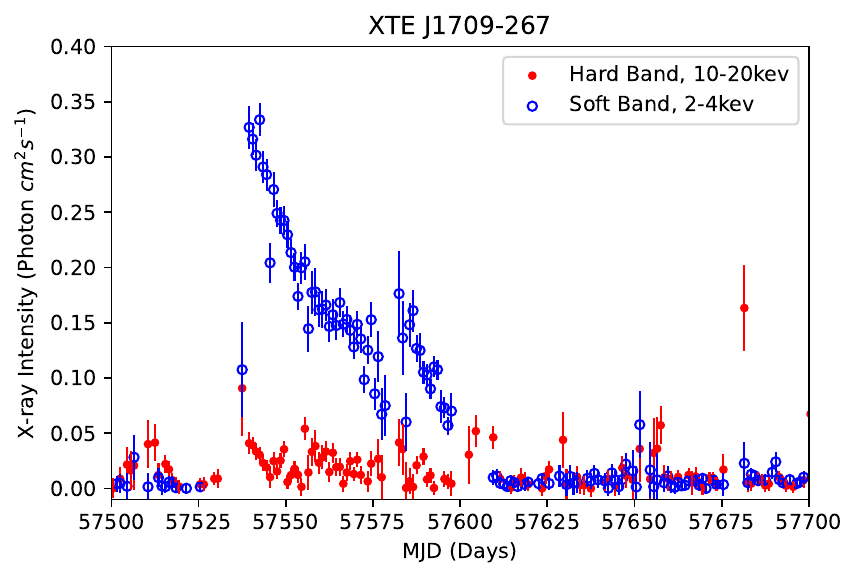}
    \includegraphics[scale=0.60]{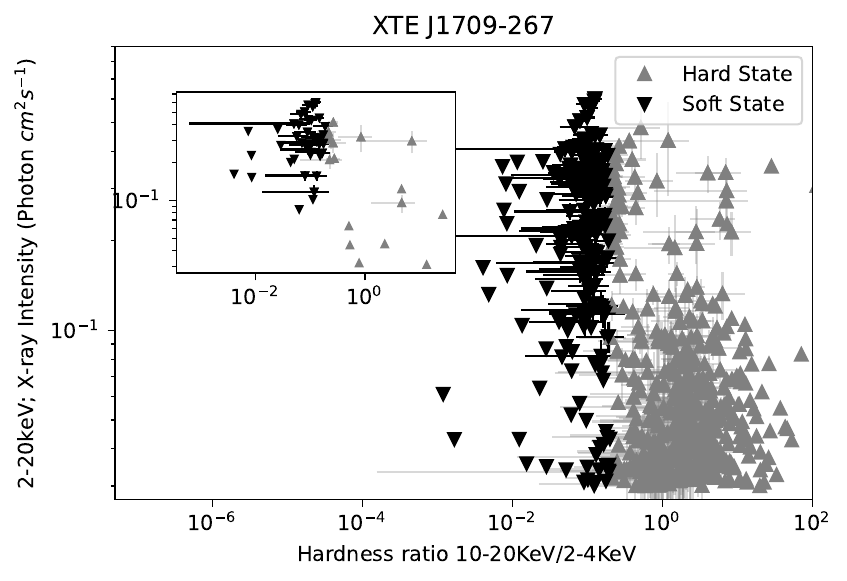}

    \captionsetup{labelformat=empty}  % Disable automatic "Figure X." prefix
\caption{\textbf{Figure 46. }{\it Lightcurves and HID of XTE J1709–267}: 
Top left panel: Lightcurves obtained with \maxi{} in the 2–4 keV (blue open circles) and 10–20 keV (red solid circles) bands.
Top right panel: Zoomed view of the first two intensity peaks. 
Bottom left panel: Zoomed view of the third outburst, characterized by a rapid rise and gradual decay in intensity. 
Bottom right panel: Corresponding Hardness Intensity Diagram (HID) of the entire light curve, where hardness is defined as the ratio of 10–20 keV to 2–4 keV intensity and intensity as the 2–20 keV count rate. The inset shows the HID during the third outburst.}

    \label{fig: XTE J1709–267}
\end{figure*}

This is the NSXB system classified with the mass of the compact object estimated to be $\sim$1.4$M_\odot$. Multiple intensity peaks due to outbursts can be observed during the observation period (Top left panel of Fig. \ref{fig: XTE J1709–267}). Zoomed in view of an intensity peak shows a typical outburst (Bottom left panel of Fig. \ref{fig: XTE J1709–267}).

\end{document}